\documentclass[amsmath]{aastex702}   

\usepackage{lmodern}      
\usepackage[utf8]{inputenc}
\usepackage[T1]{fontenc}
\usepackage{bm}           
\usepackage{booktabs}
\usepackage{graphicx}
\usepackage{xcolor}
\usepackage{nicefrac}
\usepackage{microtype}
\usepackage{marginnote}   
\usepackage{enumitem}
\usepackage[normalem]{ulem}

\let\fs\undefined


\makeatletter

\DeclareRobustCommand{\todo}[1]{%
  \ifdraftmode
    \textcolor{red}{\textbf{[TODO: #1]}}%
  \else
    \PackageError{macros}{Unresolved \string\todo{} in a final-mode build:
      #1}{Resolve it, or set \string\draftmodetrue\space to build a draft.}%
  \fi}
\DeclareRobustCommand{\citeslot}[1]{%
  \ifdraftmode
    \textcolor{purple}{[CITE: #1]}%
  \else
    \PackageError{macros}{Unresolved \string\citeslot{} in a final-mode build:
      #1}{Add the citation, or set \string\draftmodetrue\space to build a draft.}%
  \fi}
\makeatother

\DeclareMathOperator{\Var}{Var}
\DeclareMathOperator{\Cov}{Cov}
\DeclareMathOperator{\sd}{sd}
\newcommand{\E}{\mathbb{E}}
\newcommand{\R}{\mathbb{R}}
\newcommand{\dd}{\mathrm{d}}
\newcommand{\Nm}{\mathbf{N}}                 
\newcommand{\Id}{\mathbb{I}}
\newcommand{\Jop}{\mathcal{J}}               
\newcommand{\Dop}{\bm{\Delta}}               
\newcommand{\kv}{\mathbf{k}}
\newcommand{\xv}{\mathbf{x}}
\newcommand{\tauv}{\bm{\tau}}
\newcommand{\tautil}{\tilde{\tau}}
\newcommand{\that}{\hat{t}}
\newcommand{\fhat}{\hat{f}}
\newcommand{\Ahat}{\hat{A}}
\newcommand{\AMF}{\hat{A}_{\mathrm{MF}}}
\newcommand{\sigMF}{\sigma_{\mathrm{MF}}}
\newcommand{\sigemp}{\sigma_{\mathrm{emp}}}
\newcommand{\signorm}{\sigma_{\mathrm{norm}}}
\newcommand{\sigor}{\sigma_{\mathrm{oracle}}}
\newcommand{\signuis}{\sigma_{\mathrm{nuis}}}
\newcommand{\sigCRLB}{\sigma_{\mathrm{CRLB}}}
\newcommand{\etaB}{\eta_{B}}
\newcommand{\etaA}{\eta_{A}}
\newcommand{\etawi}{\eta_{\mathrm{wi}}}

\newcommand{\etalin}{\eta_{\mathrm{lin}}}
\newcommand{\fs}{f_{s}}
\newcommand{\xirms}{\xi_{\mathrm{rms}}}
\newcommand{\Scut}{S_{\mathrm{cut}}}
\newcommand{\Smin}{S_{\mathrm{min}}}
\newcommand{\sigc}{\sigma_{\mathrm{c}}}
\newcommand{\Nbeam}{N_{\mathrm{beam}}}
\newcommand{\mjb}{\mathrm{mJy\,beam}^{-1}}
\newcommand{\mum}{\ensuremath{\mu\mathrm{m}}}
\newcommand{\dnds}{\mathrm{d}N/\mathrm{d}S}
\newcommand{\avf}[1]{\langle #1\rangle_{\fhat}}   
\newcommand{\Varf}[1]{\Var_{\fhat}\!\left[#1\right]}
\newcommand{\Pbar}{\bar{P}}
\newcommand{\Phat}{\hat{P}}
\newcommand{\rung}[1]{\texttt{#1}}          
\newcommand{\code}[1]{\texttt{#1}}

\newcommand{\mWhite}{white noise}
\newcommand{\mRed}{red ($1/f^{3}$) noise}
\newcommand{\mReal}{red\,+\,white noise}
\newcommand{\mAniso}{fixed-direction anisotropy}
\newcommand{\mTilt}{spectral-tilt mixture}          
\newcommand{\mTiltS}{slope-only tilt mixture}       
\newcommand{\mTiltA}{amplitude-only mixture}        
\newcommand{\mRot}{random-orientation anisotropy}   
\newcommand{\mMedian}{row-median residual}          
\newcommand{\mCrossS}{scan crossings (symmetric)}   
\newcommand{\mCrossP}{scan crossings (positive)}    
\newcommand{\mGlitch}{sub-threshold glitches}       
\newcommand{\mPCA}{PCA leakage}                     
\newcommand{\mConf}{source confusion}               

\newcommand{\ie}{i.e.}

\newcommand{\PaperII}{Paper~II}

\newif\ifdraftmode
\draftmodefalse         

\definecolor{linknavy}{rgb}{0.05,0.25,0.55}
\definecolor{citegreen}{rgb}{0.05,0.35,0.20}
\definecolor{urlplum}{rgb}{0.45,0.10,0.35}
\hypersetup{
  linkcolor=linknavy,
  citecolor=citegreen,
  urlcolor=urlplum,
  pdftitle={Beyond the BLUE I: the advantage ceiling},
  pdfsubject={astro-ph.IM, astro-ph.CO, stat.ML},
  pdfauthor={Kaustuv Basu},
  pdfkeywords={matched filter, Cramer-Rao bound, convolutional neural networks, non-Gaussian noise, submillimeter surveys},
}

\shorttitle{Beyond the BLUE I: the advantage ceiling}
\shortauthors{Basu}
\submitjournal{ApJS}

\begin{document}

\title{Beyond the BLUE I: the advantage ceiling -- how much can any estimator beat the matched filter in mm/submm survey data?}

\author[orcid=0000-0001-5276-8730,gname=Kaustuv,sname=Basu]{Kaustuv Basu}
\affiliation{Argelander Institute for Astronomy, University of Bonn, Auf dem H\"ugel 71, D-53121 Bonn, Germany}
\email[show]{kbasu@uni-bonn.de}

\begin{abstract}
Papers claiming that convolutional neural networks outperform the matched filter are now common in astronomy. Since that filter is the best linear unbiased estimator (BLUE) for any noise of given covariance, and the minimum-variance unbiased estimator under Gaussian noise, such a claim requires that the noise covariance varies from image to image or the noise is non-Gaussian. We introduce the \emph{advantage ceiling} $\eta\ge1$, which quantifies both cases: the Fisher information for the amplitude in units of the matched filter's, computable from noise-only simulations before any network is trained, and a bound on the variance of any estimator unbiased at each true amplitude. We compute $\eta$ for a suite of millimeter/submillimeter survey noise models and for compact and extended source templates, using a variational score-matching ladder whose rungs are estimator classes of increasing statistical order, from linear to convolutional. With an instrumental white-noise floor, Gaussian noise returns $\eta=0.94$--$1.06$ ($\pm0.04$); covariance mixtures give exact ceilings of $10.2$ (spectral tilt, extended source), $2.1$ (leaked components) and $1.1$--$1.8$ otherwise; residual artifacts give $\eta\le1.09$. Source confusion gives the largest advantage, $\eta\ge3.8$ (extended) and $\ge2.4$ (compact). Three unmodelled channels -- a zeroed DC mode, a missing white floor and floating-point rounding -- can manufacture spurious advantages of up to two orders of magnitude. A ResNet regressor is $6$--$15$\,\% \emph{less} efficient than the matched filter on Gaussian noise once prior shrinkage is divided out, and realizes a factor $1.8$ of the $10.2$ available on the spectral-tilt mixture; a companion paper tests such regressors against these ceilings.
\end{abstract}

\keywords{\uat{Astrostatistics techniques}{1886} --- \uat{Fisher's Information}{1922} ---
  \uat{Convolutional neural networks}{1938} ---
  \uat{Astronomy image processing}{2306} --- \uat{Millimeter astronomy}{1061} ---
  \uat{Submillimeter astronomy}{1647} --- \uat{Cosmic microwave background radiation}{322}}

\section{Introduction}
\label{sec:intro}

\subsection{The question}
\label{sec:intro_question}

Extracting the amplitude of a source of known shape from a noisy map is one of the oldest problems of
observational astronomy, and the matched filter (abbreviated MF in some domains) is its oldest
and most used solution. Its optimality is a theorem with two nested scopes. For \emph{any} noise of a
given covariance it is the best linear unbiased estimator (BLUE) -- the minimum-variance member of the
linear class, by the Gauss--Markov theorem. When that noise is Gaussian with a \emph{known} covariance
the restriction to linear estimators can be lifted, and it becomes the minimum-variance unbiased
estimator (MVUE) outright, attaining the Cram\'er--Rao bound
\citep{Kay1993,VanTrees1968,LehmannCasella1998}. Going beyond the BLUE therefore means leaving one of
those two hypotheses behind, and this paper is about how far that can be taken.

In the setting of this paper, a millimeter or submillimeter survey, the source is a beam-convolved point source or an extended
profile, and the noise is whatever the atmosphere, the detectors and the map-maker leave behind. The
matched filter was brought into this setting early, for cluster detection through the Sunyaev--Zeldovich (SZ)
effect \citep{HaehneltTegmark1996}, and developed since into multifrequency, multiscale and all-sky
forms \citep{Herranz2002,Melin2006,Schaefer2006}. This is the estimator that every catalog pipeline runs
first: it is what built the SZ cluster catalogs of ACT, SPT and \emph{Planck}
\citep{Hasselfield2013,Bleem2015,PlanckXXVII2016}, and it extends without modification to joint detection
across surveys \citep{Tarrio2018}.

The same estimator is the workhorse of gravitational-wave astronomy, where banks of matched filters are run over the
strain in low latency \citep{Cannon2012,Brown2012}. It is there that the most careful comparisons with
neural networks have been made, demonstrating an equivalence rather than an advantage: a convolutional
classifier reproduces the sensitivity of a matched-filter search on the same data \citep{Gabbard2018},
and another reaches comparable sensitivity at a small fraction of the cost \citep{GeorgeHuerta2018},
the gain there being latency and robustness to glitches rather than variance. Both results are
evidence for the centrality of matched filtering.

Nevertheless, in the last few years this classical tool has acquired a large family of nonlinear
competitors, and the claims made for them are less uniform. Convolutional networks are now trained to
find and to measure point sources in microwave maps \citep{Bonavera2021,Casas2022}, to detect sources in
interferometric images \citep{Sadr2019}, to detect clusters and to extract the thermal Sunyaev--Zeldovich
signal \citep{Bonjean2020,Pratt2025}, to find clusters and groups in optical surveys
\citep{Grishin2023,Ma2025}, to classify X-ray cluster candidates \citep{Kosiba2020}, and to generalize the
matched filter itself \citep{Yan2022}. Many of these papers report that their network beats a
filter-based baseline on some figure of merit. We do not adjudicate the individual claims here -- they
differ in estimand, in noise baseline and in what is held fixed, and a fair audit of each is a paper of its own
-- but the confusion they collectively produce is real, and it is what the current paper is written to address.
The optimality theorem states 
 that this is possible only if one of its hypotheses fails: the noise must vary in covariance from
image to image, or it must be non-Gaussian, or the comparison must not be between unbiased estimators.
The last of these is a confound, not an advantage -- a network trained by least squares learns the
posterior mean, which is shrunk toward the prior mean and has a smaller mean-squared error than any
unbiased estimator for reasons that have nothing to do with the noise \citep{JamesStein1961,Efron2011}.
The first two are the genuine channels, and the question any astronomical survey should ask before it trains anything
is therefore not ``does a network beat the matched filter?'' but ``how much is there to gain, from these
two channels, for this noise and this source, and by what class of estimator?''

Our two-paper series answers that question in the context of mm/submm survey data, using the noise as the sole
discriminator between the linear and the neural estimator: the template is known exactly, the estimand
is a single amplitude at a known position, and every other route to a sub-optimal matched filter --
template mismatch, position uncertainty, the faint-end deboosting problem
\citep{Eddington1913,HoggTurner1998,Coppin2006,Vieira2010} -- is deliberately held fixed. The current paper is the theory and
the prediction; \PaperII\ is the test from a convolutional network.

\subsection{This paper}
\label{sec:intro_thispaper}

We introduce a single number, the \emph{advantage ceiling} $\eta \ge 1$: the Fisher information for the
amplitude in units of the matched filter's. It is computable from noise-only simulations before any
network is trained; it bounds the variance reduction available to any estimator that is unbiased at each
true amplitude; and its excess over unity is the projection of an \emph{excess-score operator} -- the
departure of the noise's score from that of a Gaussian of the same covariance -- onto the source
template's band in Fourier space. That last property makes the advantage template-dependent:
the same noise can leave nothing for a compact source and a factor of ten for an extended one.
$\eta$ can also be read as a survey-time factor: wherever the noise integrates down, an ideal nonlinear
estimator is worth $\eta$ times the survey time for that source class -- time spent on area at fixed
depth, not on depth (\S\ref{sec:eta_def}) -- with the exception of source confusion, which does not
integrate down at all.

\begin{figure}[t]
  \centering
  \includegraphics[width=\textwidth]{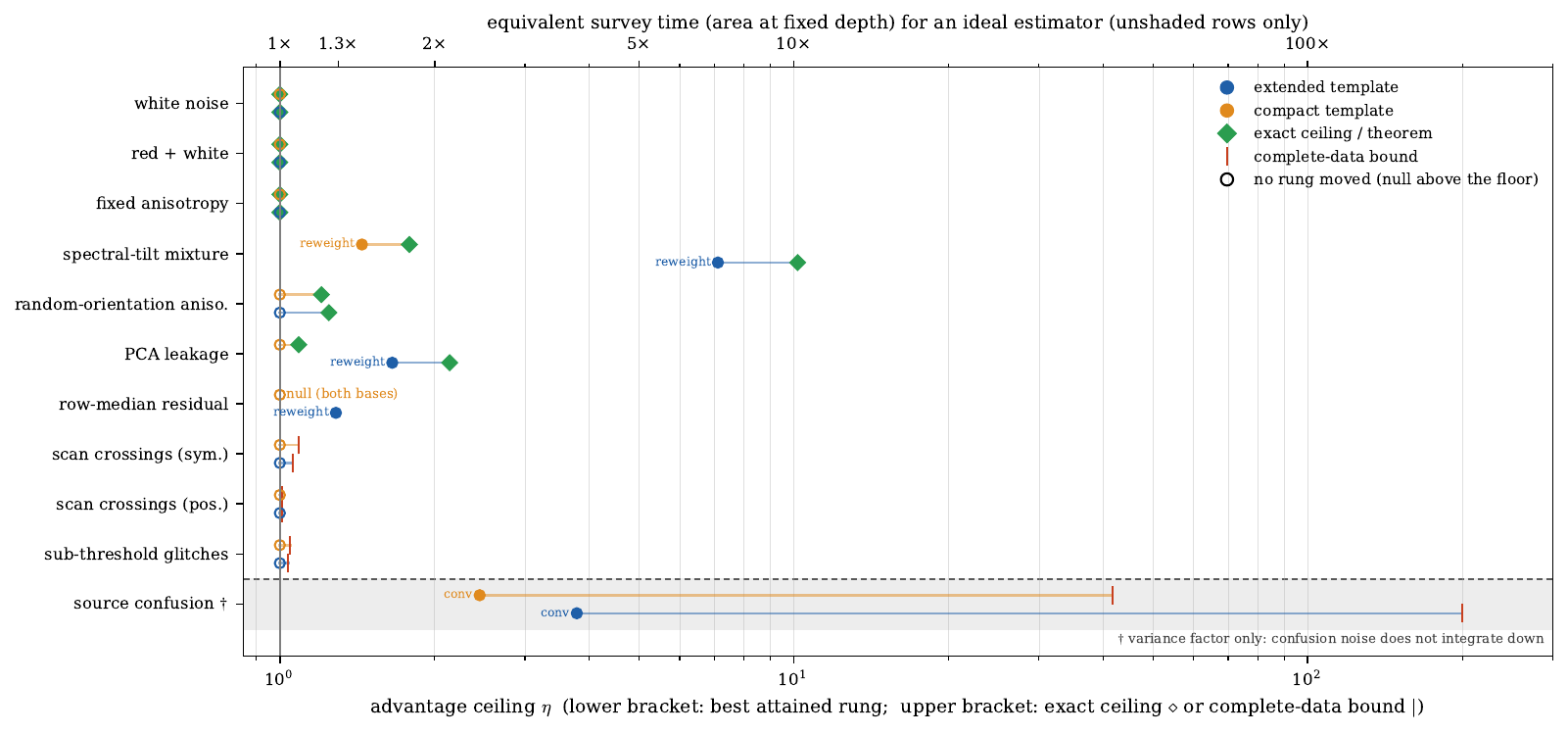}
  \caption{Every ceiling of the realistic configuration on one axis (three of the four Gaussian controls are
  shown). For each model and template the
  upper bracket is the exact ceiling (diamond; a theorem for the Gaussian controls) or the
  complete-data bound (bar), and the lower bracket the best attained rung (filled circle, labeled) or
  the anchor where no rung exceeded the $1.05$ sensitivity floor (open circle; the random-orientation
  \rung{reweight} gains of 3--6\,\% fall below it). The upper axis gives the survey reading of \S\ref{sec:eta_def} -- the equivalent survey time (area at
  fixed depth) -- which applies only to noise that integrates down; the shaded, daggered confusion rows are excluded from it, and there $\eta$ is a
  variance factor that no amount of integration buys. The
  row-median compact cell is labeled a null on both bases: its ceiling statement rests on the ladder
  alone, and the calibration of \S\ref{sec:sensitivity} makes it an informed null rather than an
  uninformative one. Where a cell's upper bracket is distant or absent, its ceiling is bounded loosely
  or not at all.}
  \label{fig:ceilingchart}
\end{figure}

We compute $\eta$ for a suite of noise models built from 
 real-world examples of mm/submm pipelines --
atmospheric red noise, scan-synchronous anisotropy, patch-to-patch spectral variation, row-median
destriping residuals, sub-threshold glitches and scan crossings, leaked principal components, and source
confusion -- for a compact and an extended source, by methods of increasing generality: exact
latent quadratures where the noise is a mixture of Gaussians, complete-data bounds and polyspectral
expansions where it is non-Gaussian, and a variational score-matching ladder whose rungs (linear,
mixture matched filter, quadratic, cubic, convolutional) correspond to estimator classes of increasing
statistical order and whose value is a certified lower bound on $\eta$. The results are summarized in one
chart (Fig.~\ref{fig:ceilingchart}): Gaussian noise of known covariance returns $\eta = 1$ on
full-size maps; covariance mixtures give exact ceilings of $10$, $2$ and $1.1$--$1.8$, most of which a
mixture matched filter of $1.2$--$1.8$\,k parameters already attains; residual artifacts give nothing; and
source confusion, the one model whose non-Gaussian component \emph{is} the noise, gives the largest
measured advantage, $\eta \ge 3.8$ (extended) and $\ge 2.4$ (compact), half of it already at the
bispectrum rung and the rest only above it. Along the way we find that three unphysical numerical channels can each manufacture a
spurious advantage, ranging from a factor of a few to two orders of magnitude, and we set out the
reporting framework that keeps them out of a network-versus-filter comparison.

Although everything here is worked out for the matched filter of a single map, the estimator class is
wider than that. The internal linear combination used for multi-frequency component separation is the
same object with the source template replaced by a frequency spectrum and the noise covariance by the
per-mode frequency--frequency covariance. The correspondence is exact and has been drawn from both
sides: the constrained matched filter \citep{Erler2019} is mathematically identical to the constrained
ILC \citep{Remazeilles2011}, and the multifrequency matched filter is recovered by substituting ILC
weights into it \citep{Zubeldia2023}. The construction of this paper therefore carries over to the ILC family unchanged
-- the same ceiling, the same two channels, the same ladder. Its \emph{numbers} do not: the ceilings
computed here are those of mm/submm survey noise on a single map, and a foreground-dominated
multi-frequency ensemble is a different noise problem, which we will address in future work.

\subsection{What this paper builds on}
\label{sec:intro_literature}

The individual pieces of this framework are old and well-established, 
and part of the value of this paper is in the assembly. The
observation that Gaussian noise is the worst case for estimation -- that the matched filter's variance is
an upper bound on the attainable one -- is the Cram\'er--Rao inequality applied to the matched filter
itself: as an unbiased estimator its variance cannot fall below the inverse Fisher information, with
equality exactly when the score is linear in the data, \ie\ when the noise is Gaussian
\citep{Kay1993,LehmannCasella1998}. 
Our $\eta$ is that inequality turned into a ratio and projected onto a
template; in information theory, the same fact appears as a corollary of Stam's inequality
\citep{Stam1959,CoverThomas2006}. The estimator that attains it in the small-signal
limit, the score-function nonlinearity followed by the matched filter, is the locally optimal detector of
the classical theory of signal detection in non-Gaussian noise \citep{Kassam1988}, with Fisher
information as the metric of locally optimal processing \citep{Duan2012}.

Our ladder's Volterra rungs are truncations of that nonlinearity; the same construction appears in astronomy as the
non-Gaussian matched filter for transit searches \citep{RobnikSeljak2021}.
The nonlinear estimators of CMB science are its most successful instances: the quadratic estimators for
lensing \citep{HuOkamoto2002,HirataSeljak2003} and for primordial non-Gaussianity \citep{KSW2005} are the
quadratic rung of our ladder built by hand, and the field-level Cram\'er--Rao bound
\citep{ChenGreenLee2026} is the same ceiling used for a different estimand.
Forecasting what an experiment can
achieve with a Fisher matrix, before any data exist, is older still in CMB work \citep{Tegmark1997};
our $\eta$ applies it to an estimator class rather than to a parameter set. The ceiling also has a reading
in information units: in the flat-prior, many-modes limit, $\tfrac12\ln\eta$ nats is the expected
information gain that the matched filter leaves unclaimed, by the classical correspondence between Fisher
information and expected information gain \citep{Lindley1956} and in the ``achieved information gain''
language of \citet{Ensslin2025AIG}; see \S\ref{sec:verdict}.

On the estimation side, 
the methodology is built from three relatively recent strands of papers. Learned estimators can be
constrained to be unbiased, and the bias-corrected loss we use for the network of \S\ref{sec:teaser}
follows \citet{DiskinEldarWiesel2023}; the biased Cram\'er--Rao bound with which we certify a network's
shrinkage is from \citet{Eldar2004}. Score-based generative models allow the Fisher information, and
therefore a Bayesian Cram\'er--Rao bound, to be estimated from samples \citep{SongErmon2019,ScopeCrafts2025},
and a Cram\'er--Rao bound has been learned from data before any estimator is built \citep{Habi2023}.
Our variational
identity does this with a score-matching objective \citep{Hyvarinen2005,Vincent2011}, restricted
to the one direction (the amplitude) -- a projected objective in the spirit of sliced score matching
\citep{Song2020sliced} -- making it cheap enough to run for a whole noise
suite; that the score is the compression of the data that preserves Fisher information is also the basis
of score-based data compression in cosmology \citep{AlsingWandelt2018}. And the locally optimal nonlinearity for weak-signal detection under arbitrary noise has
been learned directly \citep{Zschetzsche2026},
which is the detection-side counterpart of our conv rung. The
shrinkage confound is Stein's paradox, in its empirical-Bayes form \citep{Robbins1956,Efron2011}, and the
identity that lets a posterior mean be read as a score -- Tweedie's formula \citep{Miyasawa1961} -- is
what connects the regression network's output to the same score operator that bounds it.

Source confusion has a long history in astronomical image analyses \citep{Scheuer1957,Condon1974}. 
The confusion model in our paper is a marked
Poisson process \citep{DaleyVereJones2003} whose cumulants are given by Campbell's theorem, and the
ceiling we compute quantifies how much a nonlinear estimator can gain from the non-Gaussianity of
confusion. The atmospheric and
scan-synchronous models draw on \citet{Sayers2010} and \citet{Aguirre2011}; the row-median and
leaked-component models on the map-making and destriping practice of ACT \citep{Dunner2013,Naess2025}. The comparison with the
gravitational-wave literature is instructive in this regard: there the matched filter is compared
against networks on a background that is often taken to be Gaussian, stationary and of known
covariance, and the framework of this paper says in advance that nothing but speed can be gained in
that setting -- which is what has been found on simulated Gaussian noise \citep{Gabbard2018}. When real
detector data are used the background is neither stationary nor Gaussian, $\eta > 1$ becomes possible in
principle, and the robustness to glitches reported there \citep{GeorgeHuerta2018} is of the kind this framework
would allow for such a background, though we have not applied it there.

\subsection{What is new}
\label{sec:intro_novelty}

Three things are new here, as far as we are aware. First, the framework is a
\emph{data-independent prediction}: $\eta$ is a property of the noise ensemble and the template alone,
prior-free and architecture-free. Thus, the \emph{ceiling} of any estimator -- a network, a Volterra filter, a
hand-built mixture filter -- is fixed before it is built, and the certified band on which a network's
advantage may legitimately be reported is fixed with it. Second, the excess-score decomposition orders
the noise properties into tiers that we have not seen used in the astronomical literature: Tier~0, Gaussian
of known covariance, where $\eta = 1$ is a theorem; Tier~1, Gaussian covariance mixtures, where $\eta$ is
exact by quadrature and splits into a coherent channel (band-level variation, removable by a weight
map) and an incoherent one (within-band shape, requiring a per-image filter); and Tier~2, non-Gaussian
noise, where $\eta$ is bracketed between a ladder and a complete-data bound. 
The tiers factorize: the ceiling of a Tier~2 component on a Tier~1 background is, to
leading order, a product. This allows our suite of twelve noise models to reduce to 
three mechanisms -- a coherent channel, an incoherent channel, and polyspectral structure -- each model being one or two of them in combination.
Third, the numbers themselves: to our knowledge no previous work has computed the
attainable advantage over the matched filter for the noise of mm/submm surveys, and the two major conclusions --
that source confusion is the only place where a deep nonlinear estimator earns a factor above two, and
that most of the covariance-mixture advantage is taken by a filter with two thousand parameters -- are
a practitioner's take-away from the paper.

\subsection{Outline, and Paper II}
\label{sec:intro_outline}

\S\ref{sec:framework} sets up the estimation problem, the three flavors of unbiasedness that a
network-versus-filter comparison must distinguish, and the reference ladder against which a network is
to be reported. \S\ref{sec:ceiling} defines the ceiling, proves its properties, and develops the Tier~1
and Tier~2 theory, including the variational identity behind the ladder. \S\ref{sec:computing} describes
the computational methods and the null battery that validates them; \S\ref{sec:noise_models} the
noise suite and the white-floor convention that makes it realistic. The results follow by tier:
Gaussian controls in \S\ref{sec:results_gaussian}, with a first ResNet regressor evaluated under the
paper's own certification as a teaser of \PaperII\ (\S\ref{sec:teaser}); covariance mixtures in
\S\ref{sec:results_mixtures}, including the attainability of their ceilings by a constructed estimator
and by the same network; non-Gaussian artifacts in \S\ref{sec:results_nongaussian}; and source confusion
in \S\ref{sec:results_confusion}. \S\ref{sec:unmodelled} is a cautionary chapter on the three unmodelled
channels and on the calibration of the ladder against an exactly solvable model. \S\ref{sec:discussion}
gives the verdict in survey time, the pre-registration workflow, and the reporting rules;
\S\ref{sec:conclusions} concludes. The appendices collect derivations (\ref{app:derivations}), the
pipeline and its failure modes (\ref{app:pipeline}), the null battery (\ref{app:nullbattery}), the
confusion model (\ref{app:confusion}) and the master table of every ceiling computed (\ref{app:master}). 
The entire codebase of this paper -- including the $\eta$-computation framework, noise-generation scripts, and the null-test battery -- is available in a public GitHub repository\footnote{\url{https://github.com/kmbasu/Beyond-the-BLUE-Paper-I}}
and archived on Zenodo \citep{BeyondTheBLUE2026code}.

\PaperII\ trains convolutional neural networks 
against the ceilings presented here, 
with the bias-corrected training and the certified-band reporting that \S\ref{sec:reporting}
prescribes. Its questions are correspondingly sharper than the one the literature has been asking: does a
generic network reach an exact Tier~1 ceiling, and does it do so by discovering the mixture filter
unaided; and what does it find in confusion above the bispectrum rung, where the ceiling of this paper is
a lower bound. The two extensions of \S\ref{sec:intro_thispaper} -- the internal linear combination and the matched
multifilter -- are set out in \S\ref{sec:conclusions} and left to a future paper.

\section{Framework: the matched filter, the network, and three meanings of ``unbiased''}
\label{sec:framework}

This section presents the basic notation and states the \emph{four results} on which everything else depends: (i) the
matched filter is the unique minimum-variance unbiased estimator under Gaussian noise of known
covariance (\S\ref{sec:mf}); (ii) a network trained with a squared-error loss estimates a posterior mean,
so that most of what looks like an advantage is prior shrinkage (\S\ref{sec:cnn}); (iii) ``unbiased'' has
three distinct meanings, and that only one of them is the deployment requirement
(\S\ref{sec:flavours}); and (iv) the reference estimators against which a network must be judged form a
ladder with one, and only one, unconditional floor (\S\ref{sec:ladder}). Each result is stated first in
words and then in the form we use later. Readers who are well-versed with matched filters may find
\S\ref{sec:mf} familiar; \S\ref{sec:flavours} is the part that is not in the textbooks, and the
certification built on it in \S\ref{sec:cnn} is the core of \PaperII.

\subsection{Model and notation}
\label{sec:notation}

A flattened image $d \in \R^{N}$ of $N = N_{\rm pix}$ pixels is modeled as
\begin{equation}
  d \;=\; A\,\tauv \;+\; n, \qquad \tauv \ \text{known}, \quad n \sim p_n(\cdot),\quad
  \E[n] = 0, \quad \Cov(n) = \Nm .
  \label{eq:model}
\end{equation}
The scalar $A \ge 0$ is the source amplitude to be estimated, $\tauv$ the known source template
(a point-spread function for a compact source, a beam-convolved profile for an extended one), and $n$ the
noise, whose distribution $p_n$ is the subject of this paper. Throughout, the template is assumed known
exactly: template mismatch and template-family uncertainty are a separate route to matched-filter
sub-optimality (the filter-bank or matched-multifilter problem) and are excluded here. Consequently
\emph{the noise is the only channel through which any estimator can beat the matched filter}.

For stationary noise the covariance is diagonal in Fourier space, with entries $P(\kv)$, the noise power
spectrum, and $\tautil(\kv)$ is the template transform. The matched filter and its variance are
\begin{equation}
  \AMF \;=\; \frac{\tauv^{\dagger}\Nm^{-1} d}{\tauv^{\dagger}\Nm^{-1}\tauv}
        \;=\; \frac{\sum_{\kv}\tautil^{*}_{\kv}\tilde d_{\kv}/P(\kv)}{\sum_{\kv}|\tautil_{\kv}|^{2}/P(\kv)},
  \qquad
  \sigMF^{2} \;=\; \big(\tauv^{\dagger}\Nm^{-1}\tauv\big)^{-1}
        \;=\; \Big(\sum_{\kv}\frac{|\tautil_{\kv}|^{2}}{P(\kv)}\Big)^{-1}.
  \label{eq:mf}
\end{equation}
Two functions of the true amplitude characterize any estimator $\Ahat$: the \emph{conditional bias}
$g(A) \equiv \E[\Ahat\,|\,A] - A$ and the \emph{conditional variance} $V(A) \equiv \Var(\Ahat\,|\,A)$, both
averaged over noise realizations at fixed $A$. Where the noise is a mixture -- a different covariance on
every image -- $z$ denotes the per-image \emph{latent} (the spectral slope and amplitude of the
atmospheric residual, the orientation of a scan ridge, the positions and amplitudes of glitches) and
$\Nm_{z}$ the covariance conditional on it.

\begin{figure}[t]
  \centering
  \includegraphics[width=\textwidth]{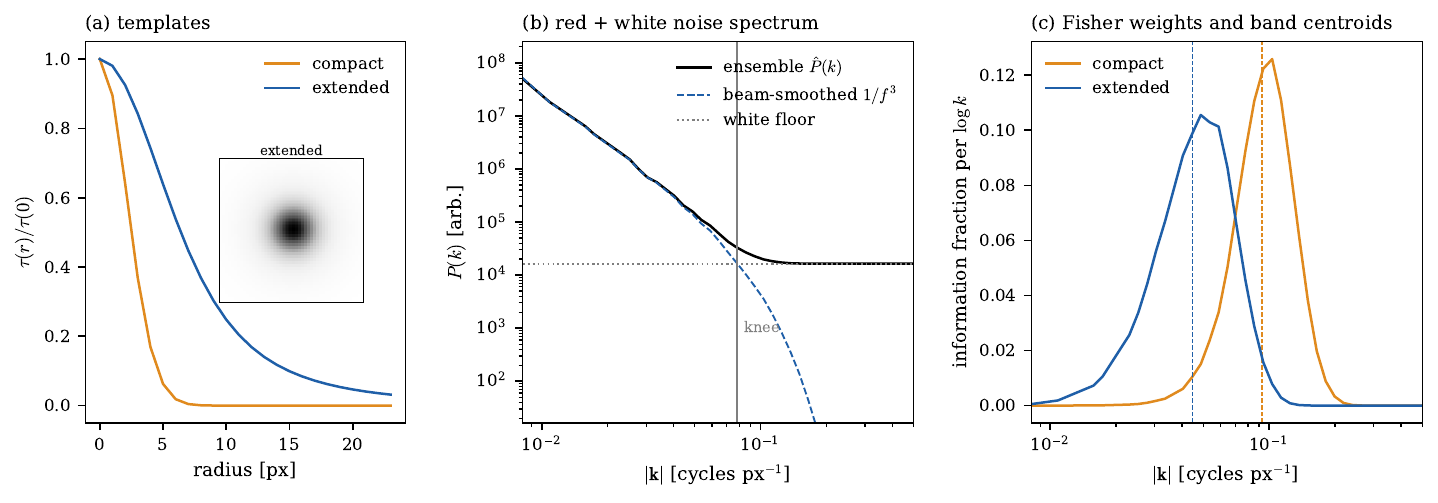}
  \caption{The two source templates and where they draw their information. \emph{(a)} Radial profiles
  of the compact ($\delta\circledast$beam) and extended ($\beta$-model $\circledast$ beam, $r_{c} = 7$\,px)
  templates. \emph{(b)} The ensemble power spectrum of the red\,+\,white noise model as the pipeline
  estimates it from 1500 maps, with its beam-smoothed $1/f^{3}$ and white components; the vertical line is
  the knee where the white floor takes over (\S\ref{sec:floor}). \emph{(c)} The Fisher weights
  $\fhat_{\kv}$ of Eq.~(\ref{eq:fisherweight}) on that spectrum, shown as the fraction of the amplitude
  information per logarithmic interval of $|\kv|$; dashed lines mark the band centroids
  $\exp\avf{\log|\kv|}$. The compact template's weight rises with $|\kv|$ until the floor terminates it;
  the extended template's sits an octave lower. The two templates look at different parts of the same
  noise.}
  \label{fig:fisherband}
\end{figure}

\paragraph{Fisher weights and the Fisher band.}
The quantity that organizes the whole paper is the fraction of the amplitude information carried by each
Fourier mode,
\begin{equation}
  \fhat_{\kv} \;\propto\; \frac{|\tautil_{\kv}|^{2}}{P(\kv)}, \qquad \sum_{\kv}\fhat_{\kv} = 1 .
  \label{eq:fisherweight}
\end{equation}
We call $\fhat_{\kv}$ the \emph{Fisher weight}, write $\avf{\cdot}$ and $\Varf{\cdot}$ for the mean and
variance over modes with those weights, and call the set of modes carrying most of the weight the
template's \emph{Fisher band}. For a compact (beam-shaped) source in red noise the band sits at high
$|\kv|$; for an extended source it sits at low $|\kv|$. In our simulations
(\S\ref{sec:noise_models}; beam FWHM 5\,px, $\beta$-model core $r_{c} = 7$\,px, red\,+\,white background)
the band centroids are $\exp\avf{\log|\kv|} \simeq 0.093$ and $0.045$\,cycles\,px$^{-1}$ respectively, and
Fig.~\ref{fig:fisherband} shows them against the noise spectrum. Almost every template-dependent result
in this paper reduces to the statement that the two templates look at different parts of the same
noise.

\subsection{The matched filter as the minimum-variance unbiased estimator}
\label{sec:mf}

\paragraph{Three optimalities that coincide under Gaussian noise.}
For $n \sim \mathcal N(0,\Nm)$ with $\tauv$ and $\Nm$ known, the matched filter is simultaneously the
best linear unbiased estimator (BLUE, by the Gauss--Markov theorem), the maximum-likelihood estimator, and
the minimum-variance unbiased estimator (MVUE). Only the first survives the loss of Gaussianity: the
BLUE property needs the covariance alone and holds for any noise distribution with that covariance,
while the other two are properties of the Gaussian likelihood. The third is the strongest of the
three, and it rests on sufficiency and completeness.

\paragraph{Sufficiency, completeness, and Basu's theorem.}
The Gaussian log-likelihood factorizes as
\begin{equation}
  \log p(d\,|\,A) \;=\; -\tfrac12\, d^{\dagger}\Nm^{-1}d \;+\; A\,\tauv^{\dagger}\Nm^{-1}d
  \;-\; \tfrac12 A^{2}\,\tauv^{\dagger}\Nm^{-1}\tauv ,
  \label{eq:loglike}
\end{equation}
so by the Neyman--Fisher factorization theorem the single number $\AMF$ is a \emph{sufficient statistic}
for $A$: every bit of information the image carries about the amplitude is in it, with
$\AMF\,|\,A \sim \mathcal N(A,\sigMF^{2})$. As a one-parameter exponential family the statistic is also
\emph{complete}, and the Lehmann--Scheff\'e theorem then makes the matched filter the \emph{unique} MVUE:
any other unbiased estimator either coincides with it almost surely or has strictly larger variance at
every $A$ \citep{LehmannCasella1998}. Basu's theorem supplies the geometry \citep{Basu1955}: a complete
sufficient statistic is independent of every \emph{ancillary} statistic (one whose distribution does not
depend on $A$), and the part of the data orthogonal to $\tauv$ in the $\Nm^{-1}$ metric is ancillary.
In words: under Gaussian noise of known covariance, everything in the image other than $\AMF$ -- every
pixel value, every correlation, every visually striking structure -- is provably independent of the
amplitude. There is nothing left for a network to read.

\paragraph{Correlations cannot be exploited.}
It is worth making this concrete, because the intuition that a network could ``learn the noise
correlations in the periphery'' is a plausible origin of many advantage claims (it appeared several times
in our own early analyses). Three facts close that possibility. 
(i) The matched filter already uses the correlations, optimally: the $\Nm^{-1}$ weighting
\emph{is} the correlation structure. (ii) Whitening the data, $\tilde d = \Nm^{-1/2}d = A\,\tilde\tauv +
\tilde n$ with $\tilde n \sim \mathcal N(0,\Id)$, the matched filter is the projection of $\tilde d$ onto
$\tilde\tauv$ and the orthogonal complement $\tilde d_{\perp} = \tilde n_{\perp}$ is pure noise;
orthogonal components of a spherical Gaussian are independent, so $\AMF$ and $\tilde d_{\perp}$ are
independent. The long-range correlations of $1/f^{3}$ noise do couple central and peripheral pixels in the
original space, but only through the ancillary subspace. (iii) A Gaussian is fully specified by its
first two moments; once $\Nm$ is used, the noise is accounted for. The operative condition is a
\emph{known} covariance, not stationarity: non-stationary Gaussian noise with a known covariance is
equally beyond improvement.

\paragraph{When the matched filter reaches the Cram\'er--Rao bound, and when it cannot.}
Write the noise \emph{score} $s(n) = -\nabla_{n}\log p_{n}(n)$ and the \emph{Fisher operator}
$\Jop = \E[s\,s^{\dagger}]$ (both defined in words in \S\ref{sec:score}). Because $A$ enters
Eq.~(\ref{eq:model}) as a location shift, the Fisher information for the amplitude is
$I = \tauv^{\dagger}\Jop\,\tauv$, independent of $A$, and the Cram\'er--Rao lower bound (CRLB) for any
estimator unbiased at $A$ is $V(A) \ge 1/I$. Three cases:
\begin{itemize}[leftmargin=1.5em,itemsep=2pt]
  \item \emph{Gaussian, known covariance:} $\Jop = \Nm^{-1}$, $I = \sigMF^{-2}$, and the matched filter is
    efficient -- it attains the bound, up to the error in estimating $P(\kv)$ from the data.
  \item \emph{Non-Gaussian, same covariance:} the matched filter is still unbiased with variance
    $\sigMF^{2}$, so the Cram\'er--Rao inequality gives $1/I \le \sigMF^{2}$ -- equivalently
    $\Jop \succeq \Nm^{-1}$: among all distributions with a given covariance the Gaussian has the
    \emph{least} Fisher information \citep{Stam1959,CoverThomas2006} -- with equality if and only if the
    noise is Gaussian.
    A nonlinear estimator using the full likelihood can be unbiased and still have smaller variance than
    the matched filter.
  \item \emph{Per-image-varying covariance:} the deployable matched filter must use an ensemble-averaged or
    estimated covariance and is handicapped even when each image is conditionally Gaussian. A mixture of
    Gaussians is not Gaussian, so this case is a special case of the previous one; we keep it separate
    because it behaves differently in practice (\S\ref{sec:flavours}, \S\ref{sec:tier1}).
\end{itemize}
The ratio $\eta \equiv I\,\sigMF^{2} \ge 1$ that these cases suggest is the subject of
\S\ref{sec:ceiling}.

\paragraph{Prior-agnosticism.}
The matched filter's form depends only on $(\tauv,\Nm)$ and its sampling law at fixed $A$,
$\mathcal N(A,\sigMF^{2})$, is invariant to the source population. It naturally returns negative
amplitudes for sub-threshold sources, which is essential for unbiased faint-source statistics, stacking
and $P(D)$ work. Both properties are lost by any estimator that has been trained on a source population,
which is the subject of the next subsection.

\subsection{What a network trained with squared error actually estimates}
\label{sec:cnn}

A regressor minimizing the mean squared error against the true amplitudes converges to the posterior
mean $\E[A\,|\,d]$ under the training prior $\pi(A)$. By sufficiency, under Gaussian noise
$\E[A\,|\,d] = \E[A\,|\,\AMF]$, so the best any such network can do is the one-dimensional posterior mean
on the matched-filter statistic $y \equiv \AMF$,
\begin{equation}
  \Ahat_{\rm Bayes}(y) \;=\; \frac{\int A\,\pi(A)\,\mathcal N(y;A,\sigMF^{2})\,\dd A}
                                  {\int \pi(A)\,\mathcal N(y;A,\sigMF^{2})\,\dd A}
  \;=\; y + \sigMF^{2}\,\frac{\dd}{\dd y}\log m(y), \qquad m = \pi * \mathcal N_{\sigMF},
  \label{eq:tweedie}
\end{equation}
which is Tweedie's formula \citep{Robbins1956,Efron2011}. Shrinkage toward the prior scales with the
prior's informativeness; for a steeply falling $\dnds$ it is the familiar Eddington bias or deboosting
correction \citep{Eddington1913,HoggTurner1998}. Any mean-squared-error advantage of such an estimator
over the raw matched filter is therefore prior-dependent and, without a stated prior, numerically
arbitrary.

There are three distinct ways an apparent advantage over the matched filter can arise, and this paper
is organized around separating them.
\begin{enumerate}[label=(\Alph*),leftmargin=2em,itemsep=2pt]
  \item \emph{Prior shrinkage.} Available to any estimator given the prior; fully captured by
    Eq.~(\ref{eq:tweedie}); a Bayes-versus-frequentist statement, not a network-versus-filter one.
  \item \emph{Information beyond the matched-filter statistic.} Requires non-sufficiency, \ie\
    non-Gaussian noise, which includes covariance mixtures.
  \item \emph{Latent-dependent reweighting.} If the noise level varies from image to image and an
    estimator can read that level off the image, it can down-weight noisy images and up-weight quiet ones.
    This needs no prior on $A$, and it is not information about $A$ in any single image: it is
    inverse-variance weighting across the ensemble, done inside the per-object estimator. It is
    legitimate, it is prior-free, it is formally a sub-case of (B), and -- as \S\ref{sec:coherent_in_A}
    shows -- it fades away for sources brighter than the noise. It deserves its own name because it
    behaves so differently from (B) in practice.
\end{enumerate}

\paragraph{The global-bias trap.}
The posterior mean has \emph{exactly} zero bias averaged over the prior (by the tower property) and beats
$\sigMF$ in prior-averaged root-mean-square error, on pure Gaussian noise, while extracting nothing beyond
the matched filter. A small ``overall bias'' is therefore generically the cancellation of a positive bias
at the faint end against a negative one at the bright end, not pointwise unbiasedness. The correct
variance bound for a biased estimator is the biased Cram\'er--Rao bound \citep{Kay1993,Eldar2004},
\begin{equation}
  V(A) \;\ge\; \frac{[\,1 + g'(A)\,]^{2}}{I} \;=\; [\,1 + g'(A)\,]^{2}\,\sigMF^{2}
  \quad\text{(Gaussian, known covariance)},
  \label{eq:biasedcrlb}
\end{equation}
so wherever an estimator shrinks ($g' < 0$) its variance may legally sit below $\sigMF^{2}$. This is a
natural origin of sub-$\sigMF$ root-mean-square errors, and it is not a violation of anything.
Any comparison between a trained estimator and the matched filter must therefore be made
\emph{conditionally on $A$}, and the quantity to compare is the \emph{shrinkage-cleaned variance ratio}
\begin{equation}
  R(A) \;\equiv\; \frac{V(A)}{[\,1 + g'(A)\,]^{2}\,\sigMF^{2}} \;\ge\; \frac{1}{\eta},
  \label{eq:Rdef}
\end{equation}
which divides the shrinkage credit out explicitly: $R = 1$ is a Gaussian-efficient estimator, $R > 1$ an
inefficient one, and $R < 1$ is possible only with non-Gaussian Fisher information, up to the ceiling
$1/\eta$ of \S\ref{sec:ceiling}. The comparison is made in a range of amplitudes where $g(A)$ is consistent with
zero within a stated tolerance -- the \emph{certified band} -- and the faint end, within $\sim 2\sigMF$ of zero, is excluded by
construction because a non-negative source population forces $g(0) > 0$ on any regressor. \PaperII\
implements this certification; here we need only $R$ and the fact that the band exists and excludes the
faint end, because the faint end is precisely where channel (C) lives.

\subsection{Three meanings of ``unbiased''}
\label{sec:flavours}

The word ``unbiased'' is used for three distinct statements (Table~\ref{tab:flavours}), 
and much of the confusion in the
network-versus-filter literature can be traced to conflating their meaning. 
The classes are nested: every estimator unbiased conditional
on the latent is unbiased at every $A$, and every estimator unbiased at every $A$ is unbiased on average
over the prior. The variance floors go \emph{up} as the classes shrink, which is what makes the
distinction consequential.

\begin{table}[t]
  \caption{Three flavors of unbiasedness, the class of estimators each defines, and the variance floor
  of that class. The middle row is the deployment requirement and the class that the ceiling $\eta$
  bounds.}
  \label{tab:flavours}
  \centering
  \small
  \begin{tabular}{p{2.2cm}p{5.3cm}p{4.3cm}p{2.9cm}}
    \toprule
    flavor & statement & variance floor & who lives there \\
    \midrule
    global & $\E_{\pi}[\Ahat - A] = 0$, averaged over the source prior &
      none (the constant $\Ahat = \E_{\pi}[A]$ has zero variance) & useless as a certification \\
    \textbf{marginal} & $g(A) = \E[\Ahat\,|\,A] - A = 0$ for every $A$, averaged over noise \emph{and} latent &
      $\sigMF^{2}/\eta$ (the Cram\'er--Rao bound from the marginal noise density) &
      \textbf{the deployment requirement} \\
    conditional-on-latent & $\E[\Ahat\,|\,A,z] - A = 0$ for every $A$ \emph{and} every $z$ &
      $\sigor^{2}$, the oracle matched filter (conditionally Gaussian case) & every normalized fixed linear filter \\
    \bottomrule
  \end{tabular}
\end{table}

The middle class is strictly larger than the bottom one. An estimator may be systematically low on noisy
images and systematically high on quiet ones and still be unbiased at every $A$ once one averages over
which kind of image was drawn. That freedom is exactly channel (C) of \S\ref{sec:cnn}. An astronomer's
picture: a hundred nights of varying transparency, one flux measured per night. Weighting the modes \emph{within}
each night is what the matched filter does. Additionally shrinking the bad nights and boosting the good
ones, so that the distortions cancel across the hundred, makes every individual flux systematically wrong
but leaves the catalog unbiased at every true flux -- and it lowers the scatter. A network that has seen
many nights' data does the second thing internally. It is the population-level inverse-variance
weighting that a survey would do downstream, performed prematurely inside the per-object estimator.

Two consequences follow from this. First, any bias-correction scheme applied during training, and any
certification of the form ``$g(A) = 0$ in a band'', tests the \emph{middle} flavor and nothing more: a
reweighting estimator $\Ahat' = b(\hat z)\AMF$ with $\E[b] = 1$ has $g(A) = A(\E[b] - 1) = 0$ identically and
passes every such test while being systematically wrong on every individual class of image. That is not a
defect of the tests -- marginal unbiasedness \emph{is} the deployment requirement -- but it means the
oracle matched filter, which is optimal only in the bottom class, is not the ceiling for a deployable
estimator. Second, ``the network beat the oracle'' does not, by itself, indicate leakage or
over-fitting: it is in fact expected at small $A$. The only unconditional impossibility line is the
one in the next subsection.

\subsection{The reference estimators, and the one unconditional floor}
\label{sec:ladder}

Table~\ref{tab:refladder} lists the five reference variances against which a trained estimator is read.
Two are operational -- things a real pipeline can run -- and three are interpretive, computable only in
simulation where the noise was generated.

\begin{table}[t]
  \caption{Reference estimators. The guaranteed orderings are
  $\sigCRLB \le \sigemp$, $\sigor \le \sigemp$ and $\sigor \le \signuis$; the relation between
  $\sigCRLB$ and $\sigor$ is \emph{not} fixed (see text).}
  \label{tab:refladder}
  \centering
  \small
  \begin{tabular}{p{2.5cm}p{5.8cm}p{6.7cm}}
    \toprule
    symbol & what it is & role \\
    \midrule
    $\sigemp$ & ensemble matched filter: one global filter from the ensemble-mean power spectrum &
      the default baseline; ``does the network beat what a pipeline uses?'' \\
    $\signorm$ & the same filter shape, each image rescaled by its own measured noise level &
      what a hit/weight map buys for free; any pipeline that ships a weight map is already here \\
    $\sigor$ & matched filter built from the \emph{true} per-image covariance (``oracle'') &
      best possible \emph{linear} estimator (Gauss--Markov); optimum of the conditional-on-latent class;
      not deployable \\
    $\signuis$ & CRLB with the latent treated as a nuisance parameter estimated per image &
      the price of not being told $z$ \citep[\S3.5]{Kay1993}; optional \\
    $\sigCRLB = \sigMF/\sqrt{\eta}$ & CRLB from the \emph{marginal} noise density &
      \textbf{the only unconditional floor}; bounds every marginally unbiased estimator \\
    \bottomrule
  \end{tabular}
\end{table}

The ordering that is \emph{not} guaranteed is the interesting one. Two effects compete. Fisher
information is convex in the density \citep{Cohen1968}, so mixing loses information,
$I_{\rm marg} \le \E_{z}[I_{z}]$, which pushes $\sigCRLB$ up. Jensen's inequality pushes the other way,
$1/\E_{z}[I_{z}] \le \E_{z}[1/I_{z}] = \sigor^{2}$. In the \emph{well-inferable limit} -- the latent is
read off the image with negligible error, which is the case for a spectral slope measured from thousands
of modes -- the first effect vanishes and
\begin{equation}
  \sigCRLB \;\to\; \big(\E_{z}[I_{z}]\big)^{-1/2} \;\le\; \sigor .
  \label{eq:wellinferable}
\end{equation}
The oracle is handed the true covariance but is required to be unbiased on every individual image
separately. If we give up that requirement, in the way \S\ref{sec:flavours} describes, then it can be beaten.
The amplitude-only mixture of \S\ref{sec:results_mixtures} is the verified instance: there the ensemble
and oracle filters are the \emph{same estimator} (the matched filter is invariant under $\Nm \to c\Nm$), so there is
no linear headroom whatsoever, and yet $\eta = \E[\sigma^{2}]\,\E[\sigma^{-2}] > 1$ by the Cauchy--Schwarz
inequality.

The adjudication that follows from Table~\ref{tab:refladder} is short. $\sigma_{\rm net} \approx \sigemp$:
no advantage. $\sigemp > \sigma_{\rm net} \ge \sigor$: the advantage is bounded by the best linear filter,
\ie\ better noise modeling. $\sigor > \sigma_{\rm net} \ge \sigCRLB$: the estimator beats every linear
filter -- for fixed-covariance non-Gaussian noise this is the unambiguous higher-order signature; for a
covariance mixture it may instead be channel (C), and the $A$-dependence decides if that is the case 
(\S\ref{sec:coherent_in_A}). $\sigma_{\rm net} < \sigCRLB$ inside the certified band: \emph{impossible},
hence over-fitting, leakage or residual bias. That last line is the reason the ceiling matters
operationally: for non-Gaussian noise there is no impossibility line without a computable $\eta$, and an
advantage that cannot be tested against one has to be taken on trust.

\section{The advantage ceiling}
\label{sec:ceiling}

\S\ref{sec:framework} established that a genuine, prior-free advantage over the matched filter can
exist only along two axes -- a covariance that varies from image to image, or noise that is not
Gaussian -- and that the Cram\'er--Rao bound from the marginal noise density is the one floor no
marginally unbiased estimator can cross. What it did not supply is a number computable in advance: given
a noise model and a template, how large can the advantage possibly be? Without such a number every noise
scenario requires training and validating over a full regression network merely to discover that the
available gain was, say, three percent. This section supplies the number, and the 
physical reasoning behind it. 

\subsection{The score, the Fisher operator, and the excess-score operator}
\label{sec:score}

For a noise field $n \in \R^{N}$ with probability density $p_{n}$, the \emph{score} is
\begin{equation}
  s(n) \;=\; -\nabla_{n}\log p_{n}(n) \;\in\; \R^{N}.
  \label{eq:score}
\end{equation}
It is the density's own sensitivity map: it points in the direction in which the observed realization is
most surprising, with a magnitude set by how fast the log-probability falls. For zero-mean Gaussian noise
the score is \emph{linear}, $s(n) = \Nm^{-1}n$, and this is the inverse-covariance weighting inside the matched
filter. For any non-Gaussian density it is a nonlinear function of the map, and the content of this
paper can be summarized in one sentence: \emph{the network-versus-filter question is the question of how
nonlinear the noise score is, as seen by the template.}

The score governs amplitude estimation because $A$ enters Eq.~(\ref{eq:model}) only as a location
shift: $p(d\,|\,A) = p_{n}(d - A\tauv)$, so $\partial_{A}\log p(d\,|\,A) = \tauv^{\dagger}s(d - A\tauv)$,
and the \emph{Fisher information for the amplitude} is
\begin{equation}
  I(A) \;=\; \E\big[(\tauv^{\dagger}s(n))^{2}\big] \;=\; \tauv^{\dagger}\Jop\,\tauv ,
  \qquad \Jop \;\equiv\; \E[\,s\,s^{\dagger}\,] ,
  \label{eq:fisherop}
\end{equation}
with $\Jop$ the \emph{Fisher operator} of the noise, a property of the noise distribution alone. Because
the noise is additive, $I$ does not in fact depend on $A$, and we write it simply as $I$ from here on. For Gaussian noise $\Jop = \Nm^{-1}$ exactly. In
general we define the \emph{excess-score operator}
\begin{equation}
  \boxed{\;\Dop \;\equiv\; \Jop - \Nm^{-1} \;\succeq\; 0\;}
  \label{eq:delta}
\end{equation}
with equality if and only if the noise is Gaussian. The proof is one line and is the Gaussian
optimality theorem in operator form: Stein's identity gives $\E[s\,n^{\dagger}] = \Id$, so for any
direction $a$, $0 \le \E[(a^{\dagger}s - a^{\dagger}\Nm^{-1}n)^{2}] = a^{\dagger}\Dop\,a$
(Appendix~\ref{app:derivations}). This is the Cram\'er--Rao inequality for the best linear unbiased
estimator in direction $a$, written as an operator statement; in the information-theory literature the
same fact -- at fixed covariance the Gaussian has the least Fisher information -- is a corollary of
Stam's inequality \citep{Stam1959}.

The physical reading gives us valuable intuition. 
A Gaussian field is the maximum-entropy field at fixed
covariance; it ``looks like nothing''. Noise that looks like something -- glitches, stripes, leaked modes,
a per-image brightness, the granularity of unresolved sources -- has $\Dop \ne 0$, concentrated on the
modes and at the statistical orders where that structure lives. \emph{Any structure beyond the two-point
function is extra information, and $\Dop$ is the map of where it is.}

\subsection{Definition, three roles, and the survey-time reading}
\label{sec:eta_def}

With $\sigMF^{2} = (\tauv^{\dagger}\Nm^{-1}\tauv)^{-1}$, the \emph{advantage ceiling} is
\begin{equation}
  \boxed{\;\eta \;\equiv\; I\,\sigMF^{2} \;=\; 1 + \sigMF^{2}\;\tauv^{\dagger}\Dop\,\tauv \;\ge\; 1 .\;}
  \label{eq:eta}
\end{equation}
In whitened coordinates -- $x = \Nm^{-1/2}n$, $t = \Nm^{-1/2}\tauv$, $\that = t/\|t\|$, so that
$\sigMF^{2} = 1/\|t\|^{2}$ -- it reads $\eta = \E[(\that^{\dagger}s_{x}(x))^{2}]$: the Gaussian part of the
score is $x$ itself and contributes exactly $1$, and everything above $1$ is the nonlinear part of the
score projected on the template. All the numerical methods of \S\ref{sec:computing} work in these
coordinates.

The ceiling plays three roles.
\begin{enumerate}[label=(\roman*),leftmargin=2em,itemsep=2pt]
  \item \emph{Ceiling.} Any estimator unbiased at $A$ satisfies $V(A) \ge \sigMF^{2}/\eta$; inside a
    certified band this bounds a network's conditional variance, and outside it the biased form
    $V \ge (1+g')^{2}/I$ applies.
  \item \emph{Impossibility line.} Where $\eta$ is known exactly -- Tier~0 by theorem, Tier~1 by
    quadrature, an exactly solvable model by its closed form -- a measured advantage exceeding $\eta$
    cannot be information; it is shrinkage, leakage or bias. This generalizes the Gaussian ``no
    advantage'' theorem to every such noise class. Where only a certified lower bracket is available
    (Tier~2, from the ladder of \S\ref{sec:ladder_var}), the pre-registered number is a floor, not a line:
    an estimator that exceeds it has measured the ladder's looseness, and the impossibility line is the
    complete-data bound of \S\ref{sec:tier2}, where one exists.
  \item \emph{Forecast.} $\eta - 1$ is, to first approximation, the fraction of noise power in the
    template's Fisher band that is carried by identifiable, removable structure (made precise in
    \S\ref{sec:tier2}). It decides whether a noise scenario deserves a training campaign at all.
\end{enumerate}

Three properties make $\eta$ the right instrument for the question posed in \S\ref{sec:intro}. It is
built from the noise density and the template and nothing else: no prior on $A$ (the Fisher information
of a location family is prior-free), no training set, no architecture. It reduces to exactly $1$ for
Gaussian noise of known covariance, reproducing the negative result of \S\ref{sec:mf} as $\Dop = 0$. And
it is invariant under any invertible linear preprocessing of the map -- a linear filter transforms
$\Jop$, $\Nm^{-1}$ and $\tauv$ covariantly -- which is the operational statement of the rule that one must
filter the data, not the noise. The one thing it is not prior-free about is trainability: the \emph{design
ratio} $r = \sigMF/\mathrm{std}(A)$ of \PaperII\ depends on the flux prior and controls whether a network
can be trained on the same model at all -- a large $r$ means the sources are faint against the noise, so
that the network has little signal to learn from. Changing the prior leaves every $\eta$ untouched and
moves every certification number, which is why we keep the two decisions separate.

\paragraph{The survey-time reading, and its condition.}
An estimator that reaches the ceiling multiplies the Fisher information by $\eta$. A survey multiplies
its Fisher information by the same factor when it collects $\eta$ times as many independent samples of
the \emph{same} noise ensemble -- $\eta$ times the area at fixed per-map depth -- so $\eta$ is the factor
by which an ideal nonlinear estimator multiplies the effective survey time for that source class, and
$\sqrt{\eta}$ is the factor in $\sigma$. The reading carries one condition: the trade must leave the noise ensemble unchanged, so the time is
spent on area, not on depth. Enlarging the area at fixed depth does leave it unchanged. Deepening the same field does not: longer integration averages down not only
the instrumental noise but also the per-map scatter of a covariance mixture, which is what makes
$\eta > 1$ for the Tier-1 models, so the deeper survey has a different and smaller ceiling. A map of
nonuniform depth is a union of such ensembles, each with its own $\eta$, and the reading applies to each
depth stratum, not to the map as a whole. With that caveat in mind,
a ceiling of $1.3$ is thirty percent more survey time -- worth having; a ceiling of $2$ is a survey twice
the size; a ceiling of $10$ is transformational. We quote a ceiling in both forms where the condition
holds, because the dimensionless number alone systematically under-communicates how much a modest
$\eta$ is worth, and over-communicates how much of a large one is deployable (\S\ref{sec:coherent_in_A}).

Source confusion fails the condition in a stronger way. It is an astrophysical background rather than an
instrumental one: its amplitude is set by the sky, not by the length or the extent of the observation, so
no amount of integration reduces it and the survey-time reading does not apply at all. There $\eta$ remains a
variance factor -- and, in one sense, a more valuable one, since it is the only kind of gain a larger
survey could not have bought instead. Every confusion number in this paper is therefore quoted as a
variance factor alone (\S\ref{sec:results_confusion}, \S\ref{sec:discussion}).

\paragraph{The projection law.}
Equation~(\ref{eq:eta}) makes template dependence a theorem. $\Dop$ is the universal, template-free
object, but it is an \emph{operator} -- in the stationary case a set of polyspectra -- not a scalar, and
the advantage is its inner product with the template's Fisher weight. Two noise models can concentrate
$\Dop$ on disjoint parts of $\kv$-space -- one at low $k$, the other at high $k$ -- so that an extended
source sees the first and a compact one the second. A scalar ``non-Gaussianity'' measure assigns each
model a single number and hence a single ordering, the same for every template, whereas the ordering of
their ceilings reverses between compact and extended sources. No template-free scalar advantage metric
can therefore exist; the correct universal report is one
rank higher than a scalar -- the operator itself, in practice its leading polyspectral projections --
together with the curve of $\eta$ against template extension. The results of
\S\S\ref{sec:results_mixtures}--\ref{sec:results_confusion} contain several pairs of models whose
template ordering is opposite for exactly this reason.

\subsection{Covariance mixtures: an exact ceiling with two channels}
\label{sec:tier1}

Consider noise that is Gaussian on every image but with a covariance $\Nm_{z}$ that depends on a
per-image latent $z$ drawn from $w(z)$ -- a per-patch atmospheric spectral slope, a scan-ridge
orientation, the amplitude of a leaked mode. Marginally the noise is a Gaussian mixture, and its score
is available in closed form through the latent posterior,
\begin{equation}
  p_{n}(n) = \int \dd z\; w(z)\,\mathcal N(0,\Nm_{z})
  \quad\Longrightarrow\quad
  s(n) \;=\; \E_{z|n}\big[\Nm_{z}^{-1}\big]\,n ,
  \label{eq:mixturescore}
\end{equation}
which is a \emph{posterior-weighted whitening}. With one or two latent parameters the posterior is a quadrature
over a grid, one FFT per image, and Eq.~(\ref{eq:fisherop}) follows by Monte Carlo over the noise
ensemble: minutes of compute, and an \emph{exact} number against which every approximation can be
checked (\S\ref{sec:computing}). Cheaper still is
\begin{equation}
  \etawi \;=\; \sigMF^{2}\;\E_{z}\big[I_{z}\big], \qquad I_{z} = \tauv^{\dagger}\Nm_{z}^{-1}\tauv ,
  \label{eq:etawi}
\end{equation}
the Fisher information averaged over the latent in units of the ensemble filter's, which is
Eq.~(\ref{eq:wellinferable}) read as a ceiling. We call $\etawi$ the \emph{well-inferable ceiling}: the
ceiling one would face if the latent were handed over on every image rather than inferred from it. 
$\etawi$ is not merely the limiting value of $\eta$: it bounds $\eta$ from above for every covariance
mixture. Fisher information is convex in the density, so $I \le \E_{z}[I_{z}]$ for \emph{every} covariance mixture and
\begin{equation}
  \boxed{\;\eta \;\le\; \etawi\;}
  \label{eq:etawi_bound}
\end{equation}
holds unconditionally, with equality when each map fixes its own latent exactly -- when the posterior
$w(z\,|\,n)$ collapses onto a single value, which is the well-inferable limit
(Appendix~\ref{app:wellinferable}). That is the ordinary situation whenever the latent is read off far
more modes than it has parameters: a spectral slope inferred from thousands of Fourier modes is pinned by
the map itself to a small fraction of its ensemble scatter. This is a training-free upper bound on the
ceiling of a covariance mixture: it needs one quadrature over the components, and neither the marginal
score nor an estimator of any kind. Where the latent is well inferred it is also the value, agreeing
with the exact quadrature to within the Monte Carlo error, and where it is not it is loose by the
amount the posterior is broad (\S\ref{sec:results_mixtures}).

\paragraph{Two channels.}
Where does the excess information $\eta - 1$, the excess-score term of Eq.~(\ref{eq:eta}), come from? If we write the per-image log-spectrum fluctuation
$\delta_{i}(\kv) = \log P_{i}(\kv) - \log\Pbar(\kv)$ about the ensemble-mean spectrum $\Pbar$, and split it,
\emph{relative to the template's Fisher band}, we get a band mean and a residual:
\begin{equation}
  \delta_{i}(\kv) \;=\; \underbrace{\avf{\delta_{i}}}_{\text{coherent}}
  \;+\; \underbrace{\big[\delta_{i}(\kv) - \avf{\delta_{i}}\big]}_{\text{incoherent (shape)}} .
  \label{eq:twochannels}
\end{equation}
The coherent part answers ``how loud is the noise in \emph{my} band on this image?''; the incoherent
part answers ``how does the noise shape \emph{within} my band wobble on this image?'' To second order in
$\delta$ the two contribute additively to the ceiling (Appendix~\ref{app:derivations}),
\begin{equation}
  \boxed{\;\eta - 1 \;\simeq\; \underbrace{\Var\big[\avf{\delta}\big]}_{\text{coherent}}
  \;+\; \underbrace{\E\big[\Varf{\delta}\big]}_{\text{incoherent}} ,\;}
  \label{eq:twochannel_eta}
\end{equation}
and in the regime of our simulations the split is better written multiplicatively, $\eta \simeq
\etaB\,\etaA$, with $\etaB = \E[c]\,\E[c^{-1}]$ the gain from rescaling each image by its inferred
band power $c$ (a Cauchy--Schwarz quantity, equal to $1$ only for a degenerate latent) and
$\etaA = 1 + \E[\Varf{\delta}]$ the second-order estimate of the gain from re-shaping the filter within
the band. 

The two channels behave completely differently, and Table~\ref{tab:channels} is the reference the
results sections lean on. The incoherent term is the gap between the ensemble filter and the oracle: it
is what a per-image filter can recover, it is unbiased conditional on the latent, it is flat in $A$, and
no scalar weight map can touch it. The coherent term is invisible to any linear filter (the matched
filter is scale-invariant) and to the oracle comparison (the oracle is conditionally unbiased); it is
channel (C) of \S\ref{sec:cnn}, it is removed by a weight map whenever the latent is measurable, and it
dies for sources brighter than the noise (\S\ref{sec:coherent_in_A}). The dispersion formula that
earlier treatments used for the whole Tier-1 advantage is the incoherent term and nothing else -- the
ensemble-to-oracle gap -- and it is therefore a \emph{lower} bound on $\eta - 1$, not an estimate of it.

\begin{table}[t]
  \caption{The two channels of a covariance-mixture advantage, relative to a fixed template's Fisher
  band. ``Weight map'' means a per-image scalar rescaling by a measured noise level.}
  \label{tab:channels}
  \centering
  \small
  \begin{tabular}{p{4.3cm}p{5.3cm}p{5.4cm}}
    \toprule
    & coherent (band amplitude) & incoherent (within-band shape) \\
    \midrule
    what varies per image & the loudness of \emph{my} band & the spectral shape \emph{within} my band \\
    reached by one fixed linear filter? & no (scale-invariance) & no (needs a per-image filter) \\
    reached by the oracle filter? & \textbf{no} (conditionally unbiased) & yes -- this \emph{is} the ensemble$\to$oracle gap \\
    unbiased given the latent? & \textbf{no} & yes \\
    dependence on $A$ & fades for $A \gtrsim \sigMF$ & flat \\
    survives in the certified band? & largely no & yes \\
    removed by a weight map? & yes, if the latent is measurable & no \\
    measured by (exact factor; severity coordinate) & $\etaB = \E[c]\E[c^{-1}]$; $\sd[\avf{\delta}]$ & $\eta/\etaB$; $\E[\Varf{\delta}]^{1/2}$ (the oracle gap) \\
    \bottomrule
  \end{tabular}
\end{table}

\paragraph{The lever arm: why a spectral tilt is mostly a band-amplitude variation.}
An astronomer's picture is useful once again, and the atmospheric-residual model of \S\ref{sec:noise_models}
supplies that. 
By a \emph{spectral tilt} we mean a change in the exponent of the noise power spectrum,
$P(k) \propto k^{-\alpha}$, from one map to the next. In a mm/submm survey that is what atmospheric
removal leaves behind: the atmosphere is a bright, red, spatially correlated emitter, the pipeline
subtracts a common mode or a few principal components, and how steep the surviving residual is depends on
how well that subtraction worked. Our tilt model accordingly draws both a level
and a slope per map. A change of level is purely coherent -- the whole band grows louder or quieter
together, and one number per map removes it. A change of slope is what one might call
incoherent, 
but that intuition is mostly wrong, for the following argument.

For a power-law tilt about a pivot, $\delta_{i}(k) = -(\alpha_{i} - \bar\alpha)\log(k/k_{\rm piv})$, the two
terms are
\begin{equation}
  \eta - 1 \;\simeq\; \Var(\alpha)\Big[\underbrace{\avf{\log(k/k_{\rm piv})}^{2}}_{\text{lever arm}^{2}}
  \;+\; \underbrace{\Varf{\log k}}_{\text{shape}}\Big] .
  \label{eq:leverarm}
\end{equation}
Tilting a power law about a pivot moves the curve most where one is furthest from the pivot; if a
template's Fisher band sits decades away from the pivot, a small change of slope is a large change in the
\emph{level} of the noise in that band. In our tilt model the pivot is at $k_{\rm piv} = 1$\,cycle\,px$^{-1}$,
outside the map, and the extended template's band centroid $\avf{\log k} = -3.2$ gives a squared lever arm
of $10$ against a shape term of $0.2$: the coherent channel dominates (by a factor of eight in the measured
severity coordinates of Table~\ref{tab:mixtures}, the floor adding within-band shape beyond the power-law
term), and it is strongly template-dependent, whereas the shape term depends only on the band width. Which part of a
spectral variation is called ``amplitude'' and which ``slope'' is a choice of pivot -- re-pinning the
power law moves variance from one to the other -- but the coherent/incoherent split, once a template is
fixed, is invariant. We report the latter, and we quote severity as the induced band-power scatter
$\sd[\avf{\delta}]$ per template rather than as a slope dispersion, which is not an observable.

\paragraph{What the second-order forms are for.}
Two things about Eqs.~(\ref{eq:twochannel_eta}) and (\ref{eq:leverarm}) are to be kept in mind. $\etaB$ is
exact for a purely coherent latent and is computed directly from the realized band powers; everything
else in the two equations is second order in $\delta$, and since the exact expression exponentiates a
convex function, the truncation always \emph{under}estimates -- at the campaign's severities by a factor
of two to three (\S\ref{sec:two_channels_results}). We therefore report the \emph{exact} factors $\etaB$
and $\eta/\etaB$ as the two-channel split, and use $\sd[\avf{\delta}]$ and $\E[\Varf{\delta}]^{1/2}$ only as
the two severity coordinates that locate a cell; the second-order forms explain the ordering of cells
and templates, not their values.

\subsection{Where the coherent channel lives in amplitude}
\label{sec:coherent_in_A}

$\eta$ is constant in $A$, but attainability is not. The natural estimator that realizes the coherent
channel rescales the matched filter by a factor read off the inferred noise level,
$\Ahat' = b(\hat z)\,\AMF$ with $\E[b] = 1$. It is unbiased at every $A$ and its variance is
\begin{equation}
  V_{b}(A) \;=\; \sigma_{0}^{2}\,\E[b^{2}c] \;+\; A^{2}\,\Var(b) ,
  \label{eq:Vb}
\end{equation}
where $\sigma_{0}^{2}c$ is the per-image matched-filter variance. At $A = 0$ the optimum is
$b \propto c^{-1}$ and the gain is exactly $\etaB$; the second term is the price, a scatter proportional
to the answer itself, and the two balance at $A \simeq \sigMF$. Re-optimizing $b$
at each $A$ gives the closed-form \emph{attainability envelope}
\begin{equation}
  V_{E}(A) \;=\; \Big(\E_{z}\Big[\frac{1}{\sigMF^{2}\,c(z) + A^{2}}\Big]\Big)^{-1} - A^{2},
  \qquad V_{E}(0) = \frac{\sigMF^{2}}{\etaB} ,
  \label{eq:methodE}
\end{equation}
which is a family envelope -- each $b(\cdot)$ is a real, implementable, marginally unbiased estimator,
so every point on the curve is an \emph{achievable} variance -- and it brackets the truth from above
while $\sigMF^{2}/\eta$ brackets it from below. For a purely coherent latent the envelope touches the
floor at $A = 0$; for a purely incoherent one it is flat at $\sigMF^{2}$ even when $\eta$ is large, because
a scalar rescaling cannot re-shape a filter. Since a certified band starts at $A \gtrsim 2\sigMF$, the
coherent channel is largely outside the region where anyone is willing to make claims: a marginal
ceiling of ten and a modest in-band advantage are two consistent facts about the same noise model
(\S\ref{sec:results_mixtures}). \emph{The magnitude of $\eta$ says nothing about whether an advantage is
deployable; the channel decomposition does.}

\subsection{Non-Gaussian noise: polyspectra, marked point processes, and removability}
\label{sec:tier2}

\paragraph{Weak non-Gaussianity.}
When the noise is weakly non-Gaussian, the score admits an Edgeworth expansion in the connected
higher-order statistics and, to quadratic order in the cumulants,
\begin{equation}
  \boxed{\;\eta - 1 \;\simeq\; \tfrac12\,\big\|\tilde\kappa_{3}(\that,\cdot,\cdot)\big\|_{F}^{2}
  \;+\; \tfrac16\,\big\|\tilde\kappa_{4}(\that,\cdot,\cdot,\cdot)\big\|_{F}^{2}\;}
  \label{eq:perturbative}
\end{equation}
-- the Frobenius norms of the whitened third- and fourth-cumulant tensors contracted once with the
unit whitened template. In Fourier space these tensors are the
bispectrum and trispectrum, familiar objects in CMB non-Gaussianity analysis.
Three things are worth reading off the formula. The contraction with $\that$ \emph{is} the
source--departure overlap: compact and extended templates select different slices of the same
polyspectra. The expression has the structure of the Fisher information of a primordial-bispectrum
amplitude with one leg carrying the template's Fisher weight, so the optimal weakly nonlinear estimator
is the matched filter plus a bispectrum-weighted quadratic term -- the analog of a KSW estimator
\citep{KSW2005}. 
And the formula is a norm over the full tensors, not over the pixel histogram: a field can have a nearly
Gaussian one-point distribution while its phase-coherent structure carries large contracted norms, so
``the residuals look Gaussian'' systematically understates $\eta$. The single-pixel reduction is
$\eta - 1 \simeq \gamma_{3}^{2}/2 + \gamma_{4}^{2}/6$ in terms of skewness and excess kurtosis.
Equation~(\ref{eq:perturbative}) is the leading term of an expansion in the cumulants, not a bound: it can
err in either direction once the higher cumulants are not small (for a standardized Student-$t$ with
$\nu = 10$ it gives $1.17$ against an exact $1.06$), and it fails outright once individual structures
become significant. The rigorous versions of its two terms are the \rung{quadratic} and \rung{cubic}
rungs of \S\ref{sec:ladder_var}, which are lower bounds whatever the cumulants.

\paragraph{Marked point processes.}
Several of our models share one generative skeleton: a superposition of identical profiles at Poisson
positions with random amplitudes, $n_{\rm struct}(\xv) = \sum_{i} a_{i}\psi(\xv - \xv_{i})$. For such a
marked Poisson process -- shot noise; the engine behind confusion and $P(D)$ analysis since the 1950s
\citep{Scheuer1957,Condon1974} -- Campbell's theorem gives every cumulant linear in the
density $\lambda$ \citep{DaleyVereJones2003}:
\begin{equation}
  P_{\rm struct} = \lambda\langle a^{2}\rangle|\tilde\psi|^{2}, \qquad
  B(\kv_{1},\kv_{2}) = \lambda\langle a^{3}\rangle\tilde\psi_{1}\tilde\psi_{2}\tilde\psi^{*}_{1+2}, \qquad
  T(\kv_{1},\kv_{2},\kv_{3}) = \lambda\langle a^{4}\rangle\tilde\psi_{1}\tilde\psi_{2}\tilde\psi_{3}\tilde\psi^{*}_{1+2+3} .
  \label{eq:campbell}
\end{equation}
Faint glitches (one-signed marks, bispectrum-led), source confusion (profile equal to the beam, marks
from the number counts, strongly skewed) and leaked principal components (a Poisson number of modes in
eigenmode space with heavy-tailed amplitudes) are all of this form; the scan-residual family is not
Poisson but is sign-symmetric, which annihilates all odd cumulants and makes it a clean trispectrum-led
case. Sign symmetry of the noise is thus a prediction: the quadratic rung of the ladder below must return
a null, and the leading term must appear at the cubic rung.

\paragraph{The exact score for structure plus Gaussian noise, and the complete-data bound.}
For noise of the form $n = c(z) + g$ with $g \sim \mathcal N(0,\Nm_{g})$ and $z$ the latent structure,
the score has an exact closed form -- a multivariate Tweedie formula \citep{Miyasawa1961,Efron2011},
\begin{equation}
  s(n) \;=\; \Nm_{g}^{-1}\big(n - \E[c(z)\,|\,n]\big):
  \label{eq:tweediescore}
\end{equation}
the optimal score is the Gaussian score applied after MMSE subtraction of the structured component. The
Fisher information is that of a matched filter operating on the map with the structure optimally
denoised away, which makes the heuristic ``$\eta - 1$ equals the removable power fraction'' rigorous and
yields the \emph{complete-data upper bound}, the perfect-removal limit,
\begin{equation}
  \eta \;\le\; \frac{\sum_{\kv}|\tautil_{\kv}|^{2}/P_{g}(\kv)}{\sum_{\kv}|\tautil_{\kv}|^{2}/P_{\rm tot}(\kv)}
  \qquad\text{(stationary case)} .
  \label{eq:cdbound}
\end{equation}
Between the perturbative formula (structures individually insignificant) and this bound (structures
individually obvious) lies the whole severity axis of a non-Gaussian model. Two rules govern its use, and
both are results from this paper. It is valid only on a band where the per-mode signal-to-noise is
regularized -- physically, by an instrumental white floor -- because without one the compact template's
Fisher weight climbs to whatever mode a numerical mask happens to stop at (\S\ref{sec:unmodelled}). And
it is the wrong ceiling for structure that is a scale mixture rather than an additive contaminant,
because perfect removal is not what a reweighting estimator can do; there the exact quadrature of
\S\ref{sec:tier1} is the ceiling, and the bound over-states the room (\S\ref{sec:results_mixtures}).

\paragraph{Two regimes.}
For structure carrying a fraction $\fs$ of the noise power in the template's Fisher band, built from
events of individual matched-filter significance $\xi$, the ceiling follows a two-regime law. For
$\xi \lesssim 1$ the Campbell cumulants in Eq.~(\ref{eq:perturbative}) give $\eta - 1 \propto \fs^{2}\xi^{2}$,
quadratic in both knobs and hence small; for $\xi \gg 1$ the events are individually identifiable, the
Tweedie score removes them, and $\eta - 1$ saturates at the removable-to-irreducible power ratio, linear
in $\fs$. This is why ``worse contamination gives the bigger nonlinear advantage'': brighter structure
is easier to identify, so its power converts to information rather than to variance. The artifact models of
\S\ref{sec:results_nongaussian} sit in the saturated regime and still have tiny ceilings, because once
the template's band is regularized they carry almost no removable power where the template looks.

\paragraph{Independent pixels are template-blind.}
One exact statement is needed for source confusion. If the whitened pixels are independent and
identically distributed, the joint score is separable, $s_{p}(x) = s_{1}(x_{p})$, hence
$\Jop = I_{1}\Id$ and $\eta = I_{1}$ for \emph{every} unit template, where $I_{1}$ is the Fisher
information of the one-pixel distribution. The sum statistic $\that^{\dagger}x$ does Gaussianize by the
central limit theorem, but the optimal estimator applies its nonlinearity per pixel \emph{before}
averaging and loses nothing: there is no central-limit suppression of $\eta$ for extended templates on
independent noise.

\subsection{The variational identity and the Volterra ladder}
\label{sec:ladder_var}

The universal numerical method rests on an exact variational identity. For any scalar test function
$f(x)$ of the whitened map with mild regularity,
\begin{equation}
  \boxed{\;\eta \;=\; \sup_{f}\Big\{\,2\,\E\big[\that^{\dagger}\nabla_{x}f(x)\big] - \E\big[f(x)^{2}\big]\Big\},\;}
  \label{eq:variational}
\end{equation}
attained at $f^{\star} = \that^{\dagger}s_{x}(x)$; Stein's integration by parts turns the bracket into
$\E[(\that^{\dagger}s)^{2}] - \E[(f - \that^{\dagger}s)^{2}]$ (Appendix~\ref{app:derivations}). Three
properties make this the workhorse of the paper.
\begin{itemize}[leftmargin=1.5em,itemsep=2pt]
  \item \emph{Every test function certifies a lower bound.} Plug in any $f$ and the bracket is a
    guaranteed lower bound on $\eta$, up to Monte Carlo error. The error from an imperfect $f$ is
    therefore one-sided rather than merely unbiased: a poor test function understates the ceiling and can
    never overstate it, so a search that stops too early underclaims rather than misleads. This is the
    estimation-side form of score matching \citep{Hyvarinen2005}, the objective
    behind modern diffusion models, restricted here to a single direction -- which is why a small network
    on noise-only data suffices. The corollary is equally important and is stated as a rule: \emph{a test
    function that fails to move proves nothing about the noise; it proves that the searched class found
    nothing.} Upgrading a null to a statement about $\Dop$ needs a Gaussianity theorem, an exact
    computation, or an independent upper bound.
  \item \emph{The matched filter is the linear solution.} Restricting $f$ to linear functions, the
    optimizer returns $f = \that^{\dagger}x$ with value exactly $1$. The entire network-versus-filter
    question is the variational headroom above linear $f$, and the linear value doubles as the mandatory
    sanity check of every numerical run.
  \item \emph{Capacity restriction is statistical order.} Restricting $f$ to polynomials of increasing
    degree produces the \emph{Volterra ladder}, the functional Taylor series of a nonlinear map of a whole
    image: for a CMB audience, the hierarchy of the quadratic estimator \citep{HuOkamoto2002} and
    its iterative continuation \citep{HirataSeljak2003}. The rungs align with Eq.~(\ref{eq:perturbative}):
    the best quadratic $f$ captures the bispectrum term, the cubic rung adds the trispectrum term, and a
    generic convolutional test function tops the ladder; the gap between the top rung and the cubic rung
    measures information beyond four-point statistics.
\end{itemize}

Between the linear and quadratic rungs there is a physically distinguished family that polynomials reach
only inefficiently: filters whose \emph{weights} depend nonlinearly on the image's own band powers,
\begin{equation}
  f(x) \;=\; \sum_{j}\beta_{j}(\hat g)\,\big(\that_{j}^{\dagger}x\big) + \sum_{m}d_{m}\,\hat g_{m},
  \label{eq:reweight}
\end{equation}
with $\that_{j}$ band- or sector-restricted templates and $\hat g$ the measured band powers. With a
$1/\hat g$ feature family the optimal band-diagonal per-image filter $\tauv^{\dagger}\hat\Nm(\hat z)^{-1}x$
lies inside the head's linear span, so reaching a covariance-mixture ceiling becomes a regression problem.
We call this the \emph{mixture matched filter}, the \rung{reweight} rung; it is the constructed estimator
that demonstrates attainability in \S\ref{sec:results_mixtures}. The full ladder is therefore
\rung{linear} $\to$ \rung{reweight} $\to$ \rung{quadratic} $\to$ \rung{cubic} $\to$ \rung{conv}, and
Table~\ref{tab:rungs} in \S\ref{sec:computing} gives the parameter count and the statistical target of
each.

\paragraph{The Needlet ILC (NILC) is the \rung{reweight} rung, in the component-separation variable.}
\S\ref{sec:intro_thispaper} noted that the internal linear combination is the matched filter with the
template replaced by a frequency spectrum and the noise covariance by the frequency--frequency one. The
correspondence extends one rung up. A global ILC takes its weights from a single ensemble-averaged
covariance and is therefore the \rung{linear} anchor of that problem, while the Needlet ILC
\citep{Delabrouille2009} estimates the covariance locally, per needlet scale and per sky domain, and
weights by it -- which is Eq.~(\ref{eq:reweight}) with the needlet domain in the role of the measured
band powers $\hat g$. The two constructions share the conclusions that follow. 
Both are linear given their
weights and nonlinear as maps of the data. Both are confined to a Tier-1 ceiling, because localizing a
second moment gives no access to third- and fourth-order structure at fixed local covariance -- which is
what Figs.~\ref{fig:methodE} and \ref{fig:ladders} show the \rung{reweight} rung doing and not
doing. And both pay the same price for adaptivity: the ILC bias incurred when a needlet domain is shrunk
until its covariance is estimated from too few modes is the component-separation form of the result of
\S\ref{sec:two_channels_results}, that a per-image rescaling inferred from too little information can
cost more than it returns. 

\paragraph{The rungs are estimators, not only bounds.}
What about estimators between
``linear'' and ``a neural network''? Each rung's optimal $f$ defines an estimator of its own class, the
one-step score estimator
\begin{equation}
  \Ahat_{f} \;=\; A_{0} + \sigMF\,\frac{f(x_{0})}{\E[\that^{\dagger}\nabla f]}, \qquad x_{0} = \Nm^{-1/2}(d - A_{0}\tauv),
  \label{eq:onestep}
\end{equation}
started from a preliminary amplitude $A_{0}$ that is independent of the map being estimated -- a split,
a held-out map, or a fixed reference amplitude (a matched-filter start from the same map makes the
correction vanish for a scale mixture; Appendix~\ref{app:variational}). For such a start its local
slope is unity by construction and its local efficiency in units of $\sigMF^{-2}$ is exactly the rung's
variational value. Here the claim is made at a single reference amplitude only; deploying a rung over an
unknown amplitude range, with bias and variance certified as in \S\ref{sec:cnn}, is the subject of
\PaperII. The
\rung{reweight} rung is the mixture matched filter, the \rung{quadratic} rung a bispectrum estimator, the
\rung{cubic} rung a trispectrum estimator (each to leading order: a test function of degree $n$ responds
first to the cumulant of order $n+1$, but its objective involves all cumulants up to order $2n$); the ladder does not merely bound the intermediate classes, it
builds them. The finding of \S\S\ref{sec:results_mixtures}--\ref{sec:results_confusion} is that on the covariance
mixtures the \rung{reweight} rung takes almost all of the available advantage, and that only source
confusion carries a signal that climbs the whole ladder to the top rung. Hand-crafted members of the same hierarchy
exist in the literature -- the locally optimum detectors \citep{Kassam1988}, the Gaussianized matched
filter \citep{RobnikSeljak2021}, the learned nonlinearity feeding a linear statistic
\citep{Zschetzsche2026} -- and the ladder is the systematic version of that program.

\section{Computing the ceiling}
\label{sec:computing}

Six computational routes to $\eta$ were built, validated against each other and against a battery of
null tests, and run on every noise model of \S\ref{sec:noise_models}. They deliver statements of
different logical strength -- exact values, upper bounds, certified lower bounds, and attainable
variances -- and the results sections always say which kind of statement a number is. This section
gives the what and the why; Appendix~\ref{app:pipeline} gives the how.

\subsection{Common infrastructure}
\label{sec:infrastructure}

Every method works on \emph{noise-only} ensembles: $6000$ maps of $128^{2}$ pixels per model (12--20\,k for
the exact quadratures of record, which have no fitting step and benefit from the larger sample; the
slope-only and amplitude-only variants of Table~\ref{tab:mixtures} use the 1500-map EVAL split), generated from a
registry that fixes the seed of every model so that any map is reconstructable from (model, seed, index).
Each ensemble is divided once into four disjoint splits, named for their roles: a power-spectral-density
(PSD) split (25\,\%) from which the ensemble power spectrum $\Phat(\kv)$ is estimated, a FIT split
(40\,\%) on which test functions are fitted, a VAL split (10\,\%) on which they are validated for early
stopping, and an EVAL split (25\,\%) on which every reported number and its bootstrap error are
evaluated. The separate VAL split exists because early stopping on the reporting split
selects upward fluctuations of a quantity we intend to quote as a bound; reusing PSD-split images inside
an estimator biases the whitening. We therefore keep the four splits strictly disjoint throughout.

Whitening is by the data-estimated spectrum, $\tilde x_{\kv} = \tilde n_{\kv}\sqrt{n_{\rm pix}/\Phat(\kv)}$,
so that $\|t\|^{2} = 1/\sigMF^{2}$ and $\that = \sigMF t$. Two consequences are worth noting. First, the
matched filter that all ceilings are referred to is the \emph{deployable} one -- built from an estimated,
ensemble-averaged spectrum -- not an oracle. Second, the ``linear rung'' of the ladder, which should be
exactly $1$, is in practice
\begin{equation}
  \etalin \;=\; 2 \,-\, \E_{\rm EVAL}\big[(\that^{\dagger}x)^{2}\big] .
  \label{eq:linrung}
\end{equation}
The leading $2$ is not an anomaly of the ladder but a property of the identity it evaluates: setting
$f = \that^{\dagger}x$ in Eq.~(\ref{eq:variational}) makes $\nabla_{x}f = \that$, so the gradient term
contributes $2\,\that^{\dagger}\that = 2$ exactly because $\that$ is a unit vector, and the second term
subtracts the measured second moment. Equation~(\ref{eq:linrung}) therefore returns $1$ only when the
EVAL split's realized second moment matches the PSD split's estimate; on heavy-tailed ensembles the
measured anchors range over $0.84$--$1.21$ floorless and $0.90$--$1.06$ floored (Tables~\ref{tab:master_floored}
and \ref{tab:master_floorless}), tracking the sample kurtosis. 
The departure from $1$ is estimator noise in the normalization, it sets the floor of every claim in a cell,
and every rung of a cell carries the same offset.

\subsection{Six methods}
\label{sec:methods}

\paragraph{Method A -- the variational ladder (universal; certified lower bounds).}
The identity of Eq.~(\ref{eq:variational}) is evaluated by Monte Carlo with a test function from one of
the families of Table~\ref{tab:rungs}, trained on FIT, early-stopped on VAL, and reported on EVAL with a
bootstrap error over images and the across-restart spread as the training systematic. Each rung is
initialized at the linear optimum (a zero-initialized head on top of $\that^{\dagger}x$) and the
initial value is included in the early-stopping selection, so a reported value can never fall below the
linear rung; every fit records its VAL trajectory and a flag \code{moved\_off\_init}, without which a
silent return of the initialization is indistinguishable from a null. Restarts matter more than epochs:
four to six restarts are used for every quoted cell.

\paragraph{Method B -- contracted cumulants (perturbative cross-check).}
The two norms in Eq.~(\ref{eq:perturbative}) are estimated directly from pairs of independent whitened
images, $\|\tilde\kappa_{3}(\that)\|^{2}_{F} = \E_{\rm pairs}[y\,y'\,(x^{\dagger}x')^{2}]$ with
$y = \that^{\dagger}x$, and a corresponding fourth-order expression that subtracts $O(M)$ disconnected
pieces from an $O(M)$ raw term. At $M = 16\,384$ pixels the fourth-order estimator is diagnostic-grade
only; the method is used to validate the ladder on weakly non-Gaussian test cases, not for quoted
numbers.

\paragraph{Method C -- Campbell analytics (closed forms).}
For marked Poisson models the cumulants of Eq.~(\ref{eq:campbell}) give the perturbative $\eta$ and the
one-point moments in closed form, with no simulation. In this paper Method C is used as a
\emph{validator}: the confusion ensemble's mean, rms, skewness and excess kurtosis match the Campbell
values to $0.006$\,\%, $0.014$\,\%, $0.031$\,\% and $0.32$\,\% and its power spectrum matches $P_{\rm conf}$ to
$0.9993 \pm 0.011$ (\S\ref{sec:results_confusion}) -- a check no other non-Gaussian model can run,
because the artifact models' mark distributions have no closed-form moments worth trusting.

\paragraph{Method D -- latent quadrature (exact; covariance mixtures).}
Equation~(\ref{eq:mixturescore}) evaluated on a latent grid: per image one FFT gives $|\tilde n_{\kv}|^{2}$,
each grid node costs one weighted sum for $\log\mathcal L(z)$, the posterior is a softmax over nodes, and
$\eta = \sigMF^{2}\,\E[(\sum_{z}W_{z}\,\tauv^{\dagger}\Nm_{z}^{-1}n)^{2}]$. Two refinements changed numbers
during our campaign and are part of the method: for amplitude-type latents the likelihood factorizes and
the amplitude axis can be made effectively continuous (a 41-node grid overshot the exact
$\E[a^{2}]\E[a^{-2}]$ limit; the factorized form lands on it to four digits), and heavy-tailed mixtures need
ensembles of $10^{4}$ maps because the EVAL-split value fluctuates by tens of percent on 1500. We learned a rule
the hard way: every exact method must be checked against an attainable lower bound, because a
``ceiling'' below an attainable value is a self-evident error.

\paragraph{Method E -- the attainability envelope (training-free; achievable variances).}
The envelope $V_{E}(A)$ of Eq.~(\ref{eq:methodE}) is evaluated from the latent posteriors that Method D
has already produced. It is the only route in this list that returns achievable variances rather than
bounds on them: every point of the curve is the variance of a real, marginally unbiased, implementable
estimator, so it brackets the truth from above while $\sigMF^{2}/\eta$ brackets it from below. Its cost is
seconds once Method D has run, and it is what tells a covariance-mixture result where its
advantage actually lives (\S\ref{sec:coherent_in_A}).

\paragraph{Method F -- training-free bounds and channel diagnostics.}
Three quantities that need neither an estimator nor a training set. The complete-data bound of
Eq.~(\ref{eq:cdbound}), evaluated from the generative spectra on a regularized band, is an upper bound on
$\eta$ wherever the contaminant is additive and removable. The well-inferable ceiling $\etawi$ of
Eq.~(\ref{eq:etawi}) is the corresponding upper bound for a covariance mixture. And the two-channel
decomposition of Eq.~(\ref{eq:twochannel_eta}) splits whatever advantage exists into its coherent and
incoherent parts -- the coherent severity $\sd[\avf{\delta}]$, the incoherent $\E[\Varf{\delta}]^{1/2}$, and
the scalar-weighting gain $\etaB = \E[c]\E[c^{-1}]$ -- alongside the two $\signorm$ baselines: the best
per-image scalar rescaling, and rescaling by the map rms, which is what a weight map actually measures.
These cost seconds, and they are the interpretive backbone of every covariance-mixture result.

Table~\ref{tab:cost} collects what each step costs. The point of the workflow -- compute the ceiling
first, train only where it is high -- rests on the asymmetry it shows: the training-free methods that
decide whether a cell deserves a campaign cost seconds to minutes, the ladder is an investment of GPU-hours
made only where they flag headroom, and a regression network costs more again.

\begin{table}[t]
  \caption{Cost of each step per (model, template) cell, on one GH200 GPU unless noted
  (Appendix~\ref{app:pipeline}).}
  \label{tab:cost}
  \centering\small
  \begin{tabular}{p{4.6cm}p{5.6cm}p{4.2cm}}
    \toprule
    step & returns & cost \\
    \midrule
    Methods A--C (anchor, pairs, Campbell) & $\sigMF$, the linear anchor, closed-form moments & seconds (CPU) \\
    Method F (bounds, channels) & $\etawi$, complete-data bound, two-channel split & seconds (CPU) \\
    Method D (exact quadrature) & exact $\eta$ of a covariance mixture & minutes (CPU, $12$--$20$\,k maps) \\
    Method E (envelope) & $V_{E}(A)$ & seconds once D has run \\
    ladder, one rung & a certified lower bound & $0.07$--$26$\,s per epoch; $0.1$--$8$\,GPU-h before early stopping \\
    ladder, full cell (five rungs, 4--6 restarts) & the bracket of Fig.~\ref{fig:ladders} & $\le 7$\,GPU-h \\
    ResNet regressor (\S\ref{sec:teaser}) & one certified network & $\sim 4$\,GPU-h per run (700 epochs on four GPUs) plus the injected training set \\
    \bottomrule
  \end{tabular}
\end{table}

\subsection{The convolutional rung, and what the ladder can see}
\label{sec:convrung}

The top rung of the ladder is a small convolutional network, and it must not be confused with the
regression network of \PaperII. This network has $\sim 2.5$\,k parameters; it is trained on noise-only maps to
approximate one direction of the score, $\that^{\dagger}s_{x}(x)$, by maximizing the unbiased raw bound
$2\E[\that\!\cdot\!\nabla f] - \E[f^{2}]$; and its output is a number whose only use is
Eq.~(\ref{eq:variational}). We call it the \emph{conv rung}. A ResNet regressor is a $10^{6}$-parameter
network trained on source-injected images to return an amplitude. The first measures a ceiling; the second
tries to reach it.

Three design rules, each the answer to a failure that produced a plausible wrong number before it was
caught (Appendix~\ref{app:pipeline}), define the \emph{fiducial} ladder. 
There are \emph{no free per-pixel
parameters}: spatial weighting lives in a fixed basis of 29 smooth Fourier modes ($k_{\max} = 3$ on the
$128^{2}$ grid) and only convolution kernels are free: with a fitting set comparable to the pixel count,
unregularized pixel maps find sample-covariance null directions before finding any genuine structure in
the noise. Training
maximizes the \emph{raw} bound with a learnable global scale, rather than the scale-free ratio
$(\E[\that\!\cdot\!\nabla f])^{2}/\E[f^{2}]$ that one obtains by optimizing the overall scale away: the
ratio's minibatch estimate is biased upward by the scatter of its denominator, and so rewards directions
that merely inflate the variance of $f$. And every rung is a linear head on nonlinear features standardized by detached running
statistics, so that a single learning rate is right for all coordinates. The convolutional features use
a smooth nonlinearity that is deliberately \emph{not odd}, that is, one with $g(-u) \ne -g(u)$, which
excludes $\tanh$ and its relatives. A network built entirely from odd activations returns an odd test
function, $f(-x) = -f(x)$. However, the part of the optimal $f^{\star}$ that carries skewness is quadratic in the
map and therefore even, so such a network is blind to the bispectrum by construction. Several of our
models are skewed.

The pooling basis turned out to be the one design choice with a physical consequence, and it was found by
calibration rather than by inspection (\S\ref{sec:sensitivity}). A fixed basis of smooth Fourier maps
(29 maps, $\sim 20$-px resolution) can represent test functions that are smooth across the map but not
the \emph{template-localized} ones -- for an i.i.d. whitened field the optimal test function is
$f^{\star} = \sum_{p}\that_{p}\,g(x_{p})$, a template-weighted sum of a pointwise nonlinearity, which for a
compact template is a single whitened pixel. The fiducial basis therefore also pools with maps built
from the whitened template itself. There are six of them, each normalized to unit rms: $\that$,
$|\that|$, $\that^{2}$, and $\that$ blurred by 1, 2 and 4\,px. Taken in union with the Fourier maps, they
place $f^{\star}$ inside the span for any template, at six pooled features per channel and still with no
free per-pixel parameter. Every ladder value in the paper is a
certified lower bound; the calibration of \S\ref{sec:sensitivity} says how tight.

The rung's protocol is 200 epochs; early stopping fires once the validation bound has gone 66 epochs
without improving (a \emph{patience} of 66, in the machine-learning term), and cannot fire at all during
a 30-epoch warm-up; and four to six restarts, one GPU per cell (Appendix~\ref{app:pipeline}
records the protocol erratum that made the warm-up necessary: the zero-initialized head gives the body
no gradient until the head has drifted, so the VAL bound sits at or below its anchor for 30--70 epochs
before it moves). Two limits of the instrument are stated here and used in \S\ref{sec:unmodelled}. In
every null cell the conv rung's VAL bound decays from the anchor to $0.75$--$0.85$ within that 66-epoch
window, because the head fits batch noise from the first epoch; the rung therefore has a \emph{sensitivity
floor} of $\eta \sim 1.1$--$1.2$, and the cubic rung, whose features are fewer, is the more sensitive
instrument at $\sim 1.05$. And a rung with $2.5$\,k parameters trained on noise alone is not a proxy for a trained
$10^{6}$-parameter regressor: it certifies what a small test function can find, and \S\ref{sec:attainability}
shows a ResNet reaching a value the conv rung does not, on a cell whose exact ceiling is known.

\subsection{The null battery}
\label{sec:nullbattery}

A number whose method has not been shown to return $1$ where $1$ is a theorem, and a known value where a
value is known, is not a measurement. Table~\ref{tab:nulls} lists the nine tests (N1--N8, with N4b the
paired form of N4) that every method must pass before a final run. Six of the
nine tests run on synthetic ensembles generated on the fly; N6 and N8 audit the stored campaign results
against theorems and across mode masks, and N7 runs the fiducial ladder on an exactly solvable model. The battery stands at nine tests and 100 assertions, all passing
checks -- the count is of individual assertions rather than of tests -- on the final code and the final
result set. Appendix~\ref{app:nullbattery} provides full details on what a null test is in this context, where the hundred
checks come from, and what four specific tests help to establish.

Two of the nine carry statements that the results sections lean on directly. \emph{N6} is the one theorem
that constrains the ladder from above: every rung is a certified lower bound, so no rung may exceed the
exact ceiling where one exists, or the complete-data bound where the structure is removable. The audit
covers the 36 cells that have such a ceiling and finds no violation. It deliberately does \emph{not}
require the rungs to be ordered -- each rung is an independent optimization over its own function class,
and the conv rung's sensitivity floor lies above the cubic rung's, so a cubic value above a convolutional
one is legitimate and does occur. \emph{N7} is the only test that asks the fiducial ladder to recover a
\emph{known} $\eta > 1$ at full map size: confusion with a beam-correlated Gaussian companion
(\S\ref{sec:confusion_setup}, calibrated in \S\ref{sec:sensitivity}) whitens to i.i.d.\ pixels, so its $\eta = I_{1} = 3.5688$ is exact and
template-independent by theorem. The ladder recovers $\sim 0.9$ of it on both templates and
exceeds it nowhere, which is the calibration behind every statement in \S\ref{sec:results_confusion} that
a ladder value is a lower bound of known tightness. Its first run failed, and what that failure exposed
is recorded in Appendix~\ref{app:nullbattery}.

\begin{table}[t]
  \caption{The null battery. Each test names what it catches; six run on synthetic ensembles before
  every campaign, N6 and N8 audit the stored results, N7 recovers a known answer through the fiducial ladder. Final state: nine tests, 100 assertions, all passing, broken down by test
  in Appendix~\ref{app:nullbattery}.}
  \label{tab:nulls}
  \centering\small
  \begin{tabular}{lp{7.2cm}p{5.4cm}}
    \toprule
    \# & test & what it catches \\
    \midrule
    N1 & Gaussian null, white: every method returns $\eta = 1$, cumulant estimators return 0 & DFT normalization, split leakage, disconnected terms \\
    N2 & Gaussian null, red, through the estimated-PSD whitening path & PSD estimation and masking \\
    N3 & known-$\eta$ validation on an i.i.d. two-component scale mixture ($\eta$ by 1-D quadrature), analytic score and a trained rung & the evaluation machinery independently of training \\
    N4 & i.i.d. template independence: $\eta$(pixel) $= \eta$(extended) & the theorem of \S\ref{sec:tier2} \\
    N4b & overlap-law ordering, paired: sparse smoothed events on a white background, extended $>$ compact, bootstrapping the per-image difference & the physically transparent form of N4; the unpaired version is under-powered \\
    N5 & latent quadrature: amplitude-only $= \E[\sigma^{2}]\E[\sigma^{-2}]$; slope-only $> 1$ & grid quantization; the closed-form closure \\
    N6 & ladder $\le$ ceiling: every physical rung at or below the exact ceiling (quadrature, exact $I_{1}$) or the complete-data bound, within $2\sigma$, for every stored cell; ordering \emph{not} required & estimator bugs and leakage, by theorem; 36 cells \\
    N7 & known-$\eta$ recovery on full-size maps: confusion with a beam-correlated companion, exact $\eta = I_{1} = 3.5688$ for every template, recovered by the fiducial ladder ($\ge 80$\,\%, both templates), not exceeded, template-independent & the ladder's tightness and reach; it caught the pooling-basis flaw \\
    N8 & band regularization: $\sigMF$, band statistics and bound stable against the mode mask & the missing-floor defect \\
    \bottomrule
  \end{tabular}
\end{table}

\begin{table}[t]
  \caption{The rungs of the variational ladder (Method A). Parameter counts are for the fiducial
  architecture; the linear rung is the untrained matched filter.}
  \label{tab:rungs}
  \centering\small
  \begin{tabular}{lp{7.6cm}p{2.2cm}p{3.2cm}}
    \toprule
    rung & family & trainable parameters & statistical target \\
    \midrule
    \rung{linear} & $\that^{\dagger}x$, untrained anchor & 0 & sanity: $\eta_{\rm lin} \simeq 1$ \\
    \rung{reweight} & mixture matched filter, Eq.~(\ref{eq:reweight}): band/sector templates with weights from $\hat g$, $\log\hat g$, $1/\hat g$ & $1.2$--$1.8$\,k & covariance mixtures (Tier 1) \\
    \rung{quadratic} & basis-pooled products of convolutional features, order 2 & 255--267 & bispectrum \\
    \rung{cubic} & + order-3 products & $\sim 500$ & trispectrum \\
    \rung{conv} & basis-pooled smooth-nonlinearity convolutional features & $\sim 2.5$\,k & generic local nonlinearity \\
    \bottomrule
  \end{tabular}
\end{table}

\section{Noise models and simulation setup}
\label{sec:noise_models}

\begin{figure}[t]
  \centering
  \includegraphics[width=\textwidth]{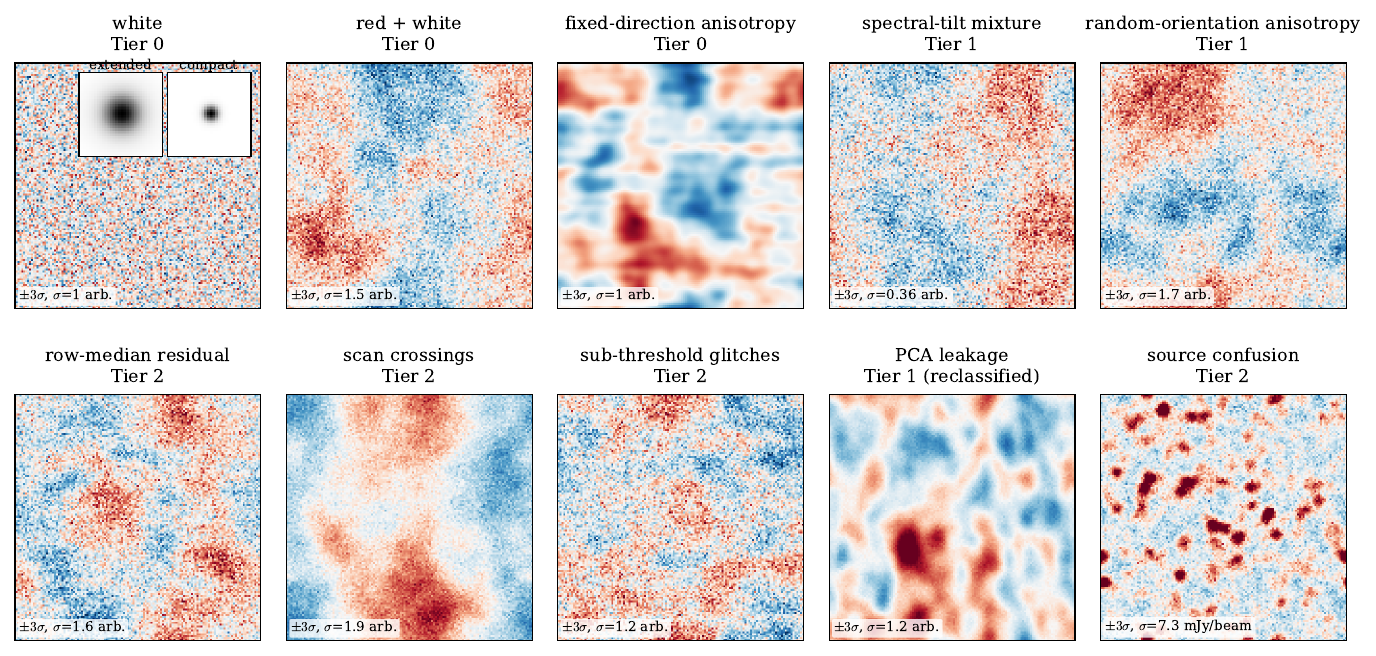}
  \caption{One realization of ten of the twelve noise models of Table~\ref{tab:taxonomy} (the floorless
  isotropic $1/f^{3}$ control and the positive-crossing variant are not shown), drawn from the campaign
  registry with its own seed, in the realistic (white-floored) configuration except for two of the Gaussian
  controls: white noise is itself a floor, and the fixed-direction anisotropy is one of the floorless
  controls of \S\ref{sec:results_gaussian}.
  All panels are $128 \times 128$-pixel maps with the same 5-px beam, \ie\ the same field in the paper's
  units; only the confusion panel carries physical units. Color scales are $\pm 3\sigma$ of each map so
  that the texture of each model is visible ($\sigma$ in the paper's arbitrary units, in $\mjb$ for
  confusion).
  Insets in the first panel: the extended ($\beta$-model $\circledast$ beam, $r_{c} = 7$\,px) and compact
  (beam) templates. The artifact models (row 2, second and third panels) are deliberately sub-threshold: their
  structure is not visible by eye, which is the point of the question the paper asks about them. PCA
  leakage is listed under Tier~1 because \S\ref{sec:results_mixtures} shows it to be a Gaussian scale
  mixture.}
  \label{fig:gallery}
\end{figure}

The noise models are modeled on the noise that an amplitude estimator encounters in a real-world millimeter or submillimeter survey map -- after the time-ordered data have been filtered, cleaned and projected -- not
textbook distributions. Each is a Gaussian background with one physically motivated departure, so that
the ceiling measures that departure and nothing else. Figure~\ref{fig:gallery} shows one realization of
each; Table~\ref{tab:taxonomy} lists their origin, class, latent, leading non-Gaussian order and floor.

\subsection{Conventions}
\label{sec:conventions}

Maps are $128^{2}$ pixels. The beam is a Gaussian of FWHM $5$\,px; the extended template is a
$\beta$-model of core radius $r_{c} = 7$\,px convolved with the beam and the compact template is the beam
itself, both centered and known exactly. Amplitudes are quoted in units of the pre-beam peak of a unit
source, which is arbitrary; only the confusion rows carry physical units ($\mjb$, \S\ref{sec:confusion_setup}).
Power spectra use the unnormalized-FFT convention, $\E|\tilde n_{\kv}|^{2} = P(\kv)$, with $\kv$ in
cycles per pixel. The pipeline order is the physical one: source, then correlated (atmospheric) noise,
then beam smoothing of both, then \emph{unsmoothed} instrumental white noise, then artifact injection,
then any nonlinear map-domain processing. The order matters twice: the artifacts enter after the optics
(readout and cosmic-ray residuals are not beam-smoothed), and the white floor enters before any
data-dependent operation (a row median of a floored map is not the floored median of a map). The DC mode
of every noise map is kept and regularized to the fundamental-frequency power (for confusion, to $\text{mean}^{2}n_{\rm pix}^{2}$,
\S\ref{sec:confusion_setup}), never zeroed
(\S\ref{sec:unmodelled}).

\subsection{The noise suite}
\label{sec:taxonomy}

\begin{table}[t]
  \caption{The noise models. ``Latent'' is the per-image quantity that makes the covariance vary;
  ``order'' the leading non-Gaussian polyspectrum; $\sigma_{w}$ the white floor of the realistic
  configuration in map units (\S\ref{sec:floor}); $\fs$ the fraction of the noise power in the
  template's Fisher band carried by the structured component (compact / extended) and $\xirms$ the
  rms matched-filter significance of a single event, both in the floored configuration.
  ($^{\ast}$: in $\mjb$.) Code names of
  the registry are in Appendix~\ref{app:master}.}
  \label{tab:taxonomy}
  \centering\footnotesize
  \begin{tabular}{p{2.3cm}p{3.2cm}p{2.05cm}p{1.4cm}p{0.95cm}p{0.9cm}p{1.5cm}p{0.8cm}}
    \toprule
    model & physical origin & class & latent & order & $\sigma_{w}$ & $\fs$ (cmp/ext) & $\xirms$ \\
    \midrule
    \mWhite & detector noise & Gaussian & -- & -- & 1 & -- & -- \\
    \mRed & atmospheric $1/f^{3}$ residual & Gaussian & -- & -- & 0 & -- & -- \\
    \mReal & beam-smoothed $1/f^{3}$ + unsmoothed white & Gaussian & -- & -- & 1 & -- & -- \\
    \mAniso & scan-synchronous striping, fixed scan direction & Gaussian & -- & -- & 0 & -- & -- \\
    \midrule
    \mTilt & patch-to-patch variation of the residual atmospheric slope and level & cov.\ mixture & slope, amplitude & -- & 0.27 & -- & -- \\
    \mRot & scan direction varying between fields & cov.\ mixture & angle & -- & 1.04 & -- & -- \\
    \mPCA & leaked principal components of atmospheric-mode removal & cov.\ mixture (scale) & leak power $c$ & (4th) & 0.10 & 0.08 / 0.50 & 16.9 \\
    \midrule
    \mMedian & row-median destriping of anisotropic noise & processing-induced & -- & 4th & 1.06 & -- & -- \\
    \mCrossS & scan-turnaround residuals, signed & sub-threshold artifacts & positions, marks & 4th & 0.51 & 0.056 / 0.049 & 3.3 \\
    \mCrossP & the same, one-signed & sub-threshold artifacts & positions, marks & 3rd & 0.48 & 0.051 / 0.058 & 4.4 \\
    \mGlitch & cosmic-ray residuals below the deglitcher & sub-threshold artifacts & positions, marks & 3rd & 0.86 & 0.042 / 0.033 & 1.36 \\
    \mConf & unresolved submillimeter galaxies & marked Poisson & positions, fluxes & 3rd & $3.22^{\ast}$ & 0.81 / 0.97 & 0.31 \\
    \bottomrule
  \end{tabular}
\end{table}

\paragraph{Gaussian controls (Tier 0).}
Four models in which the matched filter with the correct two-dimensional spectrum is optimal by the
theorem of \S\ref{sec:mf}, included so that the machinery can be shown to return $\eta = 1$ at full map
size: white noise; isotropic $1/f^{3}$ noise, the standard description of the residual atmospheric
fluctuations in a mm-wave map (a Kolmogorov--Taylor screen gives a two-dimensional power
spectrum between $-11/3$ and $-8/3$ in its thick- and thin-layer limits:  
see \citealt{Sayers2010}, and also \citealt{LayHalverson2000}, where the indices were halved for using rms rather than variance); 
the ``realistic'' combination of beam-smoothed $1/f^{3}$ with an
unsmoothed white floor; and a two-component spectrum with a scan-synchronous ridge of fixed direction,
$P \propto A_{s}^{2}/|k_{\parallel}|^{3} + A_{i}^{2}/|\kv|^{3}$, the map-domain signature of striping artifacts along
a fixed scan direction. Fixed-direction anisotropy is Gaussian; the filter must simply use the full
two-dimensional spectrum. High-pass-filtered variants of these models are also Gaussian and matched-filter
safe when the filter is applied to the data (signal and noise alike); the variant in which only the noise
is filtered manufactures a noiseless signal channel and is the first of the cautionary tales of
\S\ref{sec:unmodelled}.

\paragraph{Covariance mixtures (Tier 1).}
Three models in which each map is Gaussian but its covariance is not the same from map to map: two
built that way, and PCA leakage, which 
was reclassified (see \S\ref{sec:pca}).
\emph{\mTilt:} the slope and amplitude of the $1/f^{3}$ residual are drawn per map,
$\alpha \sim \mathcal N(3, 0.5^{2})$ and $a \sim \mathcal N(5, 0.8^{2})$ clipped below at $10^{-6}$ (a
$2 \times 10^{-10}$ event, never realized in any ensemble; the clip is what keeps $\E[\sigma^{-2}]$ of the closure in Appendix~\ref{app:nullbattery} finite), about a pivot at
$k_{\rm piv} = 1$\,cycle\,px$^{-1}$. The physical origin is less the raw atmosphere -- whose spectral shape
is stable and whose amplitude tracks the water vapor \citep{Sayers2010} -- than what the pipeline leaves
behind: the residual after common-mode or principal-component removal has a shape that depends on how well
the atmosphere was removed, and the knee scales with wind speed over scan speed
\citep{Aguirre2011}. A slope-only and an amplitude-only variant are used for the analytic closures.
\emph{\mRot:} the scan-ridge direction of the anisotropic model is drawn uniformly on $[0,\pi)$ per
map, because each field of a survey, and each cross-linked pass over one field, is scanned at its own
position angle; the ensemble spectrum is isotropized and the fixed filter is mismatched on every map.

\paragraph{Genuinely non-Gaussian survey noise (Tier 2).}
Whether map-domain noise is non-Gaussian is decided by one question: is the processing operator
data-dependent? A fixed linear operator maps Gaussian noise to Gaussian noise exactly, and only reshapes
the spectrum -- polynomial baseline removal and fixed-basis projections included. Data-dependent steps
(adaptive common-mode or PCA removal, deglitching, clipping, medians, iterative weighting) can manufacture
non-Gaussianity, and the imperfect removal of a structured non-Gaussian component leaves a non-Gaussian
residual even under a linear operator. Five models span these routes, under four headings (PCA leakage is described here, where it was built,
and counted under Tier~1, where it belongs; source confusion has \S\ref{sec:confusion_setup}). All but one
are built as a full-rank Gaussian field plus independent additive structure, so their densities exist
and are smooth and the location-family framework of \S\ref{sec:framework} applies without caveat; the
one exception is noted under its heading.
\emph{\mMedian:} the anisotropic Gaussian model with a row-wise median subtracted, a common destriping
step; the median is nonlinear and sign-symmetric, so the residual's leading non-Gaussianity is at fourth
order. In the campaign the median acts on the noise alone and the source is added afterwards, as the data
model $d = A\tauv + n$ requires; that idealization carries an exact constraint -- every row of $n$ has
median zero -- which the source does not obey, and \S\ref{sec:median} says what follows from it.
\emph{Scan crossings:} a Poisson number (mean 8) of rows carrying a structured residual of correlation
length 30\,px and $t_{2.5}$-distributed amplitude, in two sign conventions -- symmetric (signed amplitudes,
trispectrum-led) and positive (one-signed, bispectrum-led) -- on the isotropic red background. The sign is
a property of the contaminant rather than of the cleaning, and the two conventions stand for two ways a
residual can be left behind. Where a scan-synchronous \emph{drift} has been fitted and subtracted -- a
detector baseline, a common mode, the offsets a destriper solves for at crossing points -- the estimate of
it is unbiased but noisy, so what survives scatters about zero: the symmetric convention. Where a
one-signed \emph{emission} falls below the flagging threshold and is therefore never subtracted at all --
ground or far-sidelobe pickup on a few scans, warm optics, atmospheric loading a template did not
capture -- what remains keeps the sign of the source that produced it: the positive convention.
\emph{\mGlitch:} 25 cosmic-ray-like residuals per map with an exponential decay of 5\,px along the scan,
below the deglitching threshold, on the anisotropic background.
\emph{\mPCA:} a Poisson number (mean 3) of smooth anisotropic modes of correlation lengths $20 \times 40$\,px,
each a fresh Gaussian field with a heavy-tailed ($t_{2.5}$) amplitude, imitating the leakage of
atmospheric principal components back into the map whenever the truncated basis is wrong. As
\S\ref{sec:results_mixtures} shows, this model is a Gaussian scale mixture and belongs with Tier~1, where Table~\ref{tab:taxonomy} lists it;
it is described here, among the artifacts it was built with, because the reclassification is a result.
A $t_{2.5}$ amplitude has a finite variance but no finite third or fourth moment, so the population
skewness and kurtosis of the crossing and PCA artifacts do not exist; the sample values of any finite
ensemble are set by its few largest draws and do not converge with ensemble size.\footnote{In the
campaign ensembles the sample excess kurtosis of the crossing amplitudes grows from $39$ to $207$ as the
sample grows from $10^{3}$ to $10^{5}$ marks, and the sample skewness of the sign-symmetric set changes
sign between subsamples; the largest realized amplitudes are $46$ (PCA), $59$ and $101$ (crossings) in
map units, from $13\,500$ maps.} Every cumulant quoted for these two models, and the
``bispectrum-led''/``trispectrum-led'' vocabulary, therefore refers to the campaign ensembles themselves.
The results do not depend on these moments: the exact mixture ceiling of \S\ref{sec:pca} needs only
$\E[a^{2}]$ and $\E[(\sigma_{w}^{2} + \sum a^{2})^{-1}]$, both finite, and the rung values are
finite-sample optima bounded above by that ceiling or by the complete-data bound.

The artifact models are \emph{sub-threshold by design}: their per-event matched-filter significance in the
floored configuration is $\xirms = 1.4$--$4.4$ (Table~\ref{tab:taxonomy}), and their structured
component carries 3--6\,\% of the noise power in either template's Fisher band. They ask the question a
pipeline developer would ask: do the residuals that survive cleaning of time-ordered data leave anything for a nonlinear estimator?

\subsection{The white floor, and the two configurations}
\label{sec:floor}

Every real map carries an instrumental white floor that is not beam-smoothed, and the floor is what
terminates the per-mode signal-to-noise beyond the beam scale. Without it, a compact template on a
beam-smoothed background has a Fisher weight $|\tautil|^{2}/P = B^{2}/(B^{2}P_{\rm red}) \propto k^{3}$ in
which the beam cancels and the weight rises to whatever mode a numerical mask stops at, so that every
compact-template quantity reports the mask rather than the noise (\S\ref{sec:unmodelled}). The floor is
therefore a physical parameter and is set by a rule, rather than using a default. 
For each model,
\begin{equation}
  \sigma_{w}^{2} \;=\; \frac{\Pbar_{\rm bg}(k_{\rm knee})}{n_{\rm pix}}, \qquad
  k_{\rm knee} = 2\,\exp\avf{\log k}^{\rm ext},
  \label{eq:floor_rule}
\end{equation}
the smallest floor that makes the Fisher-band integral converge, with the white/red knee one octave
above the extended template's band centroid, where $\Pbar_{\rm bg}$ is the azimuthally averaged spectrum
of the Gaussian background alone. Because the rule uses the background only, the floor is independent
of the artifact severity, and because it uses the same geometric-mean statistic as the band centroid,
severity and floor share one coordinate. The calibrated values (Table~\ref{tab:taxonomy}) spread by a
factor of ten, which is physical: the anisotropic background is loud at $k \approx 0.09$ and needs a
large floor to put the knee there, while the leaked modes of the PCA model push its extended band up to
$k_{\rm ext} = 0.074$, where the isotropic background is already weak. Verified consequences: the
extended template keeps 98--99\,\% of its Fisher weight below the knee, so its ceilings move by a few
percent; the compact template's $\sigMF$ drift against the mode mask falls from 24--46\,\% (over 200\,\% for
confusion) to $0.00$\,\% in every model (N8, Appendix~\ref{app:nullbattery}). The floor is a numerical prescription calibrated on each model's own Gaussian
background, not an instrument's noise level: every ceiling in this paper is conditional on the spectra,
priors, flux cuts and floors stated here, and none is a generic survey limit.

The floor did two more things that it was not specifically designed for. It discharged the event severity -- the
glitch model's $\xirms$ fell from $20$ to $1.4$, to the boundary between the two regimes of
\S\ref{sec:tier2} -- and it made the variational estimator statistically usable, by diluting
the heavy tails of the floorless ensembles (restart spreads of $1.7\times$ became $10$\,\%).

Two configurations are kept. The \emph{floored} set is the paper's realistic configuration, and every
number in \S\S\ref{sec:results_gaussian}--\ref{sec:results_confusion} refers to it unless stated. The
\emph{floorless} set is the theoretical limit in which two analytic closures hold exactly -- the
amplitude-only identity $\eta = \E[\sigma^{2}]\E[\sigma^{-2}]$, which a non-scaling floor destroys, and
the factorized quadrature -- and it is the configuration in which two of the three unmodelled channels of
\S\ref{sec:unmodelled} are visible. Its variational values above the mixture-filter rung are not
physical and are marked accordingly in Appendix~\ref{app:master}. The two sets are never tabulated side by side.

\subsection{Source confusion}
\label{sec:confusion_setup}

The one model with physical units, and the one in which the non-Gaussian component \emph{is} the noise.
The paper's pixel convention is read as a $20''$ beam sampled at $4''$ per pixel (an $8.5'$ field), so that
$350\,\mum$ Schechter number counts fitted in \citet{BasuGuerreroBertoldi2026} apply unchanged:
$\dnds = (\phi_{*}/S_{*})(S/S_{*})^{\alpha}e^{-S/S_{*}}$ with $\alpha = -1.89$, $\phi_{*} = 7014\,{\rm deg^{-2}}$
and $S_{*} = 19.0$\,mJy. Sources are placed at Poisson positions on pixel centers with fluxes drawn between
$\Smin = 0.1$\,mJy and $\Scut = 100$\,mJy and convolved with the beam: a marked Poisson process with
$\psi$ equal to the beam, Eq.~(\ref{eq:campbell}). Analytically, $\sigc = 6.45\,\mjb$, $\Nbeam = 28$
sources per beam, pixel skewness $2.12$ and excess kurtosis $9.09$, with $\sim 16\,300$ sources per map;
no faint-end Gaussian substitution is used, so that systematic does not arise. The maps keep their
Poisson-fluctuating mean, and the matched filter treats the DC mode as noise at the level
$\text{mean}^{2}n_{\rm pix}^{2}$, which is conservative by $\sim 1$\,\% for the extended template.

The white-floor amplitude rule of Eq.~(\ref{eq:floor_rule}) does not transfer to this model: confusion puts the extended
band at $\exp\avf{\log k} = 0.020$, where the rule would demand $\sigma_{w} = 6.5\,\sigc$ and drown the
model. The fiducial row instead uses the instrument-to-confusion ratio adopted for
\emph{Herschel}/SPIRE- and CCAT-like surveys, $f_{N} = \sigma_{N}/\sigc = 0.5$, \ie\ $\sigma_{w} = 3.22\,\mjb$
-- a stated physical level rather than a convergence criterion. Three rows exist: pure confusion, which is
a diagnostic only (its whitened field is exactly i.i.d. and its $\eta$ is provably infinite,
\S\ref{sec:results_confusion}); confusion with the unsmoothed floor, the fiducial row; and confusion with
a Gaussian companion of the same map-level amplitude injected \emph{before} the beam, so that it shares
the beam and the whitened pixels are again i.i.d. -- the exactly solvable row that calibrates the ladder in
\S\ref{sec:sensitivity}. Clustering of the sources is omitted deliberately: a clustered (Cox) process adds a term
$\bar\lambda^{2}\langle a\rangle^{2}P_{\delta}(\kv)$ to $P_{\rm conf}$ and changes both the matched-filter
variance and the non-Gaussian Fisher information, and its effect on $\eta$ is not computed here
(\S\ref{sec:limitations}). Appendix~\ref{app:confusion} has the Campbell cumulants, the validation
checks and the exact solution.

\section{Results I: Gaussian controls}
\label{sec:results_gaussian}

The four Gaussian models -- white, red, red\,+\,white and fixed-direction anisotropic -- are the negative
control, and the machinery of \S\ref{sec:computing} must return $\eta = 1$ on them at full map size
before any other number can be believed. It does. Three of the four are run floorless -- the floor
changes nothing in the theorem, and white noise is itself a floor -- and red\,+\,white is the floored
control (Tables~\ref{tab:master_floored}--\ref{tab:master_floorless}). Across the eight (model, template) cells the linear
anchor and every trained rung sit at $\eta = 0.94$--$1.06 \pm 0.04$, the spread being the EVAL-split
normalization noise of \S\ref{sec:infrastructure}; no rung moved off its initialization on any restart,
and the per-fit records of \S\ref{sec:methods} -- the validation trajectory and the
\code{moved\_off\_init} flag -- certify each null rather than merely reporting it. Two of the four cases deserve a
remark each. Fixed-direction anisotropy is Gaussian: a matched filter that uses the full
two-dimensional spectrum is optimal, and a network could improve on it only if the filter were built
from an azimuthally averaged spectrum -- a mis-specified covariance, not a property of the noise. And
high-pass filtering, applied to signal and noise alike, leaves the per-mode signal-to-noise invariant and
the noise Gaussian; the ``filter the noise but not the signal'' variant is the first of the three
unmodelled channels of \S\ref{sec:unmodelled}.

What this certifies is the theorem of \S\ref{sec:mf} reproduced by measurement: under Gaussian noise
of known covariance there is nothing for a nonlinear estimator to find, and a method that claims to
find something on these models has a bug. The null battery's N1 and N2 are the same statement at small
scale; this section is the same statement at the scale, whitening path and training protocol of every
run that produced a quoted number.

\subsection{A regression network on Gaussian noise}
\label{sec:teaser}

\begin{figure}[t]
  \centering
  \includegraphics[width=\textwidth]{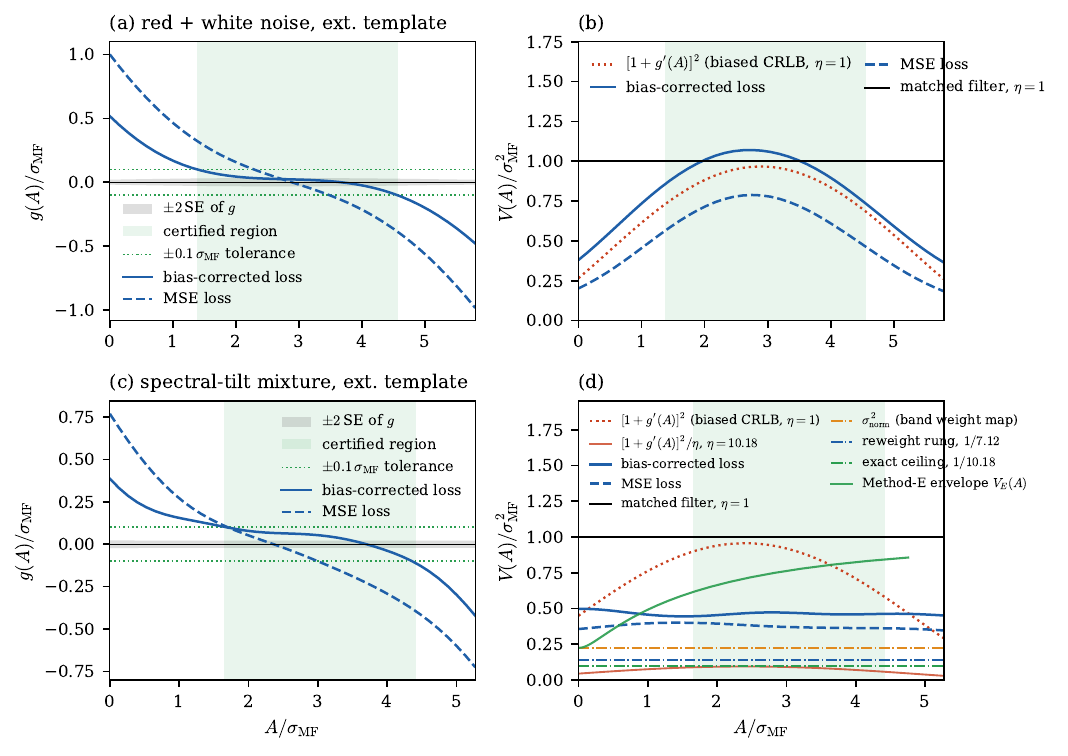}
  \caption{ResNet regressors evaluated with the certification of \S\ref{sec:cnn}, on 4000 independent
  noise realizations at 41 amplitudes (common noise across amplitudes; best-validation checkpoints).
  (a), (c): conditional bias $g(A)$ in units of $\sigMF$ with its $\pm 2$\,SE band; the shaded region is
  the certified region of the bias-corrected run, the longest run of amplitude bins with
  $|g| < \max(2\,{\rm SE},\,0.1\,\sigMF)$ (dotted lines: the $\pm 0.1\,\sigMF$ tolerance). (b), (d):
  conditional variance $V(A)/\sigMF^{2}$ with the Gaussian-normalized envelope $[1 + g'(A)]^{2}$ of the same
  run -- the biased Cram\'er--Rao bound at $\eta = 1$ -- and the matched-filter level $\eta = 1$; for the
  mixture also the actual biased bound $[1 + g'(A)]^{2}/\eta$, the band weight-map level $\signorm^{2}$, the
  \rung{reweight} rung, the exact ceiling, and the attainability envelope $V_{E}(A)$ of the best rescaled
  matched filter. $R$ of Eq.~(\ref{eq:Rdef}) is the ratio of either blue curve to the red dotted one, averaged
  over the shaded region. Solid: mean-squared error with a conditional-bias penalty; dashed: plain mean-squared
  error. (a)--(b): red\,+\,white noise (extended template, $L = 5$, $r = 0.60$): $R = 1.107 \pm 0.021$ [Eq.~(\ref{eq:Rdef})]
  ($1.150 \pm 0.021$ for MSE). (c)--(d): spectral-tilt mixture with white floor (extended, $L = 7$,
  $r = 0.66$ with $\sigMF = 1.33$ measured on the training generator): $R = 0.550 \pm 0.015$ ($0.550 \pm 0.038$).}
  \label{fig:teaser}
\end{figure}

To show the adjudication of \S\ref{sec:ladder} on a case where the answer is a theorem, a ResNet
regressor ($1.4 \times 10^{6}$ parameters, 34-layer residual architecture, dihedral augmentation) was trained
on the red\,+\,white model with the extended source injected at amplitudes uniform on $[0, L]$, $L = 5$
($r = \sigMF\sqrt{12}/L = 0.60$), with two losses: plain mean-squared error, and mean-squared error with a
conditional-bias penalty ($\lambda = 15$ over eight amplitude bins). The training maps were generated by the
campaign's own registry generator with seeds disjoint from the noise-only ensembles, and the matched-filter
noise level $\sigMF$ was re-measured on 2000 noise-only maps of the training generator ($0.863$; the analytic value to
$0.1$\,\%). Figure~\ref{fig:teaser} (top) shows the result. The certified region is the longest run of amplitude
bins in which $|g| < \max(2\,{\rm SE},\,0.1\,\sigMF)$, a stated tolerance rather than a statistical null:
with 4000 realizations the $2\,$SE band is $\pm 0.03\,\sigMF$, and the residual bias of the network --
up to $0.09\,\sigMF$ -- is resolved inside it. For the bias-corrected network that region is
$[1.4, 4.6]\,\sigMF$; its variance there is
$0.97\,\sigMF^{2}$ against a biased-Cram\'er--Rao envelope $[1 + g']^{2} = 0.88$, so the shrinkage-cleaned
ratio of Eq.~(\ref{eq:Rdef}) is $R = 1.107 \pm 0.021$. The plain-MSE network shrinks harder ($g' \simeq -0.2$), its raw variance is
$0.77\,\sigMF^{2}$ -- a number that, quoted alone, would read as a 23\,\% advantage over the matched filter --
and $R = 1.150 \pm 0.021$. Over the full amplitude grid, for every Gaussian-control run, $R$ never falls below
$1.05$ (a white-noise cell gives $R = 1.063 \pm 0.021$). This is the theorem reproduced by a trained network:
every apparent variance below $\sigMF^{2}$ is bias-gradient credit, and once it is divided out a
well-trained network is 6--15\,\% \emph{less} efficient than the matched filter, not more (11\,\% for the
bias-corrected run on red\,+\,white noise). 
The mean-squared-error runs reach their best
validation loss at epochs 30--80 and then memorize the training set, and the numbers quoted are from
best-validation checkpoints evaluated on an independent ensemble. 
\PaperII\ carries these training details and a full clarification of the ResNet architecture, including 
the definition of the locally-unbiased band.

\section{Results II: covariance mixtures}
\label{sec:results_mixtures}

Three models turn out to be covariance mixtures: the two built as such -- the spectral-tilt mixture and
the random-orientation anisotropy -- and PCA leakage, which was built as a non-Gaussian artifact model
and is reclassified here (\S\ref{sec:pca}). For all three the ceiling is an \emph{exact} number from the
latent quadrature of \S\ref{sec:tier1}, the two channels can be separated, and a constructed estimator
-- the mixture matched filter -- can be run against the ceiling. Table~\ref{tab:mixtures} collects the
numbers; Fig.~\ref{fig:ladders} shows the variance gains in each ladder. 
All values are in the white-noise floored configuration.

\begin{table}[t]
  \caption{Covariance mixtures in the realistic configuration: exact ceilings, the well-inferable limit
  $\etawi$ of Eq.~(\ref{eq:etawi}), the two severity coordinates (coherent scatter $\sd[\avf{\delta}]$,
  incoherent rms $\E[\Varf{\delta}]^{1/2}$), the two exact factors of the split (scalar-weighting gain
  $\etaB$ and the filter re-shaping remainder $\eta/\etaB$; the second-order forms of \S\ref{sec:tier1} do
  not reproduce them at these severities), the best rung of the ladder, and the survey reading of the
  ceiling; the bracketed figure is the fraction of the ceiling attained. $\sigMF$ is the empirical
  matched-filter rms on the ensemble (Tables~\ref{tab:master_floored}--\ref{tab:master_floorless} list
  the spectrum-predicted one). Errors are bootstrap over the EVAL split.}
  \label{tab:mixtures}
  \centering\small
  \setlength{\tabcolsep}{3pt}
  \begin{tabular}{p{3.4cm}lccccccccc}
    \toprule
    model & template & $\sigMF$ & $\eta$ (exact) & $\etawi$ & coh.\ & inc.\ & $\etaB$ & $\eta/\etaB$ & \rung{reweight} & area \\
    \midrule
    \mTilt & extended & 1.21 & $10.18 \pm 1.45$ & -- & 1.36 & 0.49 & 4.46 & 2.28 & $7.12 \pm 1.13$ (70\,\%) & $10\times$ \\
    \mTilt & compact & 7.06 & $1.79 \pm 0.08$ & -- & 0.50 & 0.44 & 1.26 & 1.42 & $1.45 \pm 0.08$ (81\,\%) & $1.8\times$ \\
    \mTiltS & extended & -- & $7.76 \pm 0.79$ & 8.58 & 1.33 & 0.48 & 4.21 & 1.84 & -- & $7.8\times$ \\
    \mRot & extended & 1.03 & $1.25 \pm 0.05$ & -- & 0.002 & 0.34 & 1.000 & 1.25 & $+3$\,\% over anchor & $1.25\times$ \\
    \mRot & compact & 14.9 & $1.20 \pm 0.04$ & -- & 0.001 & 0.37 & 1.000 & 1.20 & $+6$\,\% over anchor & $1.2\times$ \\
    \mPCA & extended & 1.68 & $2.14 \pm 0.05$ & 2.139 & 0.22 & 0.36 & 1.32 & 1.62 & $1.655 \pm 0.052$ (77\,\%) & $2.1\times$ \\
    \mPCA & compact & 4.32 & $1.09 \pm 0.02$ & 1.10 & 0.005 & 0.06 & 1.02 & 1.07 & anchor & $1.1\times$ \\
    \midrule
    \mTiltA & both & -- & $1.1166$ (analytic) & -- & -- & 0 & 1.117 & 1.00 & -- & $1.1\times$ \\
    \bottomrule
  \end{tabular}
\end{table}

\subsection{Exact ceilings}
\label{sec:exact_ceilings}

The spectral-tilt mixture seen by the extended template has the one large ceiling of the paper,
$\eta = 10.18 \pm 1.45$: an ideal estimator is worth ten times the survey time for extended sources
in this noise. Seen by the compact template the same noise gives $1.79 \pm 0.08$. The random-orientation
anisotropy gives $1.25$ and $1.20$ for the two templates; PCA leakage $2.14 \pm 0.05$ and $1.09 \pm 0.02$.
Three checks tie these numbers down. The well-inferable limit of Eq.~(\ref{eq:etawi}), which needs no
pivot and no perturbative expansion, agrees with the full quadrature wherever the latent is well
inferred: $8.58$ against $7.76 \pm 0.79$ for the slope-only tilt (a $1\sigma$ difference on a
quadrature whose 1500-image error is 10\,\%), $2.139$ against $2.140 \pm 0.050$ for PCA leakage. 
This is consistent with, and within Monte Carlo error of, the
bound of Eq.~(\ref{eq:etawi_bound}), and the
latent \emph{is} well inferred: the posterior width of the spectral slope is $\sd(\alpha|n) = 0.020$
against a prior width of $0.5$, and of the leak power $\sd(\log c|n) = 0.21$ against $1.9$. The
amplitude-only mixture, in which only the overall noise level varies, has the closed form
$\eta = \E[\sigma^{2}]\E[\sigma^{-2}] = 1.1166$ for our prior and measures $1.1164 \pm 0.0128$ on $20\,000$ maps ($1.14 \pm 0.05$ on the 1500-map EVAL split of
Table~\ref{tab:master_floorless}), with the
template independence a purely coherent latent requires ($0.0096 \pm 0.0174$ between the two templates).

\subsection{Two channels on the regularized band}
\label{sec:two_channels_results}

\begin{figure}[t]
  \centering
  \includegraphics[width=\textwidth]{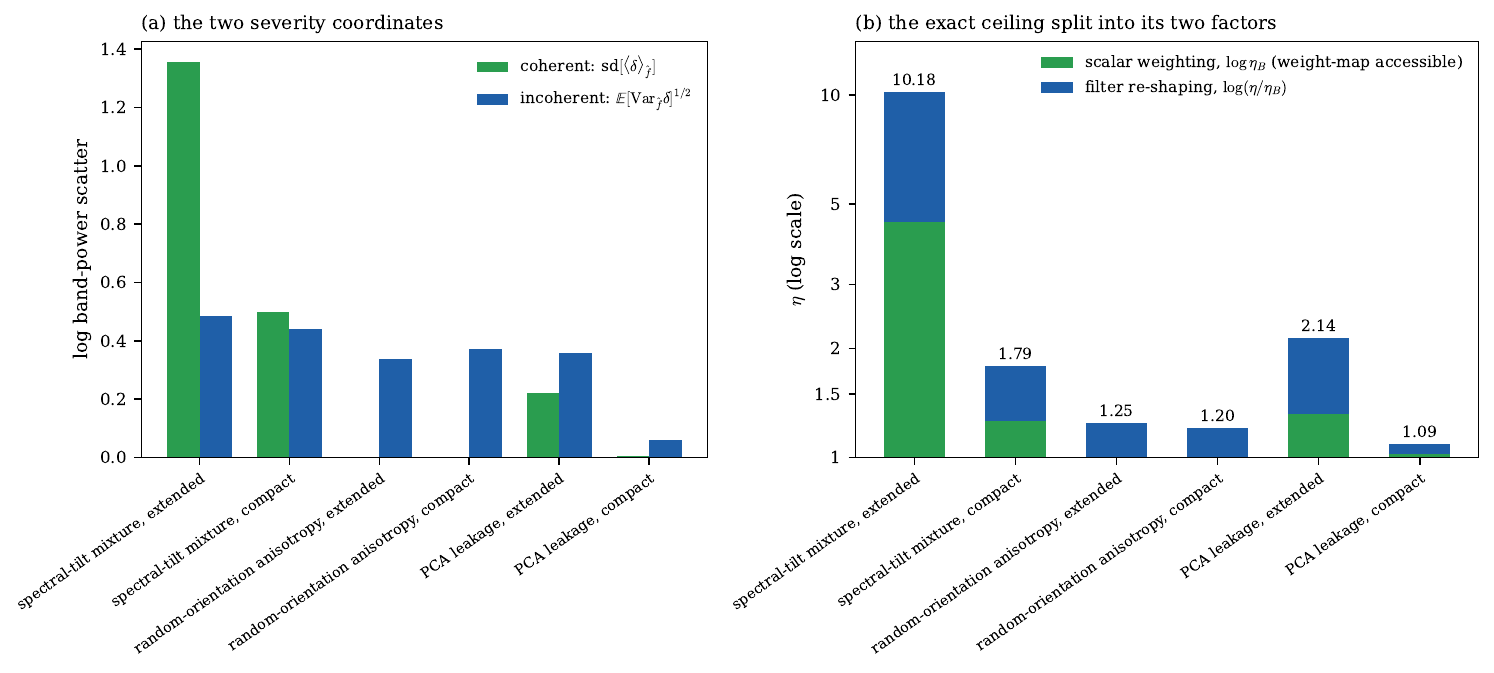}
  \caption{The two channels of the covariance-mixture advantage, floored configuration. \emph{(a)} The two
  severity coordinates per cell: the coherent scatter of the band power, $\sd[\avf{\delta}]$, and the
  incoherent within-band shape rms, $\E[\Varf{\delta}]^{1/2}$. \emph{(b)} The exact ceiling split into
  the scalar-weighting factor $\etaB = \E[c]\E[c^{-1}]$, which a weight map can take, and the filter
  re-shaping remainder $\eta/\etaB$, which needs a per-image filter. The random-orientation model has
  $\etaB = 1.000$ exactly.}
  \label{fig:channels_bar}
\end{figure}

Where does the tilt mixture's factor of ten come from? Table~\ref{tab:mixtures} shows the split. For the
extended template the coherent scatter of the band power is $\sd[\avf{\delta}] = 1.36$ in the log -- the
noise level in the template's band swings by a factor $e^{\pm 1.4}$ from patch to patch -- and the
scalar-weighting gain that captures it is $\etaB = \E[c]\E[c^{-1}] = 4.46$; the remaining factor
$\eta/\etaB = 10.18/4.46 = 2.3$ is filter re-shaping. The incoherent severity coordinate is $0.49$ in the
log, whose second-order estimate $\etaA = 1.24$ falls well short of the measured $2.3$ -- at $\delta$ of
order unity the truncated forms of \S\ref{sec:tier1} underestimate both factors (they give $3.1$ and
$5.5$ for the full ceiling), which is why the split is reported through the exact factors. The lever-arm
law of Eq.~(\ref{eq:leverarm}) explains the template ordering: the tilt is pivoted outside the map, the
extended template's band is three e-folds from the pivot, and a slope scatter of $0.5$ therefore
produces a band-level scatter of order unity for the extended template and much less for the compact
one, whose band sits closer to the pivot and is terminated by the floor.
Figure~\ref{fig:channels_bar} shows the decomposition for every mixture cell.

Two operational consequences follow: one practical leverage and one caution. Against a weight map
-- the per-image scalar rescaling by the measured band level that $\signorm$ represents -- the extended
headline drops from $10.18$ to $10.18 \times 0.224 = 2.3$: most of the largest ceiling in the paper is
available to any pipeline that already ships a weight map, and only the remaining factor of $2.3$
needs a per-image filter. However, for the compact template the naive version of that rescaling, by the map rms,
is actually detrimental: the variance ratio is $2.61$ analytically and $1.72 \pm 0.18$ empirically, because the
total map power is dominated by low $k$ and is nearly uncorrelated with the compact template's band, so
dividing by it injects noise. 

The random-orientation anisotropy has an exactly vanishing coherent channel: both templates are
circularly symmetric and the ensemble spectrum is the angle average, so a rotating ridge integrated
against a rotationally invariant Fisher weight returns the same band power at every angle
($\sd[\avf{\delta}] = 0.002$, $\etaB = 1.000$). Its entire advantage is incoherent: oracle-accessible,
unbiased given the latent, flat in $A$, immune to any weight map -- the deployable version. In the
realistic configuration it is also small, $1.25$ and $1.20$. That is a finding, and it corrects a
reading of our own floorless results: without a floor the compact-template ceiling of this model is
$9.6$, and it collapses to $1.5$ with the smallest floor tried and to $1.2$ at the calibrated one,
because the rotating ridge lives exactly in the high-$k$ region that the missing floor had left
unregularized (\S\ref{sec:unmodelled}). Where the incoherent channel stands alone, as in this model, the deployable, in-band, weight-map-immune
covariance-mixture advantage is at the 20\,\% level; where both channels are present, the residual after
the weight map, $\eta/\etaB$, reaches $2.3$ (tilt mixture, extended) and $1.6$ (PCA leakage, extended;
Table~\ref{tab:mixtures}).

\subsection{PCA leakage is a Gaussian scale mixture}
\label{sec:pca}

The PCA-leakage model was built as a non-Gaussian artifact: a Poisson number of leaked modes, each with
a heavy-tailed ($t_{2.5}$) amplitude. The ladder measured its class before the generator was re-read:
on the extended template the \rung{reweight} rung moves and nothing above it does (Fig.~\ref{fig:ladders}), which
is the signature of a covariance mixture, not of higher-order structure. The generator confirms it. Each
leaked mode is a fresh Gaussian field with one fixed anisotropic spectrum, rescaled to unit rms and
multiplied by its amplitude $a_{i}$; conditioned on the amplitudes the artifact is Gaussian with
covariance $(\sum_{i}a_{i}^{2})\,C_{\rm mode}$, so the model is a one-parameter scale mixture with latent
$c = \sum_{i \le N}a_{i}^{2}$ -- Tier~1 by construction, with a Monte Carlo prior on $c$ ($P(c = 0) = e^{-3}$,
$\E[c] = 8.1$, median $2.0$) and an exact quadrature. 
What ``exact'' means for the generator: 
each mode is rescaled to its \emph{realized} rms, so conditioned on $a_{i}$ it is a Gaussian field
constrained to a fixed norm rather than a Gaussian field. The mode filter has an effective dimension
$d_{\rm eff} = (\sum\lambda)^{2}/\sum\lambda^{2} \simeq 130$ on the $128^{2}$ map, so the constraint alters
the conditional density at order $1/d_{\rm eff}$ (a single coordinate acquires an excess kurtosis of
$-6/(d_{\rm eff}+2) = -0.05$): the quadrature is exact for the conditional-Gaussian generator and
accurate to about one percent for the actual one, below the Monte Carlo error on $2.14$, and the
ladder's $1.655$ bounds it from below whatever the generator does. 

Three consequences follow from this. (i) The ceiling is modest, $2.14$
extended and $1.09$ compact, essentially unchanged by the floor (its floor is the smallest in the set),
and the mixture matched filter attains 77\,\% of it. (ii) The complete-data bound, $4.90$,
over-states the room by a factor $2.3$: perfect \emph{removal} of the leaked modes is not an operation a
reweighting estimator can perform on structure that is only a rescaling of a Gaussian, so complete-data
bounds are to be quoted only where structure is genuinely removable (the artifact models, confusion), and
the exact quadrature is the ceiling for a mixture. (iii) The tier labels are
representation-dependent: the same generator is a
heavy-tailed additive process or a covariance mixture depending on which variable is called latent
\citep{AndrewsMallows1974}; the ceiling does not depend on that choice (Tier attribution, below). A practical rule came with this
model: on heavy-tailed
mixtures the extended template's $\sigMF$ and bound move more with the random seed than with the floor
(excess kurtosis $289$ against $45$ across two seeds -- as expected for marks without a fourth moment,
\S\ref{sec:noise_models}), so floorless-to-floored comparisons on such models must use paired seeds.

\paragraph{Tier attribution.}
Because $\eta$ is computed from the marginal density, it is agnostic about whether an advantage comes
from a covariance mixture or from higher-order structure, and the attribution is a separate step. Let
$\eta_{\rm T1}$ be the ceiling computed on the \emph{per-image-conditional Gaussianization surrogate} --
each image replaced by a Gaussian draw with its own true covariance, which keeps the mixture and destroys
all higher-order structure. Then $\eta_{\rm total} = \eta_{\rm T1}\,\eta_{\rm T2}$ defines $\eta_{\rm T2}$,
and a model whose generative description is ambiguous is adjudicated by this factorization without
settling the classification first. PCA leakage is the worked example (above).

\subsection{Attainability}
\label{sec:attainability}

The mixture matched filter of Eq.~(\ref{eq:reweight}) -- $1.2$--$1.8$\,k parameters depending on template and basis, no free pixel
map, weights that are functions of the image's own band powers -- reaches $7.12 \pm 1.13$ on the tilt
mixture's extended template (70\,\% of the exact ceiling), $1.445 \pm 0.075$ on its compact template
(81\,\%), and $1.655 \pm 0.052$ on PCA leakage (77\,\%), with four to six restarts agreeing to a few
percent. On the random-orientation model it gains only 3--6\,\% over the anchor against ceilings of
$1.2$--$1.25$: its six-sector angular basis is too coarse for a ridge whose orientation changes from map to map;
the shortfall is in the rung's basis, not in the noise model, and these cells have no headroom worth chasing. The ladder-ordering null passes on
every mixture: no rung above \rung{reweight} moves beyond it, as the theory requires of the exact classes
(the cubic rung was not run on the tilt and random-orientation mixtures, and on PCA leakage it returned
degenerate or anchor-level values, so this is a quadratic-and-conv statement). A trained rung
is a lower bound, so a quadratic or conv rung could in principle have exceeded a trained \rung{reweight}
rung by the latter's own shortfall -- 19--30\,\% here -- and none did. With the
template-pooled basis of \S\ref{sec:convrung} the conv rung does find a physical signal on PCA leakage,
$1.36 \pm 0.07$ on the extended template, below the \rung{reweight} rung and the exact ceiling as it must be:
a local network estimating the leak envelope is a coarse mixture filter.

\paragraph{A regression network on the tilt mixture.}
Figure~\ref{fig:teaser} (bottom) shows the ResNet of \S\ref{sec:teaser} trained on the floored tilt mixture,
extended template, $L = 7$ ($r = 0.66$). Both losses give $R = 0.550$ averaged over the certified region $[1.65, 4.41]\,\sigMF$
(bias-corrected: $\pm 0.015$; MSE: $\pm 0.038$; $R$ runs from $0.48$ at the center of the region to
$0.6$ at its edges), a raw variance ratio of $0.46$ (bias-corrected) and $0.39$ (MSE) -- the cleaned $R$
is the number to compare with the ceiling, since $V_{E}(A)$ and $\sigMF^{2}/\eta$ bound an estimator
unbiased at $A$ -- and, the informative part, a gain that is \emph{flat in $A$} across the whole prior.
Against the campaign's lines, matched filter $1.0 \to$ network $0.55$ (cleaned $R$) $\to$ weight map $0.224 \to$
\rung{reweight} $0.140 \to$ ceiling $0.098$: an off-the-shelf regressor realizes an advantage
of $1/R \simeq 1.8$ on a cell whose exact ceiling is $10.2$ and whose mixture matched filter reaches $7.1$ --
a genuine, certified, partial realization, short of what a per-image scalar weight map would give. The
comparison is made at matched amplitude, against $V_{E}(A)$ and against the weight-map level that a
data-derived scalar can reach -- not against the oracle filter of \S\ref{sec:ladder}, which is
conditionally unbiased and not a deployable competitor. The $A$-independence says which channel it uses. The attainability envelope $V_{E}(A)$ rises from $0.224$ at
$A = 0$ to $0.85$ at $4.7\,\sigMF$; the network sits above it below $A \simeq 1.5\,\sigMF$ and below it
above, with no crossover at $A \simeq \sigMF$. A flat gain is the signature of the incoherent,
filter-reshaping channel -- infer the tilt, adapt the filter shape, the \rung{reweight} mechanism -- not of
the coherent rescaling that fades by construction. That is a statement about this network, not about the
ceiling, and both readings sit inside the bracket $\sigMF^{2}/\eta \le V_{\rm opt}(A) \le V_{E}(A)$. It is
also the cell on which the ladder's conv rung stays at its anchor ($1.09$, below its sensitivity floor):
a $2.5$\,k-parameter test function on noise alone and a $10^{6}$-parameter regressor on source-injected
maps are different instruments, and the paper's rungs are lower brackets on what the second can reach.

\begin{figure}[t]
  \centering
  \includegraphics[width=\textwidth]{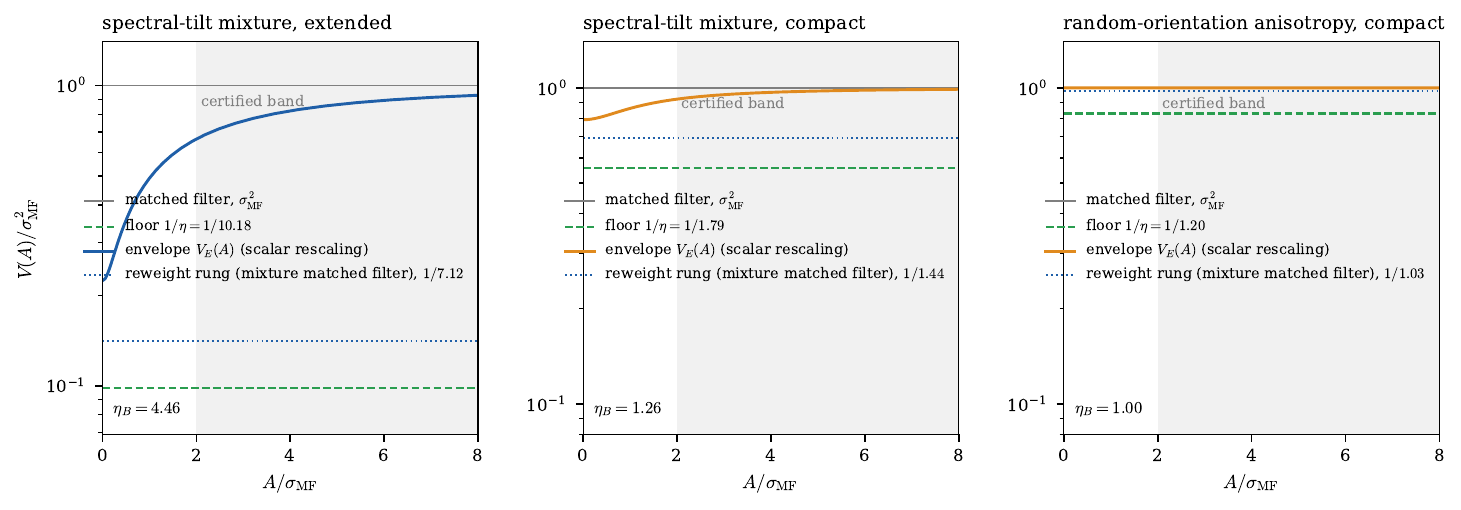}
  \caption{Where the covariance-mixture advantage lives in amplitude. Conditional variance in units of
  $\sigMF^{2}$ against $A/\sigMF$ for the spectral-tilt mixture (extended and compact templates) and the
  random-orientation anisotropy (compact template), floored configuration: the flat Cram\'er--Rao floor $1/\eta$; the
  attainability envelope $V_{E}(A)$ of Eq.~(\ref{eq:methodE}), which every point of is an achievable
  variance of a marginally unbiased estimator (a family envelope -- no single estimator follows it, since
  the optimal rescaling depends on $A$); and the mixture matched filter's rung. The certified band
  $A \gtrsim 2\sigMF$ is shaded. The envelope bounds the coherent channel only, which is why it is flat
  for the random-orientation model despite $\eta = 1.20$.}
  \label{fig:methodE}
\end{figure}

Figure~\ref{fig:methodE} shows where the tilt mixture's advantage lives in amplitude. The
attainability envelope $V_{E}(A)$ of Eq.~(\ref{eq:methodE}) touches $\sigMF^{2}/\etaB$ at $A = 0$ and
has lost most of its gain by $A \simeq \sigMF$ ($0.85\,\sigMF^{2}$ at $4.7\,\sigMF$), below the certified band where claims about a trained
estimator can be made; the flat floor $\sigMF^{2}/\eta$ and the horizontal line of the \rung{reweight}
rung bracket what a per-image filter can add. For the random-orientation model the envelope is flat at
$\sigMF^{2}$ despite $\eta > 1$, because a scalar rescaling cannot re-shape a filter. The magnitude of
The size of $\eta$ does not say whether an advantage is deployable; the decomposition into channels does
(\S\ref{sec:coherent_in_A}).

\subsection{What a practitioner takes from this section}

Of the survey-noise scenarios built here, only band-level variation seen by an extended template offers
an order-of-magnitude ceiling; a weight map takes most of it; a mixture matched
filter of $1.2$--$1.8$\,k parameters takes most of the rest; and everything else is at $1.1$--$2$, of which 77--81\,\% is again taken by
the mixture filter. The remainder -- the last 19--30\,\% on the tilt mixture, and whatever a generic
network can find without being told the latent -- is the question \PaperII\ is designed to answer.

\section{Results III: genuinely non-Gaussian survey noise}
\label{sec:results_nongaussian}

\begin{figure}[t]
  \centering
  \includegraphics[width=\textwidth]{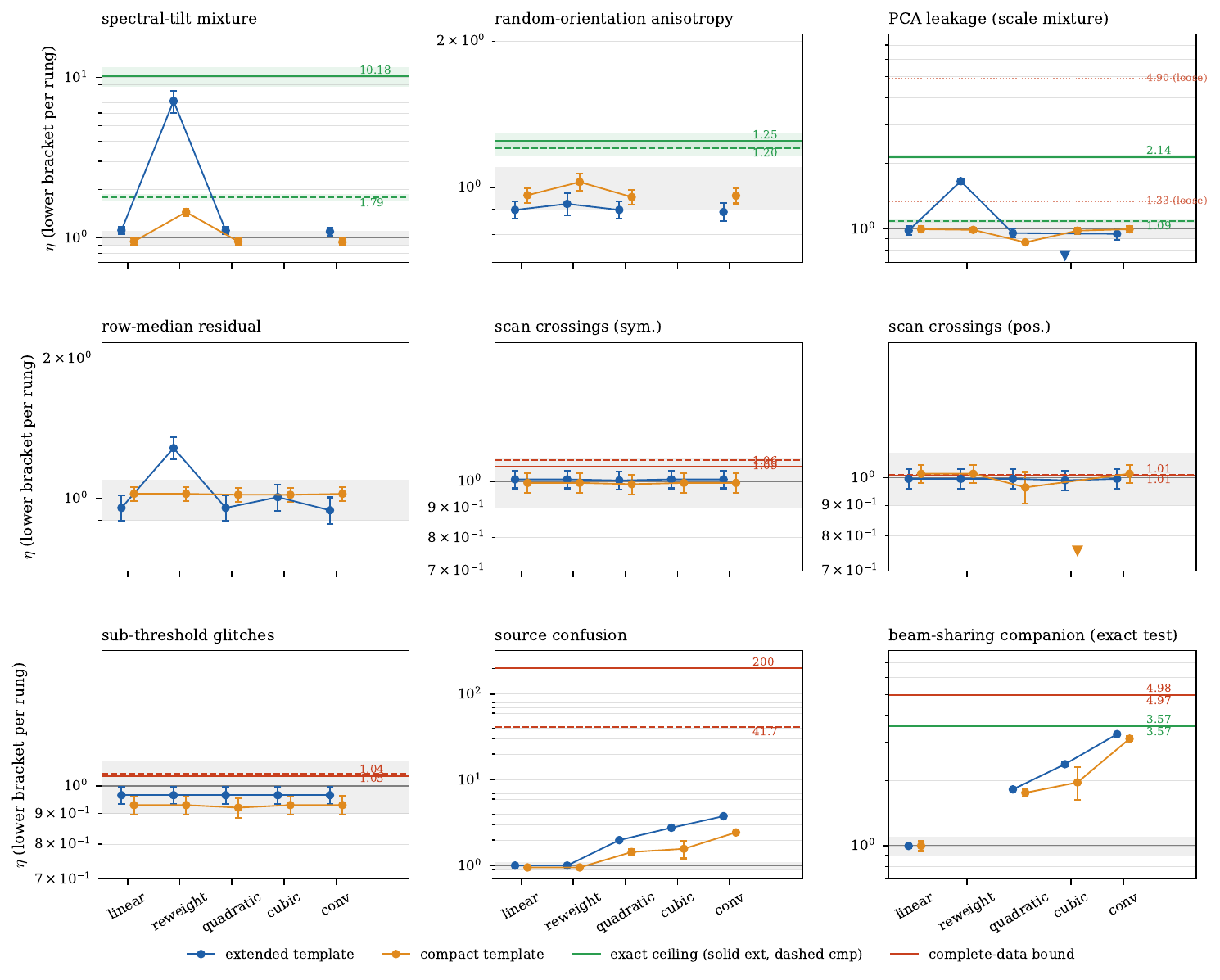}
  \caption{The variational ladder for every non-Gaussian cell of the realistic configuration, both
  templates. Each point is a rung's value on the EVAL split with its bootstrap error -- a certified lower
  bracket on $\eta$; colors distinguish the templates (legend), and for the bounds solid lines are the extended template and
  dashed lines the compact one; the green lines are exact ceilings (covariance mixtures), the red lines
  complete-data upper bounds (dotted red in the PCA panel: the loose scale-mixture bound of \S\ref{sec:pca});
  the gray band is the linear anchor's normalization noise. A rung at the anchor means the
  searched class found nothing above the ladder's sensitivity floor ($\sim 1.05$ cubic, $\sim 1.15$ conv),
  not that the noise contains nothing; the last panel calibrates that statement on a model with a known
  answer (\S\ref{sec:sensitivity}). The confusion rows use the template-pooled basis of
  \S\ref{sec:convrung} and an EVAL split of 7500 maps; the other rows the Fourier-pooled basis of the
  final pass, re-checked with the template basis where it matters (\S\ref{sec:sensitivity}). Values
  below the plotted range (a restart that returned a valid but useless negative-direction bound) are drawn
  as triangles at the bottom edge (two cells). The beam-sharing-companion panel has no \rung{reweight} point by construction
  (\S\ref{sec:sensitivity}).}
  \label{fig:ladders}
\end{figure}

The four survey-noise models that are not covariance mixtures -- \mCrossS, \mCrossP, \mGlitch\ and the
\mMedian\ -- have no exact ceiling, because none of them is a Gaussian mixture whose latent can be
integrated out by the quadrature of \S\ref{sec:tier1}. They are bracketed instead. The ladder brackets
all four from below. Three are bracketed from above by the complete-data bound of
Eq.~(\ref{eq:cdbound}): the two crossing models and the glitch model each have the additive form
$n = c(z) + g$, so one may ask what an estimator handed the contaminant $c$ exactly would gain, and that
bounds what any estimator can gain (\S\ref{sec:tier2}). The \mMedian\ has no such bound, because it is
not a contaminant added to Gaussian noise but a nonlinear operator applied to it, and there is nothing to
hand over. What stands in its place is a symmetry of the generator: a row median is sign-symmetric, so the
odd cumulants of the residual vanish identically and no estimator can find a bispectrum in it, which
removes the leading term of the expansion and is the statement \S\ref{sec:median} then measures.
Figure~\ref{fig:ladders} shows the ladders; Table~\ref{tab:nongaussian} the numbers.

\begin{table}[t]
  \caption{Non-Gaussian survey noise in the realistic configuration. ``Anchor'' is the linear rung
  (the matched filter; its deviation from 1 is normalization noise); rungs marked $\dagger$ did not exceed
  their anchor on the EVAL split ($^{\ast}$: on both bases; the gray entries of Table~\ref{tab:master_floored}
  are those whose restarts never left the anchor). Bounds are
  complete-data upper bounds.}
  \label{tab:nongaussian}
  \centering\small
  \setlength{\tabcolsep}{4pt}
  \begin{tabular}{llcccccc}
    \toprule
    model & template & anchor & \rung{reweight} & \rung{quadratic} & \rung{cubic} & \rung{conv} & bound \\
    \midrule
    \mCrossS & extended & 1.008 & $\dagger$ & $\dagger$ & $\dagger$ & $\dagger$ & 1.060 \\
    \mCrossS & compact & 0.994 & $\dagger$ & $\dagger$ & $\dagger$ & $\dagger$ & 1.088 \\
    \mCrossP & extended & 0.996 & $\dagger$ & $\dagger$ & $\dagger$ & $\dagger$ & 1.008 \\
    \mCrossP & compact & 1.016 & $\dagger$ & $\dagger$ & $\dagger$ & $\dagger$ & 1.012 \\
    \mGlitch & extended & 0.965 & $\dagger$ & $\dagger$ & $\dagger$ & $\dagger$ & 1.039 \\
    \mGlitch & compact & $0.929 \pm 0.033$ & $\dagger$ & $\dagger$ & $\dagger$ & $\dagger$ & 1.048 \\
    \mMedian & extended & 0.956 & $1.286 \pm 0.070$ & $\dagger$ & $1.008 \pm 0.066$ & $\dagger$ & -- \\
    \mMedian & compact & 1.025 & $\dagger$ & $\dagger$ & $\dagger$ & $\dagger^{\ast}$ & -- \\
    \bottomrule
  \end{tabular}
\end{table}

\subsection{Sub-threshold artifacts: nothing, from both sides}
\label{sec:artifacts}

For the scan crossings in both sign conventions and for the faint glitches, no rung exceeds its
anchor on the EVAL split, for either template, and the complete-data bounds are $1.01$--$1.09$: the
bracket $1 \le \eta \le 1.09$ is a measurement from both sides, with the instrument's own null at $0.93$
(the glitch compact anchor, normalization noise rather than a Fisher ratio; $\eta \ge 1$ is a theorem). The physical statement is the one a
pipeline developer wants: faint, numerous residual artifacts on a red background have essentially no
removable power where either template looks -- even perfect deglitching, or perfect identification of
every crossing, would buy under 9\,\% in variance (under 5\,\% in $\sigma$, or under a tenth of the survey
time). The reason is not that the events are individually insignificant. Their per-event
matched-filter significance is $\xirms = 1.4$--$4.4$, the saturated end of the two-regime law of
\S\ref{sec:tier2}, where an estimator \emph{can} identify and remove them; it is that the structured
component carries only 3--6\,\% of the noise power in the Fisher band of either template once that band
is regularized by the floor. Removing everything removable buys at most 9\,\%. This is a sharper statement than a perturbative one: it does
not rest on the events being weak -- they are not; $\xirms = 1.4$--$4.4$ is the saturated regime -- but
on where their power sits relative to the Fisher band, and so it holds for both templates and in both
regimes.

Two of these cells are also the paper's cautionary examples. Without the floor, the glitch model's
compact-template bound reads $144$ and the crossing models' $4.4$--$4.6$; with the floor they read $1.05$ and
$1.01$--$1.09$. The factor of a hundred was never in the noise (\S\ref{sec:unmodelled}). And the floored
symmetric-crossing cell, extended template, orders correctly against its bound ($1.008$ against $1.060$),
which retires a floorless ``theorem violation'' of an earlier pass ($1.14 \pm 0.06$ against $1.07$) as
heavy-tailed evaluation noise on an unrealistic configuration.

\subsection{The row-median residual: a weak signal at two orders}
\label{sec:median}

The row-median residual is the one survey-noise model with a measured signal above the matched filter,
and it is weak. On the extended template the \rung{reweight} rung reaches $1.286 \pm 0.070$ (four of four
restarts, $1.27$--$1.30$; the cubic and conv rungs of this cell were given six restarts): a data-dependent row operation makes the effective per-image covariance vary,
so the residual carries a covariance-mixture-like component that a mixture filter can use. The quadratic
rung is null, as it must be -- a sign-symmetric operation annihilates the odd cumulants, so there is no
bispectrum to find -- and the cubic rung carries a small, consistent trispectrum signal: VAL bounds of
$1.09$--$1.15$ in six of six restarts and $1.008 \pm 0.066$ on EVAL, at the sensitivity floor of the rung.
The conv rung finds nothing above its own floor. The excess-kurtosis seed control ($0.61$, $0.63$, $0.11$
across seeds and configurations) corroborates a weak but real fourth-order departure: once each row is
white-dominated, a row median is nearly linear. In the survey reading, and on the additive idealization of this cell, a per-image filter is worth
$\sim 29$\,\% more survey time on extended sources after row-median destriping; higher-order statistics add a
few percent at most.

One caveat is specific to this model and must be stated. Because the median is removed from the noise
before the source is added (\S\ref{sec:noise_models}), every row of $n$ has median exactly zero while
the rows of $d = A\tauv + n$ do not: in any row where the template is positive, $\mathrm{median}_{j}(d_{j} -
a\tau_{j})$ is a strictly decreasing function of $a$ whose unique zero is the true amplitude. The
idealized model therefore contains a noiseless amplitude channel of exactly the kind \S\ref{sec:unmodelled}
describes -- a constraint on the noise that the signal does not share -- and its Fisher information is
formally infinite. We therefore treat this cell as a reclassified one: the additive idealization has no
finite ceiling, which is why no complete-data bound is listed for it; the rung values above are lower
brackets for the estimator classes searched; and the physical model $d = R(A\tauv + n)$ -- the same
operator applied to signal and noise, in which the row medians of $d$ vanish for every $A$ and the
constraint carries nothing -- is the object to which a ceiling belongs. That model is not a location
family; its ladder requires the Jacobian template $DR(n)\,\tauv$ of the processing step in place of
$\tauv$ -- row by row, the template minus its value at the pixel that sets the row's median -- and is
left to future work.

On the compact template every rung sits at its anchor, with the Fourier-pooled basis and with the
template-pooled basis of \S\ref{sec:convrung} alike. There is no complete-data bound for this model (the
structure is processing-induced, not additive), so this is the one cell in the paper whose null rests on
the ladder alone. It is an \emph{informed} null, because the basis that finds a known compact-template
signal at 87\,\% (\S\ref{sec:sensitivity}) finds nothing here, and the physical intuition -- a
white-dominated row and a beam-sized template -- agrees.

\subsection{Reading rule}

The negative results of this section rest on upper bounds (the artifact models) or on the generator's
symmetry (the row median's quadratic null); its positive results rest on rungs that moved consistently
across restarts. A rung that did not move is, by itself, evidence about the search, not about the noise
(\S\ref{sec:ladder_var}); where the paper says ``nothing'', an upper bound is always the reason.

\section{Results IV: source confusion}
\label{sec:results_confusion}

Source confusion is the one model in the paper whose non-Gaussian component \emph{is} the noise, and the
one with physical units; it is also the one model in which a rung of the ladder above the mixture
matched filter has moved. It has its own section because it is the one astrophysical rather than instrumental noise model,
and it sits in the same figures as everything else because $\eta$ is dimensionless: a ceiling
in $\mjb$ and a ceiling in the paper's arbitrary units are comparable, even though their
$\sigMF$ are not.

\subsection{Where it sits}
\label{sec:confusion_where}

In the severity plane of \S\ref{sec:tier2} confusion occupies a corner no survey-noise model reaches:
the structured component carries $\fs = 0.81$ (compact) and $0.97$ (extended) of the noise power in the
Fisher band, against at most $0.06$ for the artifact models, and the individual events are faint,
$\xirms = 0.31$, against $1.4$--$4.4$ -- the many-faint-events regime in which the perturbative law
$\eta - 1 \propto \fs^{2}\xi^{2}$ nominally applies, although the field is far too non-Gaussian
(pixel skewness $2.1$, excess kurtosis $9.1$, and a contracted fourth-cumulant norm of order $10^{4}$
where the expansion needs it small) for the expansion to be usable. Its two severity axes are exactly
orthogonal: the number counts ($\Scut$, $\Smin$) move the cumulants only, and the floor $f_{N}$ and the
template scale move the band and the bound only, because $P_{\rm conf} \propto \sigc^{2}$ and
$\sigma_{w} = f_{N}\sigc$ cancel the counts from every second-order quantity. No other model separates the
two axes of the two-regime law.

The model's covariance is the same on every map -- there is no per-map latent -- so it has no
covariance-mixture confound at all, and it validates against a
closed form (Fig.~\ref{fig:confusion}): the Campbell cumulants of Appendix~\ref{app:confusion}
reproduce the realized mean, rms,
skewness and excess kurtosis of 8000 maps to $0.006$\,\%, $0.014$\,\%, $0.031$\,\% and $0.32$\,\%, and
$\Phat/P_{\rm conf} = 0.9993 \pm 0.011$, where $\pm 0.011$ is the scatter expected of the spectrum
estimate itself. The counts themselves are
a fitted Schechter function taken from \citet{BasuGuerreroBertoldi2026},
which is representative of the measured $350\,\mum$ counts but does not claim consistency with all
observational data. Lowering the count normalization -- $\Nbeam$ from $28$ to $18$--$20$ -- raises the
skewness by $18$--$25$\,\% and the excess kurtosis by $40$--$57$\,\% (Campbell's theorem), so a sparser
field is more non-Gaussian and the ceilings below are, in that direction, conservative.
The matched-filter
variances, measured on the ensemble as in Table~\ref{tab:mixtures}, are $\sigMF = 5.86\,\mjb$ (extended) and
$170.7\,\mjb$ (compact; the spectrum-predicted values of Table~\ref{tab:master_floored} are 5.72 and 157). The compact-template value is
enormous because with $\psi$ equal to the beam the per-mode signal-to-noise is flat and the filter has
no spectral leverage -- and its complete-data bounds, with the DC mode treated consistently
(Appendix~\ref{app:pipeline}) -- are $199.9$ (extended) and $41.7$ (compact): the largest in the paper by
a wide margin. This number \emph{is the ceiling} for perfect cataloging of every source down to $\Smin$, which is
operationally a fantasy but the right formal object.

\subsection{The ladder}
\label{sec:confusion_ladder}

With the calibrated ladder of \S\ref{sec:convrung} (template-pooled basis, six restarts, an EVAL split
of 7500 maps) the extended template reads
\begin{equation*}
  \begin{aligned}
    \text{linear } 1.006 \pm 0.020 \;&\to\; \rung{reweight}\ 1.006
    \;\to\; \rung{quadratic}\ 1.991 \pm 0.059 \\[2pt]
    &\to\; \rung{cubic}\ 2.772 \pm 0.086
    \;\to\; \rung{conv}\ \mathbf{3.783 \pm 0.071},
  \end{aligned}
\end{equation*}
with restarts within 0.4\,\%, 2.7\,\% and 2.5\,\% on the three nonlinear rungs, and the compact template
$0.956 \to 0.956 \to 1.447 \pm 0.107 \to {\sim}1.6$--$2.0 \to \mathbf{2.448 \pm 0.059}$ (the compact cubic
rung's EVAL error, $\pm 0.36$, is the heavy-tailed compact split; its six validation
trajectories sit at $1.98$--$2.03$ and the ordering quadratic $<$ cubic $<$ conv holds). Three things are
read off this ladder. First, \rung{reweight} never moves on either template -- the correct null for a
spectrum that is fixed from map to map, with no band power to reweight -- so none of the advantage is available to any
linear or reweighted filter. Second, every rung above it adds real signal, in order: the leading term is at
the bispectrum rung, as a one-signed marked Poisson process requires, and the cubic and convolutional
rungs add to it monotonically. This is the order-resolution demonstration the Volterra ladder was built
for, in the realistic configuration and immune to the rounding channel of \S\ref{sec:unmodelled}. Third,
these are lower bounds: the conv restarts peak early and then decline while their validation bound sits at
$4.05$--$4.20$ (extended) -- a representational plateau -- and the same rung recovers $\sim 0.9$ of a known
ceiling on the exactly solvable twin (\S\ref{sec:sensitivity}). The true extended ceiling is therefore
plausibly near $4$; we quote the bound and the calibration, never the extrapolation. Against the
complete-data envelope for this cell is $199.9$, and the measured $3.78$ is $1.9$\,\% of it: perfect
cataloging of every source down to the lower count limit $\Smin = 0.1$\,mJy (\S\ref{sec:confusion_setup}) is
not what a local nonlinear estimator will do.

As mentioned earlier, these are variance factors and not survey-time
factors: confusion is a property of the sky rather than of the observation, it does not integrate down, and the
translation of \S\ref{sec:eta_def} does not apply here. That makes the results even stronger: 
At the $350\,\mum$ confusion limit a nonlinear estimator divides the amplitude variance by at least $3.8$
for extended sources and at least $2.4$ for compact ones, and no survey could have reached either number
by observing longer or wider; the first factor of two is already available to a 267-parameter bispectrum
estimator.

\subsection{Which template, and why the compact one does not win}
\label{sec:confusion_template}

At the beginning of our work we expected that the \emph{compact} template would show the largest
advantage in confusion, because the matched filter has no spectral leverage there and ``the entire
discrimination is higher-order''. The premise is right and the inference is wrong, and the correction is
a major result of this work.

\begin{figure}[t]
  \centering
  \includegraphics[width=\textwidth]{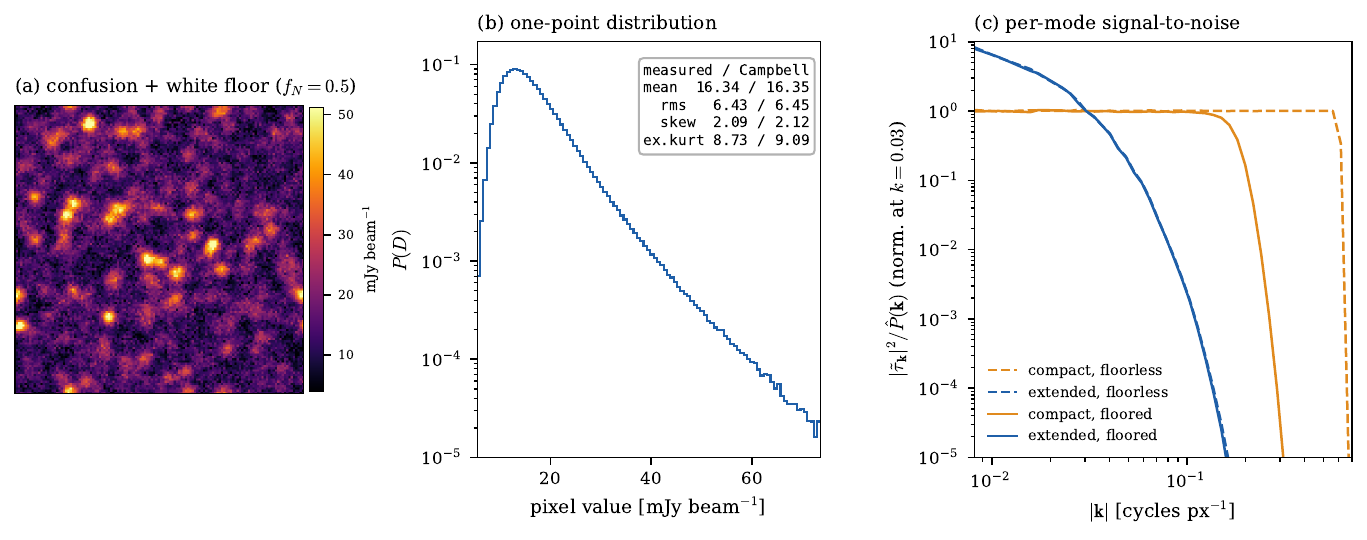}
  \caption{Source confusion. \emph{(a)} A realization of the fiducial row (confusion plus a white floor
  at $f_{N} = 0.5$, $\mjb$). \emph{(b)} The one-point distribution of the pure-confusion field, with the
  moments of a 400-map sample against the Campbell values (the $8000$-map validation is in
  \S\ref{sec:confusion_where}). \emph{(c)} Per-mode signal-to-noise $|\tautil_{\kv}|^{2}/\Phat(\kv)$ for the two
  templates: flat for the compact template without a floor (the matched filter has no spectral leverage
  and the Fisher sum is mode-count limited), rolled off past the knee with the floor. The floor hands the
  high-$k$ modes to a Gaussian component, which is why the compact template does not win.}
  \label{fig:confusion}
\end{figure}

The premise first, since it is easy to arrive at. Confusion is a white point process seen through the
beam, so by Campbell's theorem its power spectrum carries no $\kv$-dependence except the beam's,
$P_{\rm conf}(\kv) = q_{2}|\tilde B(\kv)|^{2}$, with $q_{2}$ the second moment of the counts
(Appendix~\ref{app:confusion}). A compact source is a point source seen through that same beam,
$\tautil(\kv) = \tilde B(\kv)$. The matched filter's per-mode Fisher weight is therefore
\begin{equation}
  \frac{|\tautil(\kv)|^{2}}{P_{\rm conf}(\kv)}
  \;=\; \frac{|\tilde B(\kv)|^{2}}{q_{2}\,|\tilde B(\kv)|^{2}} \;=\; \frac{1}{q_{2}} ,
  \label{eq:flatsnr}
\end{equation}
independent of $\kv$: the beam cancels, every mode carries the same information, and the filter has no
scale to prefer. Panel (c) of Fig.~\ref{fig:confusion} shows that flatness directly for the compact template without
a floor, and how the floor terminates it; the extended template's weight falls steeply with $k$ in both cases. That is what ``no spectral leverage'' means, and it is why the intuition is tempting.
It is also why the inference fails. Whitening cancels the beam by the same algebra, so the whitened map
is the bare marked lattice field, its pixels are independent, and by the theorem of \S\ref{sec:tier2}
$\eta = I_{1}$ with no template in the expression at all: perfect spectral sharing is exactly the
condition under which $\eta$ stops depending on the template. The compact template does not win here
because in this limit nothing wins -- every template is handed the same ceiling.

The floor breaks the cancellation, and that is the whole of the effect. An unsmoothed white component
$P_{w}$ does not share the beam, so the denominator of Eq.~(\ref{eq:flatsnr}) becomes
$q_{2}|\tilde B|^{2} + P_{w}$ and the weight rolls off past the knee at which the two are equal instead of
staying flat. The modes beyond that knee -- where a compact template puts most of its weight -- are
handed to a component that is Gaussian, which dilutes. Measured with the floor, the compact template
carries $\sim 40$\,\% less contracted third-cumulant norm and $\sim 33$\,\% less fourth than the extended
one, and the ladder arranges the same way: extended exceeds compact by a factor $1.55$, in the direction and
of the size the cumulant norms predicted.

A second claim,  ``confusion is intrinsically
band-regularized because the noise shares the beam'', holds for the extended template only. Flat per-mode
signal-to-noise is not convergence; with $|\tautil|^{2}/P$ constant the compact template's Fisher weight
is uniform and its $\sigMF$ is mode-count limited -- it drifts by $217$\,\% across mode cuts without a
floor and is terminated in practice by floating point, at $k \approx 0.43$ in float32 and $0.63$ in float64
(\S\ref{sec:unmodelled}) -- whereas the extended template's drifts by $0.00$\,\%.

The compact-template ceiling is bracketed between the measured $2.45$ and the bound of $41.7$, and the
floor sets it entirely: sweeping $f_{N}$ from $0.05$ to $2$ moves the compact bound from
$2.3 \times 10^{3}$ to $5.2$, and below $f_{N} \simeq 0.25$ that bound is not even mask-stable, drifting by
$5$--$13$\,\% against the mode cut where at the fiducial value it drifts by $0.41$\,\%. The
compact-to-extended bound ratio rises monotonically from $0.11$ to $0.38$ across the same range, which is
the quantitative form of the dilution above. The compact ceiling is therefore a statement about the
confusion \emph{and} the floor chosen with it, not about the confusion alone.


\section{What the ladder can and cannot see}
\label{sec:unmodelled}

This section 
serves both as a cautionary tale and as a check-list for simulation setup. 
Three times
in this program an unmodelled channel with anomalous signal-to-noise manufactured an advantage that
was not in the noise: once at $k = 0$, once at the edge of the Fisher band, and once in the mantissa of
the floating-point format -- panels (a), (b) and (c) of Fig.~\ref{fig:channels}. The three are the same
error in three places, and one principle cures all three. The section closes with the calibration of the ladder's sensitivity on a model with a known
answer, which sets the reading rule used throughout the results.

\subsection{Three unmodelled channels, one error}
\label{sec:three_channels}

\paragraph{The zeroed DC mode.}
In the first campaign of this program, the red-noise maps were mean-subtracted before the source was
injected. The matched filter, built with a $1/f^{3}$ spectrum whose DC bin was regularized to the
fundamental-frequency power, treated $k = 0$ as the noisiest mode in the map and down-weighted it; a
regression network found that the DC mode of every map was noiseless for the signal and read the
amplitude off it, with an apparent variance advantage of order $100$. A one-layer model consisting of a
global average pool reproduced the ``advantage'' without any convolutional machinery. The matched filter
was not wrong -- it was unaffected because it down-weights the lowest modes anyway -- but the
\emph{configuration} had a channel, the DC mode, in which the signal was present and the noise was
absent, and no estimator that models the noise would ever see it. The same channel is opened by any
high-pass filter applied to the noise but not to the signal (\S\ref{sec:noise_models}).

\paragraph{The missing white floor.}
The campaign's noise-only ensembles were first generated without an instrumental white floor. For the
compact template on a beam-smoothed background the beam cancels out of the Fisher weight,
$|\tautil|^{2}/P = B^{2}/(B^{2}P_{\rm red}) \propto k^{3}$, which rises to the highest retained mode; a
rounding guard placed on the analytic spectrum at $B^{2} > 10^{-8}$ then \emph{defined} the band. Measured
mask dependence of the exact tilt ceiling in an early 1500-map pass: extended flat at $9.6$ across five
decades of threshold (the ensemble of record gives $11.2 \pm 2.4$, Table~\ref{tab:master_floorless}),
compact $2.0 \to 1.7 \to 1.6 \to 1.5 \to 1.4$ with no convergence; and the complete-data bound of the
glitch model read $1.4 \to 12 \to 152$ over the same scan, with the fiducial guard's $144$
(Table~\ref{tab:master_floorless}) inside it. The configuration was
asserting that the beam removes the source's signal at high $k$ and exactly as much of the noise, leaving
the per-mode signal-to-noise unchanged, 
where a real map carries a
white floor that is not beam-smoothed and terminates it. A whole high-$k$ region was anomalous in the
same ratio as the DC bin had been. Restoring the floor (\S\ref{sec:floor}) makes every affected quantity
mask-independent to three decimals; the glitch bound becomes $1.05$, the random-orientation compact
ceiling $9.6 \to 1.2$, and the ``floorless'' case is revealed as a singular limit rather than one end of a
smooth range -- any floor will close the band, and further floor amplitude only moves the knee
logarithmically.

\begin{figure}[t]
  \centering
  \includegraphics[width=\textwidth]{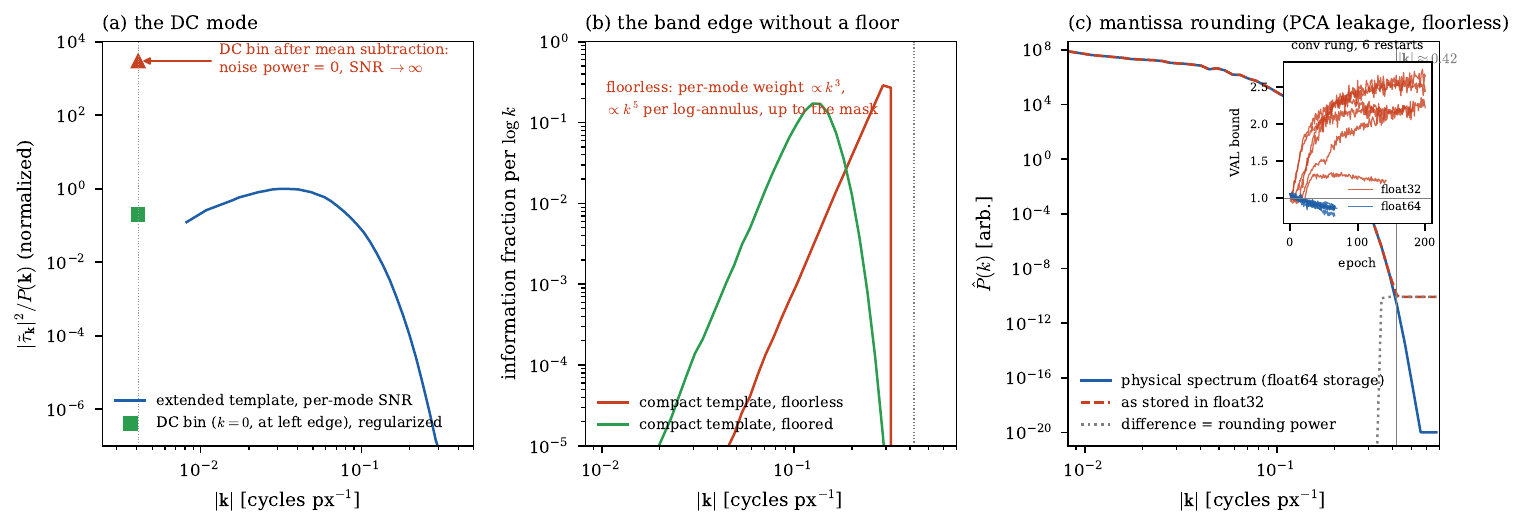}
  \caption{Three unmodelled channels, one error. Each panel shows a region of $k$-space in which the
  configuration handed an estimator signal-to-noise that the physical noise does not have: \emph{(a)} the
  DC mode after mean subtraction, against the per-mode signal-to-noise of the extended template on red
  noise and the regularized DC bin of the pipeline ($k = 0$ is drawn at the left edge of the logarithmic
  axis); \emph{(b)} the high-$k$ band of a compact template without a white
  floor, where the beam cancels from the Fisher weight; \emph{(c)} the band beyond $|\kv| \approx 0.42$
  where float32 rounding exceeds the beam-suppressed physical spectrum. Inset in (c): the validation
  trajectories of the conv rung on the floorless PCA-leakage model in float32 (five of six restarts
  climbing) and float64 (none). The white floor closes (b) and (c). It does not close (a): a mean
  subtraction annihilates $k = 0$ whatever the floor contains, and the cure there is to apply the same
  operator to the signal as to the noise -- in practice, not to subtract a mean at all.}
  \label{fig:channels}
\end{figure}

\paragraph{Mantissa rounding.}
The floorless ensembles were stored in float32, and on them the conv rung found reproducible signals
above the mixture matched filter: validation bounds of $2.3$ on PCA leakage (Fig.~\ref{fig:channels}(c),
inset; the EVAL value of Table~\ref{tab:master_floorless} is degenerate), $1.9$--$2.1$ on the row-median
residual, and a cubic rung of $1.2$ on the latter -- a textbook order-resolved ladder. All of it was rounding. Beyond
$|\kv| \approx 0.42$ cycles per pixel the physical spectrum of a beam-smoothed model is below $10^{-12}$
while the mantissa rounding of a map with rms $\sim 3$ is an absolute error $\sim |m|\,2^{-24} \approx 10^{-7}$
per pixel: white in $k$, with power $\sim 10^{-10}$, and \emph{heteroscedastic} -- its local amplitude
tracks the local map envelope. The data-estimated spectrum includes that band, whitening divides by
$\sqrt{\Phat}$, and about 40\,\% of the whitened field's modes -- a fraction of modes, not of variance;
the rounding's share of the whitened \emph{variance} on the beam-sharing companion is $\varepsilon \simeq
0.04$ (Appendix~\ref{app:confusion}) -- become unit-variance rounding whose
per-image amplitude encodes the artifact envelope. A linear filter cannot use it (the template has no
power there); the band-power \rung{reweight} rung does not need it (the leak amplitude is inferred from
the physical low-$k$ bands, which is why that rung reproduced between configurations); a nonlinear
convolutional test function reads it directly. Regenerating the same ensembles in float64 -- same seeds,
same whitening, same anchors to three decimals -- turned five of six climbing restarts into none of six
in all three cells, with VAL trajectories showing the null-cell signature. The floorless PCA conv validation bounds of
$2.3$--$2.7$ had in any case exceeded that model's exact ceiling of $2.07$, which is impossible for a
physical channel; and the confusion build measured the same threshold independently in a model with a
known answer ($\Phat/|\tilde B|^{2}$ leaves its plateau at $k \approx 0.43$ in float32 and $0.63$ in
float64). A white floor of $\sigma_{w} = 0.1$ puts $P_{w} = 164$ into every mode, twelve orders of
magnitude above the rounding, and the channel is gone: every floored cell is immune.

\paragraph{The principle.}
\emph{The configuration must contain every channel the estimator can see, at its physical level; the
numerics are part of the noise model.} A benchmark that does not model its own numerics will
manufacture an advantage, and the ladder's \rung{conv} rung is exquisitely sensitive: it found a channel
carrying $10^{-10}$ of the map power, twice, in five of six restarts. This is a critical lesson to carry: 
the floor that makes the configuration physical is also what makes the measurement honest. In
this paper the floor supplies the regularization for the analytic band and for the whitened field alike;
a physically floored ensemble is immune to the two band channels; the DC channel is closed by
regularizing (not zeroing) the mean.

\subsection{The sensitivity floor, and its calibration on a known answer}
\label{sec:sensitivity}

An estimator rung that does not move proves that the searched class found nothing, not that there is nothing
(\S\ref{sec:ladder_var}). How much can the ladder miss? The beam-sharing companion of
\S\ref{sec:confusion_setup} answers this on full-size maps: confusion plus a Gaussian companion of
the same amplitude as the fiducial floor but injected \emph{before} the beam, so that the whitened
pixels are independent and $\eta$ is known exactly and template-independent, $\eta = 3.5688$ (the
complete-data bound is exactly $(1 + \rho)/\rho = 5$, with $\rho = f_{N}^{2} = 0.25$ the
companion-to-confusion variance ratio, Appendix~\ref{app:confusion}), and the pipeline's own
float32 rounding dilutes it to a predicted $3.145$. On an i.i.d.\ whitened field a reweighting of the
image's own band powers has nothing to key on, so the \rung{reweight} rung is not run on this cell and
Fig.~\ref{fig:ladders} leaves that point blank. 

This is the one test in the campaign of whether the
framework recovers a \emph{known} $\eta > 1$ through the fiducial path, and the ladder as first built
failed it: with its fixed basis of smooth Fourier pooling maps the conv rung recovered $2.25 \pm 0.03$
on the extended template -- 63\,\% -- and nothing at all on the compact one. The diagnosis is
representational (\S\ref{sec:convrung}): on an i.i.d.\ whitened field the optimal test function is
template-localized, which smooth Fourier pooling cannot represent. Quadrupling the Fourier basis
changed little (75\,\% / 0\,\%) -- the size of the basis is not the issue, its localization is --
whereas pooling with the whitened template itself recovers $3.277 \pm 0.058$ (extended) and
$3.120 \pm 0.080$ (compact): 92\,\% and 87\,\% of the exact value, $104$\,\% and $99$\,\% of the
rounding-diluted one (the extended value sits $2.3\sigma$ above $3.145$, which the dilution model,
treating the rounding as an independent Gaussian, allows), and template-independent to $1.6\sigma$.
That both templates land on the \emph{diluted} value is an independent confirmation of the dilution
estimate of Appendix~\ref{app:confusion}.

Three consequences set the reading rules for this paper's results. 
First, every ladder value is a
lower bound; its tightness is calibrated at $0.87$--$0.92$ on the one cell with a known answer (the companion
test above) and is only roughly known elsewhere -- far lower where the basis cannot represent the optimal
statistic, as on the tilt mixture (\S\ref{sec:attainability}); results are quoted as
brackets -- an upper bound from a theorem or a complete-data envelope, a lower bound from the best
attained rung -- never as ``the ladder found nothing, so there is nothing'', and never divided by the
efficiency. Second, a compact-template null from the Fourier-pooled ladder alone was uninformative, and the
paper relies on no such null result. 
The covariance-mixture compact cells rest on exact quadratures, the artifact
compact cells on complete-data bounds of $1.01$--$1.09$, and the two cells whose ceiling statement rested on
the ladder alone were re-run with the template basis -- source confusion, where a null became
$2.45 \pm 0.06$ (\S\ref{sec:results_confusion}), and the row-median residual, which stays at its anchor and is
now an informed null (\S\ref{sec:median}). Third, the template-pooled basis moved only cells with known headroom, and nothing above an exact ceiling: the tilt mixture's compact conv rung stays at $0.97$ (exact
$1.79$, \rung{reweight} $1.45$), the glitch model's compact cell at its anchor (bound $1.05$), and PCA
leakage's extended conv rung rises to $1.36$, below its \rung{reweight} $1.655$ and exact $2.14$
(\S\ref{sec:attainability}). An earlier conclusion of this program -- that the conv rung never moved above the \rung{reweight} rung
on any noise model except through the floating-point channel of \S\ref{sec:three_channels} -- conflated
\emph{nothing to find} with \emph{cannot find}; with the template-pooled basis it has moved above the anchor on three cells -- confusion on both
templates and PCA leakage on the extended template -- and above the \rung{reweight} rung on confusion. Together with the rungs' sensitivity floors on Gaussian cells
(\S\ref{sec:convrung}) -- $\sim 1.05$ cubic, $\sim 1.1$--$1.2$ conv -- 
this confirms the intended meaning of the null values: 
a null above those floors, for statistics the basis can represent.

\subsection{Process}
\label{sec:process}

Four process failures produced plausible numbers in this program before they were caught, and each is
now a rule (Appendix~\ref{app:pipeline}): a numerical guard that silently became a physical definition
(the mode mask); a protocol constant not derived from the run's own parameters (the early-stopping
patience); a storage format treated as physics-neutral (float32); and a ladder that returned its
initialization for seven hours without saying so (the \code{moved\_off\_init} flag). We list them because
the claims of this paper 
are only as strong as the
framework's demonstrated ability to return the right answer where the answer is known, and because
each of them would, uncorrected, have appeared in a paper as a neural-network advantage.

\section{Discussion}
\label{sec:discussion}

\subsection{The verdict, in survey time}
\label{sec:verdict}

For a practitioner, the results of \S\S\ref{sec:results_gaussian}--\ref{sec:results_confusion} reduce to
five sentences. Gaussian noise of known covariance: nothing to be gained, certified at full map size. Covariance
mixtures: exact ceilings of $10$ (band-level variation seen by an extended source), $2$ (leaked-mode
power) and $1.1$--$1.8$ (everything else), \ie\ the equivalent of ten, two, and 1.1--1.8 times the
survey time (\S\ref{sec:eta_def}), of which a weight map already takes the larger part of the first and a
mixture matched filter of $1.2$--$1.8$\,k parameters takes 70--81\,\% of the three with headroom. Residual artifacts below the cleaning threshold: nothing,
bounded from both sides at the 9\,\% level ($\eta \le 1.09$). Row-median destriping residuals: 29\,\% more survey time from a
per-image filter (on the additive idealization, \S\ref{sec:median}), a few percent more from fourth-order statistics. Source confusion: a variance factor of at least
$3.8$ for extended sources and $2.4$ for compact ones -- and, alone among these models, one that no
amount of integration buys, since confusion noise does not integrate down -- none of it reachable by a
linear or reweighted filter, half of it by a bispectrum estimator, from a ladder whose tightness is
calibrated at $0.9$; the formal envelope is $53$ and $17$ times higher and is not the relevant number.

\paragraph{Information units.}
The ceiling also has a reading in bits. In the flat-prior limit, and when the template's Fisher band
contains many modes, the expected Kullback--Leibler divergence from the ideal amplitude posterior to the
matched-filter posterior is $\tfrac12\ln\eta$ nats: the matched-filter update leaves $\tfrac12\log_{2}\eta$
bits per source of achievable information unclaimed -- $1.7$ bits for the tilt mixture seen by an extended
source, at least $0.96$ (extended) and $0.65$ (compact) bits in confusion. In the achieved-information-gain language of
\citet{Ensslin2025AIG} this is the expected remaining information gain of the matched-filter update;
the identity follows from the classical correspondence between Fisher information and expected
information gain \citep{Lindley1956,ClarkeBarron1990} and holds asymptotically. The bit count is a translation of $\eta$, not a
route to it: it takes the Fisher information as given, whereas $\eta$ is how that information is computed.
It also keeps $\eta$'s prior independence -- the information-gain fidelity ratio of \citet{Ensslin2025AIG}
changes with the assumed source population (the design ratio $r$ of \S\ref{sec:eta_def}), while $\eta$
does not.

\subsection{Which noise scenarios deserve a training campaign}
\label{sec:preregistration}

The workflow this paper proposes -- compute the ceiling first, train only where it is high -- was
exercised on its own models and changed the plan. Of the eighteen floored cells with a ladder run (Table~\ref{tab:master_floored}), five have ceilings
above $1.5$: the tilt mixture on both templates, PCA leakage on the extended template, and confusion on
both. Confusion is the one
where a $2.5$\,k-parameter test function certifies $\ge 3.8$ and the ceiling is plausibly
higher. Glitches and crossings, originally slated for full training campaigns, were removed from the plan
by bounds of $1.01$--$1.09$; the random-orientation model was demoted by its floor; confusion was promoted by its
ladder. The first ResNet results of \S\ref{sec:teaser} and \S\ref{sec:attainability} show what the
pre-registration buys: a network that reads as 23\,\% better than the matched filter on Gaussian noise is
6--15\,\% worse once the shrinkage credit is divided out, and a network on the tilt mixture realizes a
certified $1.8$ of an available $10.2$, through the filter-reshaping channel rather than the coherent one. \PaperII\ trains ResNet regressors preferentially on these cells, against the ceilings published here,
with a smaller set of runs on cells where the ceiling says there is nothing to find -- to show a network
reproducing a null, and to show how a comparison made without the ceiling misreads shrinkage as advantage --
and its questions are correspondingly sharper than ``does the network beat the matched filter'': does a
generic network reach the exact ceiling on a covariance mixture, and does it do so by \emph{discovering}
the mixture filter unaided; does it find the last 19--30\,\% that the constructed filter leaves on the tilt
mixture; and what does it find in confusion above the quadratic rung, where the ceiling is not known. The
value of a network in this framing is model-agnosticism -- reaching, from data, an estimator that we had to
build by hand knowing the generator -- and the advantage is reported inside the certified band and the
ceiling overall as two different numbers.

\subsection{What to report, and what not to conclude}
\label{sec:reporting}

Several habits of the network-versus-filter literature become, in the light of \S\ref{sec:framework} and
\S\ref{sec:unmodelled}, avoidable mistakes, and the framework supplies the replacement for each. Never write
``unbiased'' without saying which of the three flavors is meant; a global bias number generically hides
a positive faint-end bias canceling a negative bright-end one, and the posterior mean has zero global
bias while extracting nothing. A network below the oracle matched filter at small $A$ is legal on a
covariance mixture and is not a leakage flag; a network below $\sigMF/\sqrt{\eta}$ inside a certified band
is -- where $\eta$ is exact. Where the pre-registered number is a ladder bracket, as for confusion and
the row-median residual, a network below $\sigMF/\sqrt{\eta_{\rm ladder}}$ is a measurement that the
ladder was loose, and the line that cannot be crossed is the complete-data bound
(\S\ref{sec:results_confusion}) where one exists 
(e.g., the row-median residual has no upper bound). A large $\eta$ does not mean a network is under-performing: the channel decomposition and the
attainability envelope say how much of it is deployable, and for the tilt mixture most of it is not
(\S\ref{sec:two_channels_results}, \S\ref{sec:attainability}). Report against $\signorm$ as well as against
the ensemble filter, or the achievement of the network is overstated whenever the noise level varies.
Quote the floor and the mode mask with every ceiling, and treat any quantity that moves with the mask as
undefined rather than uncertain. A larger \emph{relative} advantage is not better science: for extended
sources in red noise the absolute $\sigMF$ is intrinsically poor, and a large ratio may only reflect a
more handicapped baseline. And a sample power spectrum that fluctuates from map to map is not a latent: for a
fixed-covariance model those fluctuations are $\chi^{2}$ sampling about one common spectrum, and carry
nothing a filter can act on. The two-channel split of \S\ref{sec:tier1} applies only where the
covariance of the \emph{generating process} itself differs between images -- where there is a latent
$z$ that changes $\Nm_{z}$, so that a filter built for one image is genuinely the wrong filter for the
next -- and not where the same $\Nm$ has merely been drawn from twice.

\subsection{Limitations and scope}
\label{sec:limitations}

The template is known exactly throughout, both its profile and its position on the map. Template and
profile uncertainty is a separate route to matched-filter sub-optimality, 
which we address below in \S\ref{sec:blind} in more detail. 
In our framework, the estimand is a single amplitude at that
known position; detection, cataloging, deboosting and the faint end are downstream of it, and the faint end ($A \lesssim 2\sigMF$)
is where the coherent channel lives and where no regression network is unbiased. The ladder's
values are lower bounds whose tightness is calibrated at $\sim 0.9$ on one exactly solvable model
(\S\ref{sec:sensitivity}); on the fiducial cells the residual inefficiency is unknown and the confusion
ceilings may sit higher than the quoted $3.8$ and $2.4$. 
The confusion noise model ignores clustering of sources, which potentially alters 
the matched-filter variance and the Fisher information in ways that need not share a sign, and its effect
on the confusion ceilings is left to future work. On heavy-tailed ensembles the EVAL-split statistics limit
the error bars, not the values, and the extended-template complete-data bounds may carry a few-percent
high bias from the noisy low-$k$ end of a 1500-map spectrum estimate (the exactly solvable companion's
bound read $5.3$ against $5.0$ before the DC correction and $4.98$ after it). Finally, the noise models
are simulations built from the known properties of mm/submm pipelines, rather than from a specific instrument's
time-ordered data, even though the framework is designed to apply unchanged to real noise ensembles. 

\subsection{Unknown position, the faint end, and filter banks}
\label{sec:blind}

\paragraph{Unknown position.}
Surveys rarely know where their sources are, but for stationary noise this costs the ceiling
nothing at high significance. Stationarity makes $\Dop$ diagonal in Fourier space, so
Eq.~(\ref{eq:eta}) is the average of a per-mode gain $1 + P(\kv)\Dop(\kv)$ under the
matched-filter weights $|\tautil_{\kv}|^{2}/P(\kv)$, identical at every pixel: one computation
serves the whole map, and the optimal statistic becomes a scan -- the template correlated with the
score field rather than with $\Nm^{-1}d$, which for Gaussian noise is the matched-filter map itself.
When the position is fitted rather than given, amplitude and position are orthogonal parameters for
any symmetric template, since their Fisher cross term vanishes in the matched-filter metric and in
$\Jop$ alike; for a well-localized source the amplitude ceiling with unknown position is therefore the
known-position $\eta$. The position has a ceiling of its own, the same per-mode gain weighted by a
further $k^{2}$, which leans on the modes where a white floor leaves least to gain: for the fiducial
confusion row the complete-data envelopes are $14$ (compact) and $157$ (extended) for the position,\footnote{The
envelopes and the Monte Carlo of this subsection are computed with \code{sec11\_filter\_bank\_envelopes.py} and
\code{sec11\_mc\_threshold.py} in the code release \citep{BeyondTheBLUE2026code}, on the fiducial confusion row
with a $\beta$-profile bank.}
against $42$ and $200$ for the amplitude. A nonlinear estimator, therefore, would improve 
photometry more than astrometry. 

\paragraph{The faint end.}
Maximizing the
filtered map over position (and filter scale) increases the chance of picking up a noise fluctuation: 
in Gaussian noise the
expected squared significance grows by one unit per fitted parameter, which is the origin of the
relation $\langle\xi\rangle = \sqrt{\zeta^{2} + 3}$ between the maximized and the true-position
significance, for two position parameters and one filter scale, used in SPT cluster selection
\citep{Vanderlinde2010}. The resulting upward bias, of order $p/2\zeta$ in units of $\sigMF$ for $p$
fitted parameters, is of higher order than the Fisher matrix sees \citep{CoxSnell1968} and is not
small at $\zeta \approx 4$--$5$. Closer to threshold, noise peaks in the search region rival the
source, the fit jumps to them, and the error departs from the Cram\'er--Rao bound altogether -- the
threshold effect of nonlinear estimation \citep{VanTrees1968,RifeBoorstyn1974}. 
As $A \to 0$, the position is not
identified at all, and the maximum is distributed as the extreme of a random field rather than as a
$\chi^{2}$ variable \citep[e.g.,][]{Davies1987}. Detection thresholds sit, by construction,
where the maximum of the noise field over the survey lies, i.e.\ at the edge of this regime. In a
Monte Carlo of a blind position-and-scale search in white noise, the amplitude error is $2.0$ and
$1.3$ times its Cram\'er--Rao bound at $\zeta = 3$ and $4$; global errors disappear only above
$\zeta \approx 6$, and the joint fit still leaves the error $15$--$20$\,\% above the bound up to
$\zeta = 15$. At the faint end of a blind catalog neither the matched filter nor a nonlinear estimator
works at its Fisher level, $\eta$ is not a forecast, and the comparison has to be made in completeness
at fixed purity, which the tails of each statistic's null distribution decide.

\paragraph{Filter banks.}
When the template belongs to a family -- the core radius of a $\beta$-model for clusters, 
the size of an extended galaxy -- two questions must be kept apart.
The conditional ceiling $\eta(\theta)$ of a member assumed known needs no ladder per member: $\eta$ is
a mode-weighted average of one template-free function, so mapping that function once
yields the whole bank, while $\sigMF$, the complete-data bound and the cumulant norms are cheap enough
to compute for every member directly. The result is then the curve $\eta(\theta)$ rather than a single
number; in the fiducial confusion row the complete-data envelope alone rises from $42$ for a point
source to $222$ for a $20$-pixel core. When $\theta$ is instead fitted jointly, as in a catalog,
amplitude and scale are not orthogonal, and the ceiling is the ratio of the amplitude variances after
marginalizing $\theta$ in the two metrics. It equals $\eta(\theta)$ when the per-mode gain is flat but
not in general, and it depends on the estimand: in the same envelopes, fitting the size lowers the
bound on the central amplitude of a $2$-pixel core from $113$ to $31$ but leaves that on the total flux
unchanged. The fair baseline for a network trained across a family is accordingly the matched
multifilter with the scale fitted \citep{Herranz2002,Melin2006}. 
A single fixed filter would fold the template-adaptation route of
\S\ref{sec:limitations} into the noise channel and credit the network with an advantage even in
Gaussian noise.

\section{Conclusions and outlook}
\label{sec:conclusions}

We have introduced the advantage ceiling $\eta$ -- the Fisher information for a source amplitude in
units of the matched filter's, the projection of the excess-score operator $\Dop = \Jop - \Nm^{-1}$ onto
the template's Fisher band -- and computed it for a suite of millimeter and submillimeter survey
noise. Six conclusions.

\begin{enumerate}[leftmargin=2em,itemsep=3pt]
  \item \emph{The framework validates itself.} Eight Gaussian cells return $\eta = 0.94$--$1.06$ ($\pm 0.04$) with
    every nonlinear rung at its anchor; two analytic closures hold to four digits; and a battery of nine null
    tests passes all of its 100 assertions. 
    An exactly solvable confusion companion calibrates the ladder at $\sim 0.9$ of a
    known ceiling, after exposing -- and fixing -- a representational flaw that every other null test had
    missed: a pooling basis too smooth to represent the template-localized test function that an
    i.i.d.\ whitened field calls for (\S\ref{sec:convrung}).
  \item \emph{Covariance mixtures have exact, understood ceilings, and a constructed estimator attains
    most of them.} Spectral-tilt variation seen by an extended source gives $10.2$; PCA leakage -- a
    Gaussian scale mixture, as the ladder measured before the generator was read -- gives $2.1$; the
    random-orientation ridge and the compact-template cells give $1.1$--$1.8$. The advantage has two
    channels, coherent (band level; large, weight-map-removable, dying above $A \simeq \sigMF$) and
    incoherent (within-band shape; $\sim 20$\,\% where it stands alone, up to a factor $2.3$ as the residual
    after the weight map, and deployable). A mixture matched filter of
    $1.2$--$1.8$\,k parameters attains 70--81\,\% of the three ceilings with headroom.
  \item \emph{Sub-threshold artifacts leave nothing.} Glitches and scan crossings on a red background are
    bounded at $\eta \le 1.09$ from above and sit at the anchor from below: even perfect removal buys at most
    9\,\%, because the removable power is not where either template looks. 
    This is a statement about residuals: both models represent what survives the time-domain cleaning of a
    typical mm/submm pipeline -- artifacts below the deglitching and flagging thresholds -- and it is in that
    post-cleaning form that they carry so little removable power.
  \item \emph{Source confusion carries the largest measured non-Gaussian advantage}, $\eta \ge 3.8$
    (extended) and $\ge 2.4$ (compact), on a monotone, order-resolved ladder that begins at the
    bispectrum rung and on which no linear or reweighted filter moves. The compact template does not win.
    For pure confusion the beam cancels from the Fisher weight and $\eta$ becomes template-independent,
    and the component that breaks that symmetry is the instrumental white floor: it is not beam-smoothed,
    so it owns the high-$k$ modes where a compact template puts most of its weight, and it is Gaussian,
    so the modes it owns carry no advantage.
  \item \emph{The numerics are part of the noise model.} A zeroed DC mode, a missing white floor and
    float32 mantissa rounding each manufactured an advantage, from a factor of a few to two orders of
    magnitude. The lesson is that a simulated ensemble must contain every channel an estimator can see,
    at its physical level; a physically floored ensemble is immune to the two band channels, and the DC
    channel is closed by regularizing (not zeroing) the mean.
  \item \emph{Compute the ceiling before training.} It is prior-free, architecture-free, and computable
    from noise-only maps in minutes to hours; it demoted the glitch and crossing models from our own training plan,   
    promoted confusion, and supplies the impossibility line where the ceiling is exact and a floor where it is
    a ladder bracket (both taken up in Paper II). 
    Read as survey time, wherever the noise integrates down, it
    tells a survey what an ideal nonlinear estimator is worth for each source class before a network is
    built and trained.
\end{enumerate}

A first ResNet regressor, certified as this paper prescribes, is 6--15\,\% less efficient than the matched
filter on Gaussian noise once its shrinkage is divided out, and realizes a factor $1.8$ of the
$10.2$ available on the spectral-tilt mixture. \PaperII\ trains such regressors on the cells this paper
singles out, against these pre-registered ceilings, and asks whether a generic network discovers the
mixture filter and the bispectrum estimator unaided. Two extensions of the current framework follow directly.
First, the internal linear combination used for
component separation in multi-frequency CMB maps is the same estimator with the template replaced by a
frequency spectrum, $\tauv \to \mathbf a$, and the noise covariance by the per-mode frequency-frequency
covariance, $\Nm \to \mathbf C(\kv)$ \citep{Remazeilles2011,Erler2019,Zubeldia2023}. 
Hence the ceiling, the two advantage channels and the
ladder transfer as identical constructions -- though not their values, which belong to the noise models of this
paper -- with the needlet ILC reading as a hand-crafted, piecewise covariance-mixture estimator and the
Galactic foregrounds as the non-Gaussian component.
Second, the matched multifilter used for Sunyaev--Zeldovich
cluster extraction is the composition of the two, so that everything said here about compact sources
applies to it with the template's Fisher band extended to $(\kv, \nu)$. Both are the subject of future papers.

\begin{acknowledgments}
We thank Jakob Dietl, Frank Bertoldi, Thomas Reiprich, Mirko Bunse, Kevin Schmitz and
Jens Bu{\ss} for helpful discussions.
We acknowledge funding from the Collaborative Research Center 1601 (SFB 1601) funded by the
Deutsche Forschungsgemeinschaft (DFG, German Research Foundation), and from Verbundprojekt
05A2023--D-MeerKAT III funded by the German Federal Ministry of Education and Research (BMBF).
We thank the Gauss Centre for Supercomputing e.V.\ (GCS) for providing computing time on the
supercomputers JUWELS (until June 2026) and JUPITER at the J\"{u}lich Supercomputing Centre,
via the John von Neumann Institute for Computing. JUPITER is supported by the EuroHPC JU and
GCS through funding by the European Commission, the German Federal Ministry of Research,
Technology and Space, and the Ministry of Culture and Science of the State of North
Rhine-Westphalia.

\medskip
\noindent\textit{AI-usage disclosure.}
This paper was prepared with extensive use of AI assistants, specifically the Claude Opus-
and Fable-class models (Anthropic) operating in agentic, tool-using mode. The AI wrote the
analysis and figure-generation code, produced the figures, helped to refine much of the text,
and handled the \LaTeX{} formatting. The research questions, the theoretical framework and the
interpretation of the results are the author's, as is the verification of every quantitative
claim reported here. The author takes full responsibility for the work, including the
originality of the content and the accuracy of the results.

\medskip
\noindent\textit{Data availability.}
The entire codebase for this paper -- the advantage-ceiling computation framework, the
noise-generation scripts, the complete null-test battery, and the datasets needed to reproduce
the paper's results -- is publicly available on GitHub at
\url{https://github.com/kmbasu/Beyond-the-BLUE-Paper-I}; the release used for this paper
(v1.0) is archived on Zenodo \citep{BeyondTheBLUE2026code}, and the concept DOI
\doi{10.5281/zenodo.22924969} always resolves to the latest release.
\end{acknowledgments}

\software{NumPy \citep{Harris2020numpy},
          SciPy \citep{Virtanen2020scipy},
          PyTorch \citep{Paszke2019pytorch},
          Matplotlib \citep{Hunter2007matplotlib},
          h5py \citep{Collette2013h5py},
          Beyond-the-BLUE-Paper-I v1.0 \citep{BeyondTheBLUE2026code}}

\appendix
\section{Derivations}
\label{app:derivations}

Throughout, $n$ is a zero-mean noise field with density $p_{n}$, covariance $\Nm$ and score
$s(n) = -\nabla\log p_{n}(n)$; $x = \Nm^{-1/2}n$ is the whitened field with score $s_{x}$; $\tauv$ is the
template, $t = \Nm^{-1/2}\tauv$ and $\that = t/\|t\|$. Expectations are over the noise unless a subscript
says otherwise. All densities are assumed smooth enough that boundary terms in integrations by parts
vanish.

\subsection{Stein's identity and the excess-score operator}
\label{app:stein}

For any differentiable $h:\R^{N}\to\R^{N}$ of at most polynomial growth, integration by parts gives
$\E[h(n)\,s(n)^{\dagger}] = \E[\nabla h(n)]$ (Stein's identity). With $h(n) = n$,
$\E[n\,s^{\dagger}] = \Id$. For any direction $a \in \R^{N}$,
\begin{equation}
  0 \;\le\; \E\big[(a^{\dagger}s - a^{\dagger}\Nm^{-1}n)^{2}\big]
  \;=\; a^{\dagger}\Jop a - 2\,a^{\dagger}\Nm^{-1}\E[n\,s^{\dagger}]a + a^{\dagger}\Nm^{-1}\Nm\Nm^{-1}a
  \;=\; a^{\dagger}(\Jop - \Nm^{-1})\,a ,
\end{equation}
which is $\Dop \succeq 0$, Eq.~(\ref{eq:delta}), with equality for all $a$ if and only if $s = \Nm^{-1}n$
almost surely, \ie\ the noise is Gaussian. In whitened coordinates $s_{x} = \Nm^{1/2}s$ and
$\E[x\,s_{x}^{\dagger}] = \Id$, so the same argument gives $\E[s_{x}s_{x}^{\dagger}] \succeq \Id$ and
$\eta = \E[(\that^{\dagger}s_{x})^{2}] = 1 + \E[(\that^{\dagger}(s_{x} - x))^{2}] \ge 1$.

\subsection{The variational identity}
\label{app:variational}

For a scalar test function $f$ of the whitened field, Stein's identity with $h = f\,\that$ gives
$\E[f\,\that^{\dagger}s_{x}] = \E[\that^{\dagger}\nabla f]$. Hence
\begin{equation}
  2\,\E[\that^{\dagger}\nabla f] - \E[f^{2}]
  \;=\; 2\,\E[f\,\that^{\dagger}s_{x}] - \E[f^{2}]
  \;=\; \E\big[(\that^{\dagger}s_{x})^{2}\big] - \E\big[(f - \that^{\dagger}s_{x})^{2}\big]
  \;\le\; \eta ,
\end{equation}
with equality at $f^{\star} = \that^{\dagger}s_{x}$, which is Eq.~(\ref{eq:variational}). Restricted to
linear $f = w^{\dagger}x$, the bracket is $2\,w^{\dagger}\that - w^{\dagger}\E[xx^{\dagger}]w = 2w^{\dagger}\that -
\|w\|^{2}$, maximized at $w = \that$ with value $1$: the matched filter is the linear solution and its
value is exactly $1$ regardless of the noise. For the one-step estimator of Eq.~(\ref{eq:onestep}),
$\Ahat_{f} = A_{0} + \sigMF f(x_{0})/\E[\that^{\dagger}\nabla f]$ with $x_{0}$ the whitened residual at a
preliminary amplitude $A_{0}$: its expectation has unit slope in $A$ at $A_{0}$ (differentiating
$\E[f(x + (A-A_{0})\|t\|\that)]$ gives $\|t\|\,\E[\that^{\dagger}\nabla f]$, and $\sigMF = 1/\|t\|$), and its variance in units of $\sigMF^{2}$ is
$\E[f^{2}]/\E[\that^{\dagger}\nabla f]^{2}$, whose inverse is the value of the bracket at the optimal
rescaling of $f$. Both statements take $A_{0}$ to be independent of the map, so that $x_{0}$ is
distributed as whitened noise shifted by $(A - A_{0})\|t\|\that$; for $A_{0}$ the matched-filter
estimate of the same map, $x_{0}$ is constrained by $\that^{\dagger}x_{0} = 0$ and the calculation does
not apply (for a pure scale mixture the correction then vanishes identically). The rung value is
therefore the local efficiency, at an independently supplied reference amplitude, of the estimator its
test function defines.

\subsection{Covariance mixtures: the two channels}
\label{app:twochannels}

Let the noise be Gaussian given a latent with spectrum $P_{i}(\kv) = \Pbar(\kv)\,e^{\delta_{i}(\kv)}$, and
let $\fhat_{\kv} \propto |\tautil_{\kv}|^{2}/\Pbar(\kv)$ be the Fisher weights of the ensemble spectrum,
$\bar\sigma^{2} = (\sum_{\kv}|\tautil_{\kv}|^{2}/\Pbar_{\kv})^{-1}$. The ensemble matched filter applies the
fixed weights $\tautil^{*}/\Pbar$; on image $i$ its variance is
$\sigma^{2}_{{\rm ens},i} = \bar\sigma^{4}\sum_{\kv}|\tautil_{\kv}|^{2}P_{i}(\kv)/\Pbar^{2}_{\kv} = \bar\sigma^{2}\avf{e^{\delta_{i}}}$.
The oracle filter rebuilt with $P_{i}$ has $\sigma^{2}_{{\rm or},i} = (\sum_{\kv}|\tautil_{\kv}|^{2}/P_{i,\kv})^{-1}
= \bar\sigma^{2}/\avf{e^{-\delta_{i}}}$. Expanding both to second order in $\delta$,
\begin{equation}
  \frac{\sigma^{2}_{{\rm ens},i}}{\bar\sigma^{2}} \simeq 1 + \avf{\delta_{i}} + \tfrac12\avf{\delta_{i}^{2}}, \qquad
  \frac{\sigma^{2}_{{\rm or},i}}{\bar\sigma^{2}} \simeq 1 + \avf{\delta_{i}} + \avf{\delta_{i}}^{2} - \tfrac12\avf{\delta_{i}^{2}},
\end{equation}
so the ensemble-to-oracle gap is $\avf{\delta_{i}^{2}} - \avf{\delta_{i}}^{2} = \Varf{\delta_{i}}$ per image,
and $\E[\Varf{\delta}]$ on average -- the incoherent term. In the well-inferable limit the marginal Fisher
information is $\E_{i}[1/\sigma^{2}_{{\rm or},i}]$, so with $\Pbar$ the ensemble-mean spectrum --
which requires $\E[e^{\delta(\kv)}] = 1$ mode by mode, \ie\ $\E[\delta] \simeq -\tfrac12\E[\delta^{2}]$ at
this order --
\begin{equation}
  \begin{aligned}
    \eta \;&=\; \bar\sigma^{2}\,\E_{i}\big[1/\sigma^{2}_{{\rm or},i}\big]
    \;=\; \E_{i}\big[\avf{e^{-\delta_{i}}}\big]
    \;\simeq\; 1 - \E\avf{\delta} + \tfrac12\E\big[\avf{\delta^{2}}\big]
    \;\simeq\; 1 + \E\big[\avf{\delta^{2}}\big] \\[2pt]
    &=\; 1 + \E\big[\avf{\delta}^{2}\big] + \E\big[\Varf{\delta}\big]
    \;\simeq\; 1 + \Var\big[\avf{\delta}\big] + \E\big[\Varf{\delta}\big] ,
  \end{aligned}
\end{equation}
which is Eq.~(\ref{eq:twochannel_eta}). Every step after the first equality drops terms of order
$\delta^{3}$ and higher; because $e^{-\delta}$ is convex, the exact $\E[\avf{e^{-\delta}}]$ exceeds its
second-order truncation, so Eq.~(\ref{eq:twochannel_eta}) is a lower estimate of $\eta - 1$ that is
accurate only for $|\delta| \ll 1$ (\S\ref{sec:tier1} quotes the shortfall at the campaign's
severities). The coherent term $\Var[\avf{\delta}]$ is exactly what the
oracle comparison subtracts. Both factors of the multiplicative form follow by exponentiating each term
separately: for a purely coherent latent ($\delta_{i}$ constant in $\kv$, $= \log c_{i}$),
$\eta = \E[c]\,\E[c^{-1}]$ exactly (the amplitude-only closure of Appendix~\ref{app:nullbattery}), and for a
purely incoherent one $\eta = \E[\avf{e^{-\delta}}]$ with $\avf{\delta} = 0$. For a power-law tilt
$\delta_{i}(k) = -(\alpha_{i} - \bar\alpha)u(k)$, $u = \log(k/k_{\rm piv})$, the two terms become
$\Var(\alpha)\avf{u}^{2}$ and $\Var(\alpha)\Varf{u}$, Eq.~(\ref{eq:leverarm}).

\subsection{The attainability envelope}
\label{app:methodE}

Let $\Ahat' = b(\hat z)\AMF$ with $\E[b] = 1$, where on image $i$ the matched filter has variance
$\sigma_{0}^{2}c_{i}$ with $\E[c] = 1$ and $b$ is a function of the (well-inferred) latent. Then
$\E[\Ahat'|A] = A\,\E[b] = A$ and
$\Var(\Ahat'|A) = \E[b^{2}(\sigma_{0}^{2}c + A^{2})] - A^{2} = \sigma_{0}^{2}\E[b^{2}c] + A^{2}\Var(b)$,
Eq.~(\ref{eq:Vb}). Minimizing $\E[b^{2}(\sigma_{0}^{2}c + A^{2})]$ subject to $\E[b] = 1$ gives
$b \propto (\sigma_{0}^{2}c + A^{2})^{-1}$ and the envelope of Eq.~(\ref{eq:methodE}),
$V_{E}(A) = (\E[1/(\sigma_{0}^{2}c + A^{2})])^{-1} - A^{2}$. At $A = 0$, $b \propto c^{-1}$ and
$V_{E}(0) = \sigma_{0}^{2}/\E[c^{-1}] = \sigma_{0}^{2}\E[c]/(\E[c]\E[c^{-1}]) = \sigma_{0}^{2}/\etaB$, which equals
$\sigMF^{2}/\eta$ only when $\eta = \etaB$, \ie\ for a purely coherent latent. For $A \gg \sigma_{0}$ the
optimum is $b \to 1$ and $V_{E} \to \sigma_{0}^{2}$; the gain and the penalty balance where
$\sigma_{0}^{2}\E[c]\,v \simeq A^{2}v$ for small log-variance $v$ of $c$, \ie\ at $A \simeq \sigMF$.

\subsection{The well-inferable limit}
\label{app:wellinferable}

For a mixture with latent density $w(z)$, Fisher information is convex in the density
\citep{Cohen1968} -- the map $(a,b) \mapsto a^{2}/b$ is jointly convex, and
$I[p] = \int (\tauv\!\cdot\!\nabla p)^{2}/p$ inherits it -- so for the marginal
$p_{n} = \int \dd z\, w(z)\,p_{z}$,
\begin{equation}
  I_{\rm marg} \;=\; I\Big[\int \dd z\, w(z)\,p_{z}\Big] \;\le\; \int \dd z\, w(z)\,I[p_{z}]
  \;=\; \E_{z}[I_{z}], \qquad I_{z} = \tauv^{\dagger}\Nm_{z}^{-1}\tauv ,
\end{equation}
with \emph{no} condition on the posterior. Multiplying by $\sigMF^{2}$ gives Eq.~(\ref{eq:etawi_bound}),
$\eta \le \etawi$, for every covariance mixture. Equality holds when the latent posterior given a noise
map is a point mass, since then the marginal score $\E_{z|n}[\Nm_{z}^{-1}]n$ coincides with the
conditional score on every image. In that limit $\eta \to \sigMF^{2}\E_{z}[I_{z}] = \etawi$,
Eq.~(\ref{eq:etawi}), and by Jensen
$\E_{z}[I_{z}] \ge 1/\E_{z}[1/I_{z}] = 1/\sigor^{2}$, which is Eq.~(\ref{eq:wellinferable}). The
posterior widths measured in \S\ref{sec:results_mixtures} ($\sd(\alpha|n) = 0.02$ against a prior width of
$0.5$; $\sd(\log c|n) = 0.21$ against $1.9$) are what make the limit exact in practice.

\subsection{Independent pixels are template-blind}
\label{app:iid}

If the whitened pixels are independent with a common one-dimensional density $q$, then
$\log p(x) = \sum_{p}\log q(x_{p})$, the score is separable, $s_{x,p} = s_{1}(x_{p})$, and
$\Jop_{pq} = \E[s_{1}(x_{p})s_{1}(x_{q})] = I_{1}\delta_{pq}$ because the pixels are independent and
$\E[s_{1}] = 0$. Hence $\eta = \that^{\dagger}\Jop\that = I_{1}$ for every unit $\that$. The contracted
cumulant norms of Eq.~(\ref{eq:perturbative}) are likewise template-independent:
$\|\tilde\kappa_{3}(\that)\|^{2}_{F} = \sum_{p}\that_{p}^{2}\gamma_{3}^{2} = \gamma_{3}^{2}$.

\subsection{The complete-data bound}
\label{app:cdbound}

For $n = c + g$ with $g \sim \mathcal N(0,\Nm_{g})$ independent of the structure $c$, the marginal density
is $p_{n}(n) = \E_{c}[\mathcal N(n - c;0,\Nm_{g})]$, and differentiating under the expectation gives
$s(n) = \Nm_{g}^{-1}\big(n - \E[c\,|\,n]\big)$, Eq.~(\ref{eq:tweediescore}): the posterior mean of the
structure is subtracted, then the Gaussian score is applied. The bound follows from the data-processing
property of Fisher information: the amplitude information in the observed map cannot exceed the
information in the ``complete data'' $(d, c)$, and given $c$ the residual $d - c = A\tauv + g$ is Gaussian
with covariance $\Nm_{g}$, so
\begin{equation}
  I(A;\,d) \;\le\; I(A;\,d,c) \;=\; \tauv^{\dagger}\Nm_{g}^{-1}\tauv .
\end{equation}
Dividing by $\sigMF^{-2} = \tauv^{\dagger}\Nm^{-1}\tauv$ gives Eq.~(\ref{eq:cdbound}) in the stationary
case, with $P_{g}$ the spectrum of the Gaussian part and $P_{\rm tot}$ that of the whole. The bound is
attained when $c$ is perfectly recoverable from $d$ (the saturated regime of \S\ref{sec:tier2}). It is
the wrong ceiling when $c$ is not additive structure but a rescaling of the Gaussian part -- a scale
mixture -- because then ``removing'' $c$ is not an operation any estimator can perform, and the exact
mixture ceiling of \S\ref{app:wellinferable} applies instead; \S\ref{sec:results_mixtures} shows the
bound over-stating the room by a factor $2.3$ in that case.

\section{The pipeline as built}
\label{app:pipeline}

This appendix records the implementation choices behind the numbers of the paper, with emphasis on the
failure modes that were found on the way: each of them produced a plausible-looking wrong number before
it was caught, and each is now part of the stated protocol. The code -- the noise-model library \code{noise\_lib}, the $\eta$ framework \code{eta\_pipeline}, the
campaign driver, the null battery, the figure and table scripts and the diagnostic tools -- is public in a
\code{GitHub} repository\footnote{\url{https://github.com/kmbasu/Beyond-the-BLUE-Paper-I}; archived
release: \citealt{BeyondTheBLUE2026code}.} together with the result files of record, so that every table and
figure can be regenerated without a GPU. The release carries the model registry, the master seeds and the
run configurations, so that every map used in this paper is reconstructable from (model, image index);
its README gives the installation and the commands. Each subsection below ends with a pointer to the part
of the release that implements it.

\subsection{Ensembles, splits, whitening}

Every model is a row of a registry that fixes its generator, its severity parameters, its white floor
and its master seed; per-image generators are spawned from the master seed so that image $i$ of any
model, including its latents, is reconstructable from (model, $i$). Floored models are appended rows with
their own seeds, so that no floorless ensemble moved when the floor was introduced, and each floored
ensemble is its floorless twin plus a white map, image by image. Ensembles of $6000$ maps are split
25/40/10/25 into PSD/FIT/VAL/EVAL with one fixed split seed; extra FIT and VAL maps for the variational
runs are appended \emph{after} the frozen campaign ensemble, so that the PSD and EVAL splits stay
bit-identical across every method and every variational number is paired with its own quadrature and
bound by construction. The two-dimensional spectrum is the mean periodogram of the PSD split; whitening
multiplies by $\sqrt{n_{\rm pix}/\Phat(\kv)}$. Analytic spectra with strong suppressions (a beam factor
$e^{-40}$ at high $k$) are masked at $B^{2} > 10^{-8}$ in the quadratures to keep floating-point rounding
out of a division; \S\ref{sec:unmodelled} explains why this numerically necessary guard is physically
dangerous and how the floor removes the need for it.

Monte Carlo errors are bootstraps over EVAL images (images, never pixels); pair statistics (Method B)
need a two-axis bootstrap because every row shares the second batch, and bootstrapping rows alone
under-estimates the error several-fold. Heavy-tailed ensembles inflate the variance of every
moment-based estimator; where tails are extreme the variational method is preferred and the bootstrap
distributions are checked for stability.

\paragraph{Code pointer.}
The registry is the \code{REGISTRY} dictionary in \code{eta\_pipeline/models.py} (exported, with a
drift test, to \code{configs/registry.json} and \code{configs/REGISTRY.md}); image $i$ of a model is child
$i$ of \code{numpy.random.SeedSequence} of its master seed, the \code{*\_WN} floors and the splits use fixed
seed offsets stated in the README; the generators are in \code{noise\_lib/}, the spectrum estimate and
whitening in \code{eta\_pipeline/whitening.py}, and the bootstraps in the methods that report them.

\subsection{The variational ladder, round 2}

The first full-scale run of the ladder took seven hours, completed without error, and returned the
initialization for every model and template -- restart spreads exactly zero, every rung identical to full
precision -- and nothing in its output said so. Its test functions carried free per-pixel weight maps, and
with a fitting set comparable to the pixel count, unregularized training found sample-covariance null
directions before the physical solution; a second, less visible failure was that batch-mean-subtracting
the network output inside the autograd graph couples the images and breaks the integration by parts
behind Eq.~(\ref{eq:variational}), so that a network can cancel its output variance while the gradient
projection survives and the ``bound'' inflates without limit (the i.i.d. null returned $\eta = 25$ against
an exact $1.21$). The fiducial ladder is built on the rules of Table~\ref{tab:designrules}.

\begin{table}[t]
  \caption{Design rules of the fiducial ladder, each with the failure it answers.}
  \label{tab:designrules}
  \centering\small
  \begin{tabular}{p{7.4cm}p{7.4cm}}
    \toprule
    rule & because \\
    \midrule
    no free per-pixel parameters; spatial weighting in a fixed low-$k$ Fourier basis ($k_{\max} = 3$, 29 smooth maps, plus six template-pooled maps); convolution kernels free &
      $p \approx n$ pinning: pixel maps find sample-covariance null directions first \\
    train the unbiased raw bound $2\E[\text{proj}] - \E[f^{2}]$ with a learnable global scale; early-stop and report in scale-invariant form &
      the batch Rayleigh ratio is biased upward by $\Var(\text{proj})/B$ and rewards variance-inflating directions (VAL collapsed from 2.25 to 0.01 in one epoch in a smoke test) \\
    every rung is a linear head on nonlinear features standardized by detached running statistics, frozen at evaluation &
      raw feature scales spanned orders of magnitude; one learning rate was wrong for most coordinates \\
    output centering by a detached running constant, never a batch mean inside the graph &
      Stein-consistency: an in-graph batch mean breaks the integration by parts \\
    nonlinear rungs start at the linear optimum (zero-initialized head); the initial value is included in early-stopping selection &
      a reported value can never fall below the linear rung \\
    every fit records its VAL trajectory, anchor, best epoch and \code{moved\_off\_init}; a ladder in which nothing moved is stamped &
      a silent anchor return can masquerade as a result indefinitely \\
    early stopping with patience $\max(15, N_{\rm epochs}/3)$ and a warm-up $\min(30, N_{\rm epochs})$ during which it cannot fire, both derived from the run's own parameters &
      the patience erratum below \\
    a smooth, non-odd nonlinearity in the convolutional features &
      an odd network is blind to the bispectrum; several models are skewed \\
    \bottomrule
  \end{tabular}
\end{table}

\paragraph{The patience erratum.}
The first deep-rung pass called the ladder with a fixed early-stopping patience of 15 epochs that the
\code{-{}-epochs} flag did not scale. The conv rung's head is zero-initialized, so at step zero the
convolutional body receives no gradient -- $\partial f/\partial(\text{body}) \propto$ head $= 0$ -- and the
body learns only once the head has drifted; the VAL bound sits at or below its anchor for 30--70 epochs
before it moves. Every restart that did not exceed its anchor within 15 epochs was therefore killed at
epoch 15, and every conv-rung number of that pass was the value of whichever restart happened to nudge
above its anchor early. The fix is the derived patience and warm-up of Table~\ref{tab:designrules} and a
per-restart \code{stopped\_by\_patience} flag. The process lesson is general: a protocol constant that is
not derived from the run's own parameters will silently decide a result.

\paragraph{The final pass.}
Sixteen (model, template) cells of the floored configuration were run as one array job on the JUPITER
GH200 partition at the J\"ulich Supercomputing Centre, one GPU per cell and no data parallelism -- the
rungs have at most $2.5$\,k parameters and the cost is the double back-propagation through $128^{2}$
convolutions (conv $11.6$\,s per epoch, cubic $25.6$, quadratic $10.2$, \rung{reweight} $0.07$): 200
epochs, patience 66, warm-up 30, four restarts (six for the four decisive cells),
$n_{\rm FIT}/n_{\rm VAL}/n_{\rm EVAL} = 10\,800/1200/1500$, ensembles regenerated from the registry seeds so
that the splits are bit-identical to the laptop runs. No cell exceeded seven hours. The confusion cells and the
calibration row were run with the same protocol and the template-pooled basis (\code{-{}-pool both} for
the fiducial row, \code{-{}-pool template} for the calibration row), six restarts, and an ensemble of
$30\,000$ maps ($n_{\rm EVAL} = 7500$), with the heavy cells split across array tasks so that no task
approaches the 12-hour wall; per-rung checkpointing with merge-on-save was added after a cubic-only task
had overwritten a sibling's conv values with stale ones. The pooling basis itself was changed after the
calibration of \S\ref{sec:sensitivity}: the Fourier basis at $k_{\max} = 3$ has 29 maps (an earlier
docstring said 49), and the template basis adds six.

\paragraph{The float64 A/B test.}
The floorless above-\rung{reweight} values of the earlier passes were re-run on JUPITER with identical
seeds, whitening and anchors (to three decimals) and one change: the storage dtype of the maps before
whitening, float32 versus float64. In float64 the PCA-leakage conv rung, the row-median conv rung and the
row-median cubic rung all went from five of six restarts climbing to none of six moving, with VAL
trajectories showing the null-cell signature (decay to $0.76$--$0.90$ on these floorless float64 cells,
against $0.75$--$0.85$ on the floored null cells below). The float64 results are kept in a
separate directory and were deliberately not merged into the results of record; \S\ref{sec:unmodelled}
gives the mechanism.

\paragraph{Code pointer.}
The ladder of Table~\ref{tab:designrules} is \code{eta\_pipeline/variational2.py}, driven by
\code{scripts/run\_eta\_campaign.py -{}-mode variational2}; every fit's VAL trajectory, anchor, best epoch,
\code{moved\_off\_init} and \code{stopped\_by\_patience} flags are stored in the per-cell JSON files of
\code{results/eta\_results/}. The production settings of the final pass, including the pooling basis of
each cell, are the Slurm cell lists in \code{scripts/hpc/}; the float64 A/B results are kept apart in
\code{results/eta\_results\_hpc\_f64/} (the map dtype is switched with \code{ETA\_MAPS\_FLOAT64=1}).

\subsection{Sensitivity floors}

In every null cell the conv rung's VAL bound decays from its anchor to $0.75$--$0.85$ within the patience
window, because the head fits FIT-batch noise from epoch~0; a physical signal has to lift the bound above
the anchor against that decay, which sets the rung's sensitivity floor at $\eta \sim 1.1$--$1.2$. The cubic
rung, with $\sim 500$ parameters, decays less and resolves $\sim 1.05$; the \rung{reweight} rung, with no
convolutional features, is essentially noiseless on Gaussian cells. The Fourier pooling basis at $k_{\max} = 3$
resolves $\sim 20$\,px on a $128^{2}$ map against a 5-px beam, which is why the ladder as first built could
not see template-localized statistics; the template-pooled basis of \S\ref{sec:convrung} is the remedy,
calibrated in \S\ref{sec:sensitivity}.

\paragraph{Code pointer.}
The floors are read off the same result files: the Gaussian control cells' VAL trajectories in
\code{results/eta\_results/} show the decay, and the calibration row's files (the beam-sharing companion of
Appendix~\ref{app:confusion}) the recovery; N6 of \code{scripts/run\_null\_tests.py} audits the shipped
ladder values against their exact ceilings.

\subsection{Exact methods and bounds}

The latent quadrature uses a $61 \times 41$ (slope $\times$ amplitude) grid for the tilt mixtures, a
180-node angle grid for the rotating ridge, and for PCA leakage a Monte Carlo prior on the leak power
$c = \sum_{i}a_{i}^{2}$ ($10^{6}$ draws: $P(c = 0) = e^{-3}$, $\E[c] = 8.1$, median $2.0$, $q_{99} = 75$) as a
$c = 0$ atom plus a 320-node logarithmic grid. The floored quadrature is the non-factorized
two-dimensional form (the floor breaks $P = a^{2}P_{\rm base}$); at zero floor it reproduces the factorized
form to $10^{-15}$. Complete-data bounds are evaluated from the generative spectra on the regularized band
and checked for mask stability (N8). For the confusion rows the data spectrum's DC bin is dominated by
$\text{mean}^{2}n_{\rm pix}^{2}$ while the analytic background spectrum has the Poisson plateau there, so
the bound as first computed carried a DC term in its numerator that its denominator lacked; with the DC
mode treated consistently the beam-sharing-companion bound lands on its exact value ($4.98$ against $5$,
from $5.31$) and the fiducial bounds become $199.9$ (extended) and $41.7$ (compact), the values quoted
in the paper.

\paragraph{Code pointer.}
The quadratures, complete-data bounds and cumulant norms are \code{eta\_pipeline/quadrature.py},
\code{bounds.py} and \code{cumulants.py} (the one-dimensional exact cases in \code{analytic1d.py}, the
two-channel split of \S\ref{sec:tier1} in \code{channels.py}), all run by
\code{scripts/run\_eta\_campaign.py -{}-mode cheap} in minutes on a laptop;
\code{scripts/tools/compare\_with\_shipped.py} checks a re-run against the shipped files (deterministic
quantities agree to $3 \times 10^{-12}$), and the DC-mode and mask-stability checks are among the
diagnostic tools in \code{scripts/tools/}.
\section{The null battery}
\label{app:nullbattery}

\S\ref{sec:nullbattery} states the principle and Table~\ref{tab:nulls} lists the tests. This appendix
says what a null test is in this context, where the hundred checks come from, and what four of the tests
established. The battery is \code{scripts/run\_null\_tests.py} of the public release
(Appendix~\ref{app:pipeline}), which the README requires to pass before any new $\eta$ number is
produced; its final log is kept in \code{results/tests\_and\_results/}.

\subsection{What a null test is, and what it is not}

A \emph{null test} here is a configuration in which the correct answer is known before the code is run,
and on which the entire apparatus -- generator, whitening, splits, estimator, error bars -- is exercised
end to end and required to reproduce that answer. It is a test of the instrument, not of the physics.
Two kinds are needed, and a battery containing only the first kind is worthless.

The first kind has a \emph{known-null} answer. Gaussian noise of known covariance has $\eta \equiv 1$ by
the theorem of \S\ref{sec:mf}, the third and fourth cumulants of a Gaussian field vanish identically, and
a method that returns anything else on such an ensemble has a defect rather than a discovery. N1, N2 and
the Gaussian rows of N6 are of this kind. Their weakness is that they are all passed by a method that
returns $1$ unconditionally -- that is, by an instrument with no sensitivity at all, which is precisely
the failure mode a variational lower bound is prone to, since a test function that never moves off its
initialization returns the linear anchor and looks like a clean null.

The second kind therefore carries a \emph{known non-null} answer: a model contrived so that $\eta > 1$
can be computed in closed form, on which the machinery must recover that specific number and must not
exceed it. N3, N4, N4b, N5 and N7 are of this kind, and they are what make the first kind informative.
N7 is the strongest of them because it runs at full map size through the same whitening path, the same
splits and the same trained ladder as a campaign cell, so that a failure implicates the instrument as
deployed rather than a simplified version of it.

Two properties of the battery follow from this design and are used throughout the paper. First, the
tests run \emph{on the executing machine} before a final run, not once on a development laptop: three of
the defects recorded in \S\ref{sec:unmodelled} and Appendix~\ref{app:pipeline} are properties of a
storage format or of an accumulation order, and would not survive being certified elsewhere. Second, and
in the other direction, \emph{a passing null test is not a null result}. The corollary of
\S\ref{sec:ladder_var} -- a rung that does not move proves only that its class found nothing -- applies
to the battery as much as to the physics. The battery
licenses the statement ``this machinery would have seen an advantage of the size we are excluding''; it
never by itself licenses the statement ``there is no advantage''. Where the paper says ``nothing'', the
statement rests on a theorem, an exact computation or an independent upper bound, with the battery
establishing only that the instrument reporting it was working.

\subsection{Where the hundred checks come from}

The nine tests assert a hundred separate conditions, and it is the conditions that are counted.
Table~\ref{tab:nullcount} gives the breakdown. Two tests dominate because they sweep the stored results
cell by cell: N6 audits every one of the 36 cells for which an exact ceiling or a complete-data bound
exists, and N8 recomputes three quantities across several mode-mask thresholds for every cell that has
them. The remaining seven tests contribute twenty-eight conditions between them. The final state is
$100$ passed and $0$ failed, on the final code and the final result set.

\begin{table}[t]
  \caption{The hundred checks of the null battery, by test. The count is of individual assertions, not
  of tests; N6 and N8 dominate because they sweep the stored results cell by cell.}
  \label{tab:nullcount}
  \centering\small
  \begin{tabular}{lp{11.6cm}r}
    \toprule
    \# & what is asserted & checks \\
    \midrule
    N1  & white Gaussian null: $\sigMF$, three rungs, two cumulant norms            & 6 \\
    N2  & red ($1/f^{3}$ + beam) Gaussian null, same six through the PSD path        & 6 \\
    N3  & analytic score, linear anchor, pixelwise bound and $\kappa_{4}$ pair trick against the exact one-dimensional $\eta$ & 4 \\
    N4  & template independence on i.i.d.\ noise                                     & 1 \\
    N4b & paired overlap-law ordering, extended $>$ compact                          & 1 \\
    N5  & amplitude-only closure, slope-only headroom, floored-to-scaled agreement, floored legitimacy & 4 \\
    N6  & 36 audited cells, plus the assertion that rung ordering is not required    & 37 \\
    N7  & known-$\eta$ recovery: bound, recovery and non-exceedance per template, template independence & 6 \\
    N8  & mask stability of $\sigMF$ (10) and of the complete-data bound (14), floor comparisons (10), liveness (1) & 35 \\
    \midrule
    & \textbf{total} & \textbf{100} \\
    \bottomrule
  \end{tabular}
\end{table}

\subsection{Four tests in detail}

\begin{enumerate}[label=\arabic*.,leftmargin=1.6em,itemsep=4pt]

\item \emph{N5, the closed-form mixture.} For a mixture in which only the overall noise amplitude
varies, $\eta = \E[\sigma^{2}]\,\E[\sigma^{-2}]$ in the well-inferable limit of
Appendix~\ref{app:wellinferable} -- the closure is the complete-data information, and the marginal
information falls short of it by the missing-information term of \citet{Louis1982}, which vanishes as
the latent becomes inferable; here a single scalar is inferred from $128^{2}$ pixels and the term is
negligible (\S\ref{sec:results_mixtures} quotes the posterior widths). For our clipped-normal amplitude
prior the closure is $1.1166$. The $20\,000$-map fiducial quadrature returns $1.1164 \pm 0.0128$; the battery's own 1500-map
check returns $1.081 \pm 0.048$. This is a better null than the ``$\eta = 1$'' our own earlier reasoning
had predicted for this case, precisely because it is a non-trivial number: a method that returned $1$
unconditionally would fail it. It also verifies the template independence that a purely coherent latent
requires, the two templates differing by $0.0096 \pm 0.0174$.

\item \emph{N6, the one theorem that constrains the ladder from above.} Every rung is a certified lower
bound on $\eta$, so no rung may exceed the exact ceiling where one exists -- the Tier-1 and scale-mixture
quadratures, and the exact $I_{1}$ of \S\ref{sec:results_confusion} -- or the complete-data bound where
the structure is removable. The audit covers the 36 cells that have such a ceiling, within $2\sigma$, and
finds no violation. N6 deliberately does \emph{not} require the rungs to be ordered. Each rung is an
independent optimization over its own function class, and the convolutional rung's sensitivity floor
($\eta \sim 1.15$, \S\ref{sec:convrung}) lies above the cubic rung's ($\sim 1.05$), so a cubic value
above a convolutional one is legitimate and does occur; requiring an ordering would have turned a
property of the instrument into a spurious failure.

\item \emph{N7, recovery of a known $\eta > 1$ at full map size.} Confusion with a beam-correlated
Gaussian companion (\S\ref{sec:confusion_setup}, Appendix~\ref{app:confusion}) whitens to i.i.d.\
pixels, so by the theorem of \S\ref{sec:tier2} its $\eta = I_{1} = 3.5688$ is exact and the same for
every template. The fiducial ladder recovers $92$\,\% and $87$\,\% of it on the extended and compact
templates -- $104$\,\% and $99$\,\% of the $3.145$ expected once the measured floating-point dilution of the
corner modes (\S\ref{sec:unmodelled}) is applied, the extended value $2.3\sigma$ above it -- exceeds it on no rung, and returns values for the two
templates agreeing to $1.6\sigma$. This is the calibration behind every statement in
\S\ref{sec:results_confusion} that a ladder value is a lower bound of known tightness.

N7 has a history worth recording, because it is the clearest demonstration in the paper that a battery of
known-null tests is not sufficient. It replaced a tier-factorization check that had turned out to be
circular -- it compared $\eta$ on a per-image Gaussian surrogate with a quantity defined on that same
surrogate -- and which became moot in any case once the leaked-mode model proved to be an exact scale
mixture (\S\ref{sec:pca}). Its first run then failed, recovering $63$\,\% of the known value on the
extended template and $0$\,\% on the compact one. Nothing else in the battery had registered a problem:
every Gaussian null passed, every rung sat below every ceiling, and the campaign's numbers looked
internally consistent. What N7 had exposed was the pooling-basis flaw of \S\ref{sec:convrung} -- a
smooth Fourier basis at $k_{\max} = 3$ cannot represent a template-localized test function, so the ladder
was structurally blind to exactly the statistics that confusion carries. The remedy, pooling with the
whitened template itself, is the reason the confusion ceilings of \S\ref{sec:results_confusion} are what
they are.

\item \emph{N8, band regularization.} $\sigMF$, the band statistics and the complete-data bound are
recomputed at several mode-mask thresholds and required to be stable. This is the test that exposed the
missing-floor defect of \S\ref{sec:unmodelled}. Every floored cell passes it to $0.00$\,\%, while four
floorless compact cells drift by $24$--$46$\,\%, which is what keeps the diagnostic live: a test that
nothing can fail has stopped being a test.

\end{enumerate}

\section{Source confusion as a marked Poisson process}
\label{app:confusion}

\subsection{Campbell cumulants and the truncation laws}

A map of unresolved sources is $D(\xv) = \sum_{i}S_{i}B(\xv - \xv_{i})$ with $\xv_{i}$ a Poisson process of
density $\bar\lambda$ per unit solid angle and fluxes $S_{i}$ drawn independently from the counts
$\dnds$ between $\Smin$ and $\Scut$. By Campbell's theorem the cumulants of the pixel value are
\begin{equation}
  \kappa_{n} \;=\; q_{n}\,\Omega_{n}, \qquad
  q_{n} = \int_{\Smin}^{\Scut}S^{n}\,\frac{\dd N}{\dd S}\,\dd S, \qquad
  \Omega_{n} = \int B(\xv)^{n}\,\dd^{2}x = \frac{\Omega_{\rm beam}}{n}
\end{equation}
for a Gaussian beam of solid angle $\Omega_{\rm beam}$, so that the mean is $q_{1}\Omega_{\rm beam}$, the
confusion variance $\sigc^{2} = q_{2}\Omega_{\rm beam}/2$, the skewness $\kappa_{3}/\kappa_{2}^{3/2}$ and the
excess kurtosis $\kappa_{4}/\kappa_{2}^{2}$; the power spectrum, bispectrum and trispectrum are
Eq.~(\ref{eq:campbell}) with $\lambda\langle a^{n}\rangle \to q_{n}$ and $\tilde\psi \to \tilde B$. For a
power-law $\dnds \propto S^{-\gamma}$ with $1 < \gamma < 3$ the moments are dominated by the bright end,
$q_{n} \propto \Scut^{\,n + 1 - \gamma}$, so the skewness scales as $\Scut^{(\gamma - 1)/2}$ and the excess
kurtosis as $\Scut^{\gamma - 1}$: a shallower catalog cut (larger $\Scut$) makes the field more
non-Gaussian, and a deeper one Gaussianizes it, which is the $\Scut$ axis of the severity plane. The
counts fix the cumulants only; the floor $f_{N}$ and the template scale fix the band and the bound only,
because $P_{\rm conf} \propto \sigc^{2}$ and $\sigma_{w} = f_{N}\sigc$ cancel the counts from every
second-order quantity.

\subsection{Configuration and validation}

The paper's pixel convention is read as a $20''$ beam at $4''$ per pixel; the $350\,\mum$ counts are a
Schechter function of the form used by \citet{Bethermin2012}, with $\alpha = -1.8897$,
$\phi_{*} = 7014\,{\rm deg^{-2}}$ and $S_{*} = 19.012$\,mJy fitted to $350\,\mum$ data by \citet{BasuGuerreroBertoldi2026}; below $\sim 6$\,mJy the counts are an extrapolation. The flux
limits are $\Smin = 0.1$\,mJy and $\Scut = 100$\,mJy.
Analytically $\sigc = 6.449\,\mjb$, $\Nbeam = 28.25$, $\Scut/\sigc = 15.5$, skewness
$2.120$, excess kurtosis $9.091$, $16\,344$ sources per map. Because $\Nbeam < 200$ no faint-end Gaussian
substitution is made; every source is injected explicitly, so the Gaussianization-by-fiat that such
substitutions introduce, which can only bias $\eta$ downward, does not arise. The floor of the fiducial
row is $\sigma_{w} = f_{N}\sigc = 3.224\,\mjb$ with $f_{N} = 0.5$.

Eight checks are run on 8000 maps before any number is quoted. \emph{V1}, one-point closure: measured
mean, rms, skewness and excess kurtosis $16.353$, $6.450$, $2.120$, $9.120$ against the Campbell values
$16.354$, $6.449$, $2.120$, $9.091$ ($0.006$\,\%, $0.014$\,\%, $0.031$\,\%, $0.32$\,\%, computed at full precision). \emph{V2}, two-point closure:
$\Phat/P_{\rm conf} = 0.99933 \pm 0.0112$ against the sampling scatter $1/\sqrt{8000} = 0.0112$ of the spectrum estimate; the DFT
prefactors are fixed by construction, so this is a falsifiable test rather than a fit. \emph{V3},
flatness: without a floor the compact template's $|\tautil|^{2}/\Phat$ has a relative spread of $0.0112$,
equal to that sampling scatter, \ie\ flat to the precision of the measurement, while the extended
template's is $9.6$; with the floor the compact ratio rolls off past the knee. \emph{V4--V5}, the band:
the compact template's $\sigMF$ drifts by $216.7$\,\% across mode cuts without a floor and
$\sigMF\sqrt{N_{\rm modes}}$ by $0.059$\,\% -- mode-count limited -- and by $0.00$\,\% with the floor; the
extended template's by $0.00$\,\% either way. \emph{V6}, precision: floorless compact $\sigMF = 49.7$ in
float64 against $60.1$ in float32, a $20.9$\,\% dependence on the storage format; floored, $0.00$\,\%.
\emph{V7--V8}, predicted against empirical $\sigMF$ under the split discipline: agreement at $0.4$--$1.3\sigma$;
the compact template's filter output inherits the $P(D)$ tail (excess kurtosis $91$ floorless, $8.8$
floored), so its empirical $\sigMF$ has a relative standard error of $\sqrt{(g_{2} + 2)/4n}$, $10.8$\,\% on
2000 images -- a property of the model, not a defect.

\subsection{The beam-sharing companion, and its exact solution}

The third row adds to the confusion field a Gaussian companion of the same map-level variance as the
fiducial floor, $\rho = \Var[\text{companion}]/\Var[\text{confusion}] = f_{N}^{2} = 0.25$, but injected
before the beam. Then $P_{\rm tot} = (1 + \rho)\,q_{2}|\tilde B|^{2}$ is exactly proportional to
$|\tilde B|^{2}$, whitening returns $x = (D + W)\times{\rm const}$, and the pixels are independent. By
\S\ref{app:iid} $\eta$ is template-independent and equals $\Var[Y]\,I(Y)$ for the one-pixel variable
$Y = D + W$, with $D$ the compound-Poisson pixel value on the unsmoothed lattice and $W$ Gaussian; the
one-dimensional density is a convolution computable by quadrature, and $\eta = 3.568844$, converged to
$7 \times 10^{-5}$ across grid size, extent and density floor. The complete-data bound is exactly
$(1 + \rho)/\rho = 5$ for every template; on the analytic spectrum the pipeline returns $4.99997$ (extended and
sub-beam) and a compact value $0.8$\,\% low, a $10^{-20}$ guard biting at corner modes where $|\tilde B|^{2}$
has underflowed; on the data spectrum, with the DC mode treated consistently, it returns $4.98$ (extended)
and $4.97$ (compact), the values of Table~\ref{tab:master_floored}.

The row also calibrates what the floating-point channel of \S\ref{sec:unmodelled} costs when the
answer is known. Whitening by the \emph{analytic} spectrum reproduces $(D + W)\times{\rm const}$ to
$3 \times 10^{-27}$ for $k < 0.30$ and is destroyed above it (the $1/\tilde B$ inversion reaches
$10^{19}$); whitening by the \emph{estimated} spectrum, which is what the pipeline does, normalizes those
modes back to unit variance, so the cost is dilution rather than blow-up. Measured from the whitened
marginal against the exact cumulants of $Y$: skewness $8.05$ against $8.57$, excess kurtosis $151.5$
against $165.0$. If the rounding acts as an independent Gaussian of fractional variance $\varepsilon$,
those scale as $(1 - \varepsilon)^{3/2}$ and $(1 - \varepsilon)^{2}$, giving $\varepsilon = 0.0405$ and
$0.0416$ -- two routes agreeing to $2.7$\,\%, which is itself the evidence that the dilution model is the
right description. Hence $\rho_{\rm eff} = \rho + \varepsilon(1 + \rho)/(1 - \varepsilon) = 0.3035$ and
the ladder should recover $\eta = 3.145$ through the fiducial whitening, an $11.9$\,\% shortfall from
the pristine value. \S\ref{sec:sensitivity} reports what it recovers.

\section{The master tables}
\label{app:master}

Tables~\ref{tab:master_floored} and \ref{tab:master_floorless} list every cell of the campaign, generated
directly from the result files by \code{make\_figures/tabD\_master.py}. Columns: the white floor
$\sigma_{w}$ (map units; $\mjb$ for confusion); the matched-filter variance predicted from the estimated
spectrum, $\sigMF$; the exact ceiling where the model is a covariance mixture; the complete-data bound
where the structure is additive (DC-consistent for the confusion rows); the pooling basis of the ladder
run (F: the fixed Fourier basis of \S\ref{sec:convrung}; T: the template-pooled basis; F+T: both), which
\S\ref{sec:unmodelled} shows to be decisive on template-localized structure; and the five rungs of the
variational ladder on the EVAL split with bootstrap errors. The rows below the rule in
Table~\ref{tab:master_floored} are the template-pooled reruns of fiducial cells quoted in the text
(\S\ref{sec:results_mixtures}, \S\ref{sec:unmodelled}); their $\sigMF$ and bounds are those of the row
above. Rung values in gray never exceeded their
anchor on any restart within the full patience window (genuine nulls above the sensitivity floor, not
truncated searches). Values marked $^{\dagger}$ are degenerate optimizations -- a negative value, or a
bootstrap error that collapsed below $0.5$\,\% of the value -- which are still valid lower bounds but
not measurements of the rung; the collapse of an error bar by a factor of thirty against its neighbors
is the diagnostic. In the floorless table, values marked $^{\times}$ are the float32 rounding channel of
\S\ref{sec:unmodelled} and are not physical.

One floorless entry needs a word of its own. The extended conv rung of the confusion row reads
$91.4 \pm 3.5$, with every restart still climbing at the 200-epoch wall, and it is not a measurement of
anything. Without a floor the whitened one-pixel distribution of pure confusion carries an atom at
zero -- a finite chance that a beam contains no source at all -- and the Fisher information of a
distribution with an atom is unbounded, so this cell's $\eta$ is provably infinite and no finite value
could be right. What the rung is in fact climbing is the rounding channel of
\S\ref{sec:three_channels}: this row is stored in float64, and $2.5$\,\% of its modes still lie beyond
the float64 plateau at $k = 0.63$. It is listed for completeness, and is the strongest form in the paper
of the lesson that the numerics are part of the noise model. Table~\ref{tab:codenames} maps the descriptive names used
in the text to the registry's code names.

\begin{sidewaystable}
  \caption{Master table, realistic (floored) configuration.}
  \label{tab:master_floored}
  \centering\scriptsize
  \setlength{\tabcolsep}{3pt}
  \begin{tabular}{llccccccccccc}
\toprule
model & code name & tpl & $\sigma_w$ & $\sigma_{\rm MF}$ & exact $\eta$ & bound & basis & linear & reweight & quadratic & cubic & conv \\
\midrule
red + white noise & \code{T0\_RED\_REAL} & ext & 1.00 & 0.86 & -- & -- & F & 1.022 $\pm$ 0.033 & \textcolor{gray}{1.022 $\pm$ 0.033} & -- & -- & -- \\
red + white noise & \code{T0\_RED\_REAL} & com & 1.00 & 12.60 & -- & -- & F & 1.064 $\pm$ 0.032 & \textcolor{gray}{1.064 $\pm$ 0.032} & -- & -- & -- \\
spectral-tilt mixture & \code{T1\_PSRAND\_WN} & ext & 0.27 & 1.30 & 10.18 $\pm$ 1.45 & -- & F & 1.112 $\pm$ 0.058 & 7.119 $\pm$ 1.127 & 1.108 $\pm$ 0.061 & -- & 1.094 $\pm$ 0.066 \\
spectral-tilt mixture & \code{T1\_PSRAND\_WN} & com & 0.27 & 6.88 & 1.79 $\pm$ 0.08 & -- & F & 0.948 $\pm$ 0.046 & 1.445 $\pm$ 0.075 & \textcolor{gray}{0.948 $\pm$ 0.046} & -- & 0.940 $\pm$ 0.049 \\
slope-only tilt mixture & \code{T1\_PSRAND\_SLOPE\_WN} & ext & 0.27 & 1.24 & 7.76 $\pm$ 0.79 & -- & -- & -- & -- & -- & -- & -- \\
slope-only tilt mixture & \code{T1\_PSRAND\_SLOPE\_WN} & com & 0.27 & 6.80 & 1.68 $\pm$ 0.07 & -- & -- & -- & -- & -- & -- & -- \\
random-orientation anisotropy & \code{T1\_RANDOR\_WN} & ext & 1.04 & 0.98 & 1.25 $\pm$ 0.05 & -- & F & 0.898 $\pm$ 0.038 & 0.924 $\pm$ 0.047 & \textcolor{gray}{0.898 $\pm$ 0.038} & -- & 0.890 $\pm$ 0.038 \\
random-orientation anisotropy & \code{T1\_RANDOR\_WN} & com & 1.04 & 14.60 & 1.20 $\pm$ 0.04 & -- & F & 0.963 $\pm$ 0.034 & 1.026 $\pm$ 0.044 & 0.955 $\pm$ 0.034 & -- & 0.961 $\pm$ 0.035 \\
row-median residual & \code{T2\_MEDIAN\_WN} & ext & 1.06 & 0.66 & -- & -- & F & 0.956 $\pm$ 0.060 & 1.286 $\pm$ 0.070 & \textcolor{gray}{0.956 $\pm$ 0.060} & 1.008 $\pm$ 0.066 & 0.945 $\pm$ 0.062 \\
row-median residual & \code{T2\_MEDIAN\_WN} & com & 1.06 & 13.47 & -- & -- & F & 1.025 $\pm$ 0.036 & \textcolor{gray}{1.025 $\pm$ 0.036} & 1.020 $\pm$ 0.036 & 1.020 $\pm$ 0.036 & \textcolor{gray}{1.025 $\pm$ 0.036} \\
scan crossings (sym.) & \code{T2\_CROSS\_SYM\_WN} & ext & 0.51 & 0.82 & -- & 1.06 & F & 1.008 $\pm$ 0.035 & \textcolor{gray}{1.008 $\pm$ 0.035} & 1.004 $\pm$ 0.036 & \textcolor{gray}{1.008 $\pm$ 0.035} & \textcolor{gray}{1.008 $\pm$ 0.035} \\
scan crossings (sym.) & \code{T2\_CROSS\_SYM\_WN} & com & 0.51 & 8.63 & -- & 1.09 & F & 0.994 $\pm$ 0.038 & \textcolor{gray}{0.994 $\pm$ 0.038} & 0.989 $\pm$ 0.039 & \textcolor{gray}{0.994 $\pm$ 0.038} & \textcolor{gray}{0.994 $\pm$ 0.038} \\
scan crossings (pos.) & \code{T2\_CROSS\_POS\_WN} & ext & 0.48 & 0.80 & -- & 1.01 & F & 0.996 $\pm$ 0.037 & \textcolor{gray}{0.996 $\pm$ 0.037} & 0.996 $\pm$ 0.038 & 0.990 $\pm$ 0.039 & \textcolor{gray}{0.996 $\pm$ 0.037} \\
scan crossings (pos.) & \code{T2\_CROSS\_POS\_WN} & com & 0.48 & 8.03 & -- & 1.01 & F & 1.016 $\pm$ 0.034 & \textcolor{gray}{1.016 $\pm$ 0.034} & 0.964 $\pm$ 0.059 & 0.710 $\pm$ 0.285 & \textcolor{gray}{1.016 $\pm$ 0.034} \\
sub-threshold glitches & \code{T2\_GLITCH\_WN} & ext & 0.86 & 0.92 & -- & 1.04 & F & 0.965 $\pm$ 0.031 & \textcolor{gray}{0.965 $\pm$ 0.031} & \textcolor{gray}{0.965 $\pm$ 0.031} & \textcolor{gray}{0.965 $\pm$ 0.031} & \textcolor{gray}{0.965 $\pm$ 0.031} \\
sub-threshold glitches & \code{T2\_GLITCH\_WN} & com & 0.86 & 12.30 & -- & 1.05 & F & 0.929 $\pm$ 0.033 & \textcolor{gray}{0.929 $\pm$ 0.033} & 0.920 $\pm$ 0.034 & \textcolor{gray}{0.929 $\pm$ 0.033} & \textcolor{gray}{0.929 $\pm$ 0.033} \\
PCA leakage & \code{T2\_PCA\_WN} & ext & 0.10 & 1.67 & 2.14 $\pm$ 0.05 & 4.90 & F & 0.983 $\pm$ 0.048 & 1.655 $\pm$ 0.052 & 0.956 $\pm$ 0.046 & 0.266 $\pm$ 0.001$^{\dagger}$ & 0.948 $\pm$ 0.058 \\
PCA leakage & \code{T2\_PCA\_WN} & com & 0.10 & 4.31 & 1.09 $\pm$ 0.02 & 1.33 & F & 0.996 $\pm$ 0.036 & 0.989 $\pm$ 0.027 & 0.867 $\pm$ 0.015 & 0.981 $\pm$ 0.032 & \textcolor{gray}{0.996 $\pm$ 0.036} \\
source confusion & \code{T2\_CONFUSION\_WN} & ext & 3.22 & 5.72 mJy/b & -- & 200 & F+T & 1.006 $\pm$ 0.020 & \textcolor{gray}{1.006 $\pm$ 0.020} & 1.991 $\pm$ 0.059 & 2.772 $\pm$ 0.086 & 3.783 $\pm$ 0.071 \\
source confusion & \code{T2\_CONFUSION\_WN} & com & 3.22 & 157 mJy/b & -- & 41.72 & F+T & 0.956 $\pm$ 0.048 & \textcolor{gray}{0.956 $\pm$ 0.048} & 1.447 $\pm$ 0.107 & 1.576 $\pm$ 0.357 & 2.448 $\pm$ 0.059 \\
beam-sharing companion & \code{T2\_CONFUSION\_ATM} & ext & 0 & 6.37 mJy/b & 3.57 & 4.98 & T & 0.996 $\pm$ 0.020 & -- & 1.820 $\pm$ 0.035 & 2.380 $\pm$ 0.057 & 3.277 $\pm$ 0.058 \\
beam-sharing companion & \code{T2\_CONFUSION\_ATM} & com & 0 & 55.61 mJy/b & 3.57 & 4.97 & T & 0.997 $\pm$ 0.054 & -- & 1.755 $\pm$ 0.067 & 1.961 $\pm$ 0.340 & 3.120 $\pm$ 0.080 \\
\midrule
spectral-tilt mixture (rerun) & \code{T1\_PSRAND\_WN} & com & 0.27 & -- & -- & -- & F+T & 0.948 $\pm$ 0.046 & -- & 0.943 $\pm$ 0.047 & -- & 0.973 $\pm$ 0.072 \\
PCA leakage (rerun) & \code{T2\_PCA\_WN} & ext & 0.10 & -- & 2.14 $\pm$ 0.05 & -- & F+T & 0.983 $\pm$ 0.048 & -- & 0.928 $\pm$ 0.039 & 0.090 $\pm$ 0.000$^{\dagger}$ & 1.361 $\pm$ 0.068 \\
row-median residual (rerun) & \code{T2\_MEDIAN\_WN} & ext & 1.06 & -- & -- & -- & F+T & 0.956 $\pm$ 0.060 & 1.281 $\pm$ 0.074 & \textcolor{gray}{0.956 $\pm$ 0.060} & 0.947 $\pm$ 0.067 & \textcolor{gray}{0.956 $\pm$ 0.060} \\
row-median residual (rerun) & \code{T2\_MEDIAN\_WN} & com & 1.06 & -- & -- & -- & F+T & 1.025 $\pm$ 0.036 & \textcolor{gray}{1.025 $\pm$ 0.036} & 1.018 $\pm$ 0.036 & 1.015 $\pm$ 0.036 & 1.022 $\pm$ 0.037 \\
sub-threshold glitches (rerun) & \code{T2\_GLITCH\_WN} & com & 0.86 & -- & -- & -- & F+T & 0.929 $\pm$ 0.033 & -- & 0.921 $\pm$ 0.036 & \textcolor{gray}{0.929 $\pm$ 0.033} & \textcolor{gray}{0.929 $\pm$ 0.033} \\
\bottomrule
\end{tabular}

\end{sidewaystable}

\begin{sidewaystable}
  \caption{Master table, floorless configuration (theoretical limit; $^{\times}$ = not physical).}
  \label{tab:master_floorless}
  \centering\scriptsize
  \setlength{\tabcolsep}{3pt}
  \begin{tabular}{llccccccccccc}
\toprule
model & code name & tpl & $\sigma_w$ & $\sigma_{\rm MF}$ & exact $\eta$ & bound & basis & linear & reweight & quadratic & cubic & conv \\
\midrule
white noise & \code{T0\_WHITE} & ext & 0 & 0.12 & -- & -- & F & 0.947 $\pm$ 0.036 & \textcolor{gray}{0.947 $\pm$ 0.036} & -- & -- & -- \\
white noise & \code{T0\_WHITE} & com & 0 & 7.53 & -- & -- & F & 0.978 $\pm$ 0.036 & \textcolor{gray}{0.978 $\pm$ 0.036} & -- & -- & -- \\
red ($1/f^{3}$) noise & \code{T0\_RED} & ext & 0 & 0.75 & -- & -- & F & 1.013 $\pm$ 0.036 & \textcolor{gray}{1.013 $\pm$ 0.036} & -- & -- & -- \\
red ($1/f^{3}$) noise & \code{T0\_RED} & com & 0 & 0.32 & -- & -- & F & 0.944 $\pm$ 0.039 & \textcolor{gray}{0.944 $\pm$ 0.039} & -- & -- & -- \\
fixed-direction anisotropy & \code{T0\_ANISO} & ext & 0 & 0.81 & -- & -- & F & 1.001 $\pm$ 0.035 & \textcolor{gray}{1.001 $\pm$ 0.035} & -- & -- & -- \\
fixed-direction anisotropy & \code{T0\_ANISO} & com & 0 & 0.34 & -- & -- & F & 1.037 $\pm$ 0.039 & \textcolor{gray}{1.037 $\pm$ 0.039} & -- & -- & -- \\
spectral-tilt mixture & \code{T1\_PSRAND} & ext & 0 & 1.28 & 11.20 $\pm$ 2.41 & -- & F & 0.839 $\pm$ 0.119 & 4.635 $\pm$ 3.993 & -- & -- & -- \\
spectral-tilt mixture & \code{T1\_PSRAND} & com & 0 & 0.38 & 1.60 $\pm$ 0.07 & -- & F & 0.949 $\pm$ 0.055 & 1.459 $\pm$ 0.128 & -- & -- & -- \\
slope-only tilt mixture & \code{T1\_PSRAND\_SLOPE} & ext & 0 & 1.22 & 12.69 $\pm$ 4.07 & -- & -- & -- & -- & -- & -- & -- \\
slope-only tilt mixture & \code{T1\_PSRAND\_SLOPE} & com & 0 & 0.37 & 1.56 $\pm$ 0.07 & -- & -- & -- & -- & -- & -- & -- \\
amplitude-only mixture & \code{T1\_PSRAND\_AMP} & ext & 0 & 0.76 & 1.14 $\pm$ 0.05 & -- & -- & -- & -- & -- & -- & -- \\
amplitude-only mixture & \code{T1\_PSRAND\_AMP} & com & 0 & 0.32 & 1.13 $\pm$ 0.05 & -- & -- & -- & -- & -- & -- & -- \\
random-orientation anisotropy & \code{T1\_RANDOR} & ext & 0 & 0.91 & 1.28 $\pm$ 0.05 & -- & F & 0.966 $\pm$ 0.036 & 1.079 $\pm$ 0.046 & -- & -- & -- \\
random-orientation anisotropy & \code{T1\_RANDOR} & com & 0 & 1.20 & 9.64 $\pm$ 0.35 & -- & F & 0.905 $\pm$ 0.035 & 6.450 $\pm$ 0.313 & -- & -- & -- \\
row-median residual & \code{T2\_MEDIAN} & ext & 0 & 0.77 & -- & -- & F & 0.998 $\pm$ 0.035 & \textcolor{gray}{0.998 $\pm$ 0.035} & 0.992 $\pm$ 0.036$^{\times}$ & 1.222 $\pm$ 0.062$^{\times}$ & 2.089 $\pm$ 0.100$^{\times}$ \\
row-median residual & \code{T2\_MEDIAN} & com & 0 & 0.34 & -- & -- & F & 0.972 $\pm$ 0.042 & \textcolor{gray}{0.972 $\pm$ 0.042} & \textcolor{gray}{0.972 $\pm$ 0.042} & \textcolor{gray}{0.972 $\pm$ 0.042} & 0.965 $\pm$ 0.042 \\
scan crossings (sym.) & \code{T2\_CROSS\_SYM} & ext & 0 & 0.78 & -- & 1.07 & F & 1.011 $\pm$ 0.032 & \textcolor{gray}{1.011 $\pm$ 0.032} & \textcolor{gray}{1.011 $\pm$ 0.032} & 1.009 $\pm$ 0.031 & 1.142 $\pm$ 0.057$^{\times}$ \\
scan crossings (sym.) & \code{T2\_CROSS\_SYM} & com & 0 & 0.37 & -- & 4.58 & F & 0.956 $\pm$ 0.044 & \textcolor{gray}{0.956 $\pm$ 0.044} & 0.940 $\pm$ 0.045 & \textcolor{gray}{0.956 $\pm$ 0.044} & 0.927 $\pm$ 0.046 \\
scan crossings (pos.) & \code{T2\_CROSS\_POS} & ext & 0 & 0.75 & -- & 1.01 & F & 1.005 $\pm$ 0.036 & \textcolor{gray}{1.005 $\pm$ 0.036} & 0.999 $\pm$ 0.037 & 0.988 $\pm$ 0.037 & 0.982 $\pm$ 0.039 \\
scan crossings (pos.) & \code{T2\_CROSS\_POS} & com & 0 & 0.36 & -- & 4.41 & F & 0.993 $\pm$ 0.039 & \textcolor{gray}{0.993 $\pm$ 0.039} & \textcolor{gray}{0.993 $\pm$ 0.039} & 0.977 $\pm$ 0.040 & 0.969 $\pm$ 0.041 \\
sub-threshold glitches & \code{T2\_GLITCH} & ext & 0 & 0.84 & -- & 1.06 & F & 0.964 $\pm$ 0.051 & \textcolor{gray}{0.964 $\pm$ 0.051} & 0.964 $\pm$ 0.051 & 0.898 $\pm$ 0.027 & 0.959 $\pm$ 0.051 \\
sub-threshold glitches & \code{T2\_GLITCH} & com & 0 & 2.23 & -- & 144 & F & 1.213 $\pm$ 0.336 & 1.218 $\pm$ 0.640 & 1.206 $\pm$ 0.260 & 1.209 $\pm$ 0.251 & \textcolor{gray}{1.213 $\pm$ 0.336} \\
PCA leakage & \code{T2\_PCA} & ext & 0 & 1.67 & 2.07 $\pm$ 0.05 & 4.96 & F & 0.844 $\pm$ 0.167 & 1.631 $\pm$ 0.068 & 0.922 $\pm$ 0.124$^{\times}$ & 0.787 $\pm$ 0.194$^{\times}$ & 0.451 $\pm$ 0.961$^{\times}$ \\
PCA leakage & \code{T2\_PCA} & com & 0 & 0.34 & 1.02 $\pm$ 0.02 & 3.87 & F & 0.941 $\pm$ 0.046 & -0.266 $\pm$ 1.134$^{\dagger}$ & 0.642 $\pm$ 0.320 & 0.663 $\pm$ 0.286 & 0.801 $\pm$ 0.157 \\
source confusion & \code{T2\_CONFUSION} & ext & 0 & 5.69 mJy/b & -- & -- & F & 1.030 $\pm$ 0.020 & \textcolor{gray}{1.030 $\pm$ 0.020} & 1.492 $\pm$ 0.029 & 1.824 $\pm$ 0.056 & 91.370 $\pm$ 3.524 \\
source confusion & \code{T2\_CONFUSION} & com & 0 & 49.62 mJy/b & -- & -- & F & 0.991 $\pm$ 0.126 & 0.819 $\pm$ 0.220 & 0.991 $\pm$ 0.127 & \textcolor{gray}{0.991 $\pm$ 0.126} & \textcolor{gray}{0.991 $\pm$ 0.126} \\
\bottomrule
\end{tabular}

\end{sidewaystable}

\begin{table}[h]
  \caption{Descriptive names used in the text and the corresponding registry code names.}
  \label{tab:codenames}
  \centering\small
  \begin{tabular}{ll}
\toprule
descriptive name & registry code names \\
\midrule
white noise & \code{T0\_WHITE} \\
red ($1/f^{3}$) noise & \code{T0\_RED} \\
red + white noise & \code{T0\_RED\_REAL} \\
fixed-direction anisotropy & \code{T0\_ANISO} \\
spectral-tilt mixture & \code{T1\_PSRAND}, \code{T1\_PSRAND\_WN} \\
slope-only tilt mixture & \code{T1\_PSRAND\_SLOPE}, \code{T1\_PSRAND\_SLOPE\_WN} \\
amplitude-only mixture & \code{T1\_PSRAND\_AMP} \\
random-orientation anisotropy & \code{T1\_RANDOR}, \code{T1\_RANDOR\_WN} \\
row-median residual & \code{T2\_MEDIAN}, \code{T2\_MEDIAN\_WN} \\
scan crossings (sym.) & \code{T2\_CROSS\_SYM}, \code{T2\_CROSS\_SYM\_WN} \\
scan crossings (pos.) & \code{T2\_CROSS\_POS}, \code{T2\_CROSS\_POS\_WN} \\
sub-threshold glitches & \code{T2\_GLITCH}, \code{T2\_GLITCH\_WN} \\
PCA leakage & \code{T2\_PCA}, \code{T2\_PCA\_WN} \\
source confusion & \code{T2\_CONFUSION}, \code{T2\_CONFUSION\_WN} \\
beam-sharing companion & \code{T2\_CONFUSION\_ATM} \\
\bottomrule
\end{tabular}

\end{table}

\clearpage   
\bibliography{references_arxiv}{}
\bibliographystyle{aasjournalv7.1}

\end{document}